\documentclass[letterpaper]{JHEP3}
\pdfoutput=1
\usepackage{graphics}
\usepackage{epsfig,amsmath,euscript}

\newcommand{\vAi}{{\cal A}_{i_1\cdots i_n}} \newcommand{\vAim}{{\cal
A}_{i_1\cdots i_{n-1}}} \newcommand{\vAbi}{\bar{\cal A}^{i_1\cdots i_n}}
\newcommand{\vAbim}{\bar{\cal A}^{i_1\cdots i_{n-1}}}
\newcommand{\htS}{\hat{S}} \newcommand{\htR}{\hat{R}}
\newcommand{\htI}{\hat{I}}
\newcommand{\htB}{\hat{B}} \newcommand{\htD}{\hat{D}}
\newcommand{\htV}{\hat{V}}    
  \newcommand{{\vp}}{{\vec
p}} \newcommand{{\vq}}{{\vec q}} 

\newcommand{\tr}{{{\rm Tr}}} \newcommand{\beq}{\begin{equation}}
\newcommand{\eeq}[1]{\label{#1} \end{equation}} \newcommand{\half}{{\textstyle
\frac{1}{2}}} 
\newcommand{\lton}{\mathrel{\lower.9ex \hbox{$\stackrel{\displaystyle
<}{\sim}$}}} \newcommand{\ee}{\end{equation}}
\newcommand{\ben}{\begin{enumerate}} \newcommand{\een}{\end{enumerate}}
\newcommand{\bit}{\begin{itemize}} \newcommand{\eit}{\end{itemize}}
\newcommand{\bc}{\begin{center}} \newcommand{\ec}{\end{center}}
\newcommand{\bea}{\begin{eqnarray}} \newcommand{\eea}{\end{eqnarray}}
\newcommand{\beqar}{\begin{eqnarray}} \newcommand{\eeqar}[1]{\label{#1}
\end{eqnarray}} \newcommand{\bra}[1]{\langle {#1}|}
\newcommand{\ket}[1]{|{#1}\rangle}

\newcommand{\mcdot}{\!\cdot\!}

\newcommand{\e}{\mathrm{e}}

\newcommand{\eq}[1]{Eq.~\eqref{#1}}

\newcommand{\vc}[1]{\boldsymbol{#1}}

\def\bm#1{\mbox{\boldmath$#1$\unboldmath}}
\def\eps{\varepsilon}

\def\SCETG{${\rm SCET}_{\rm G}\,$}

\def\nslash{n\!\!\!\slash}
\def\bnslash{{\bar{n}}\!\!\!\slash}

\def\Dslash{D\!\!\!\!\slash} 

\def\pslash{p\!\!\!\slash}

\def\bm#1{\mbox{\boldmath$#1$\unboldmath}}
\def\eps{\varepsilon}

\def\SCETG{${\rm SCET}_{\rm G}\,$}

\def\nslash{n\!\!\!\slash}
\def\bnslash{{\bar{n}}\!\!\!\slash}

\def\Dslash{D\!\!\!\!\slash} 

\def\pslash{p\!\!\!\slash}
\def\qslash{q\!\!\!\slash}
\def\kslash{k\!\!\!\slash}

\title{An effective theory for jet propagation in dense QCD matter: jet broadening and 
medium-induced bremsstrahlung}

\author{ Grigory Ovanesyan, Ivan Vitev \\
Los Alamos National Laboratory, Theoretical Division, MS B283, Los Alamos, NM 87545, U.S.A.\\
E-mail: \email{ovanesyan@lanl.gov}, \email{ivitev@lanl.gov} }

\abstract{ Two effects, jet broadening and gluon bremsstrahlung induced by the propagation 
of a highly energetic quark in dense QCD matter, are 
reconsidered from effective theory point of view. We modify the standard Soft Collinear 
Effective Theory (SCET) Lagrangian to include Glauber modes, which are needed to implement 
the interactions between the medium and the collinear fields. We derive the Feynman
rules for this Lagrangian and show that it is invariant under soft and collinear gauge 
transformations. We find that the newly constructed theory SCET$_{\rm G}$  recovers 
exactly the general result for the transverse momentum broadening of jets. In the limit 
where the radiated gluons are significantly less energetic than the parent quark, we 
obtain a jet energy-loss kernel identical to the one discussed in the reaction operator 
approach to parton propagation in matter. In the framework of SCET$_{\rm G}$ we present 
results for the fully-differential bremsstrahlung spectrum for both the incoherent and
the Landau-Pomeranchunk-Migdal suppressed regimes beyond the soft-gluon approximation. 
Gauge invariance of the physics results is  demonstrated  explicitly by performing the 
calculations in both the light-cone and covariant  $R_{\xi}$ gauges. We also show how 
the process-dependent medium-induced radiative corrections factorize from the jet production 
cross section on the example of the quark jets considered here. }

\begin{document}
\section{Introduction}

The start up of the Large Hadron Collider (LHC) has reinvigorated strong theoretical interest in the 
physics of energetic jets~\cite{Sterman:1977wj,Feynman:1978dt}. At present, many of the new 
developments in the Quantum Chromodynamics (QCD) theory of hard jet and particle production are 
motivated by the need for precise evaluation of the Standard Model background to signature 
processes for new physics~\cite{Ellis:2007zzc,Aad:2009wy,Ball:2007zza}. Such accuracy relies
on factorization theorems~\cite{Collins:1981ta, Collins:1989gx} that allow to separate the perturbative 
hard scattering part of the cross section from the non-perturbative parton distribution and 
fragmentation  functions.

Recently, Soft Collinear Effective Theory (SCET)~\cite{Bauer:2000ew, Bauer:2000yr, Bauer:2001ct, Bauer:2001yt}
has emerged as a powerful new tool to address large $Q^2$ processes in lepton-lepton ($\ell^+ + \ell^-$),
lepton-hadron ($\ell^+ + p$) and hadron-hadron ($p+p$ or $p+\bar{p}$) collisions.  
SCET in conjunction with QCD factorization, which has been proven in this framework for a number of
processes~\cite{Bauer:2002nz,Manohar:2003vb, Becher:2006mr,Fleming:2007xt,Bauer:2008dt,Becher:2009th,Mantry:2009qz,Stewart:2009yx},  is particularly suited to improving the precision of multi-scale 
calculations\footnote{We refer here to energy, momentum or mass scales.} through the resummation of
large Sudakov type logarithms~\cite{Becher:2009th,Fleming:2007qr,Becher:2007ty,Becher:2008cf,Ahrens:2008nc,Hornig:2009vb}.

An important  multi-scale problem is presented by the production of jets in reactions with 
large nuclei, such as  lepton-nucleus  ($\ell+A$), proton-nucleus ($p+A$) and nucleus-nucleus ($A+A$)
reactions. In these processes the energetic quarks and gluons must traverse a region of dense nuclear 
matter of ${\cal O}(5\; {\rm fm})$ and their interactions in the medium induce broadening and  
a new type of radiative corrections that can significantly alter the corresponding jet 
cross sections and 
shapes~\cite{Vitev:2008rz,Vitev:2009rd,Renk:2009hv,Neufeld:2010fj}. Preliminary results from the 
Relativistic Heavy Ion Collider (RHIC) and new results from the LHC indicate that these effects 
may indeed be observable in
A+A collisions~\cite{Salur:2009vz,Lai:2009zq,Ploskon:2009zd,Aamodt:2010jd,Chatrchyan:2011sx,Aad:2010bu}. 
It will be natural to use SCET to
describe  the collisional and radiative interactions of the jet in the medium when 
the typical transverse momentum of the partons in the jet is comparable to the size of the momentum 
exchange with the medium and much smaller than the jet energy. So far, only broadening of the 
final-state parton in semi-inclusive deeply inelastic scattering (SDIS)~\cite{Idilbi:2008vm} has been 
considered in an effective theory of QCD. Reference~\cite{DEramo:2010ak} argues to present a 
less model-dependent result but its gauge invariance remains to be demonstrated explicitly.

To set up a general SCET framework that can describe the collisional and radiative processes 
induced by the propagation of an energetic parton in strongly interacting matter, to derive the 
operators that describe the momentum space evolution of the propagating quark or quark+gluon system 
and to demonstrate the gauge invariance of these results is the main goal of this paper. By comparing
our findings to previous calculations of parton broadening~\cite{Gyulassy:2002yv,Qiu:2003pm} and energy  
loss~\cite{Baier:2000mf,Gyulassy:2000er,Wang:2001ifa,Arnold:2002ja} in nuclear matter we 
identify areas where the effective theory calculation will be able to improve the accuracy of existing 
computations.

Our manuscript is organized as follows: in section~\ref{scet} we review very briefly the basic 
concepts  of Soft Collinear Effective Theory. The kinematics  of jet-medium interactions beyond the static 
scattering approximation is discussed in section~\ref{sec:kinematics}. We identify the 
regime relevant to high energy jet production in hadronic reactions with large 
nuclei and elucidate the possibility for constructing an effective
theory to describe parton propagation in dense QCD matter. The  gauge-invariant 
Lagrangian for this effective theory is constructed in section~\ref{scetg}. Feynman rules are derived
in the light-cone and covariant $R_{\xi}$ gauges. In section~\ref{collisional} we evaluate the 
transverse momentum broadening of jets, induced by their collisional interactions in the 
strongly interacting medium. Radiative processes are discussed in section~\ref{radiative}. 
Our focus here is on the soft gluon limit, when the energy  of the emitted gluon ($\omega$) is much less than the energy ($E$) of the quark that splits: $\omega \ll E$, for comparison to previous results. 
We demonstrate the gauge invariance of the jet broadening and energy loss results in 
section~\ref{invariant}.  We deduce the kernels that describe the broadening and medium-induced 
bremsstrahlung as a function of the quark interactions in the medium in section~\ref{sec:ROP}.
An application of the reaction operators for collisional and radiative processes is also 
discussed in this section.
The extension of  radiative energy loss calculation beyond soft gluon 
approximation is presented  in section~\ref{bg}. We also show how the process-dependent medium-induced radiative 
corrections factorize from the hard jet production cross section. 
Our conclusions are given in section~\ref{conclude}. We have  moved some of the background  technical
discussion to appendices.

\section{A brief  overview of SCET}
\label{scet}
SCET~\cite{Bauer:2000ew, Bauer:2000yr, Bauer:2001ct, Bauer:2001yt} is an effective theory of 
QCD which describes the dynamics of highly energetic quarks and gluons. The relevant physical 
scales in this effective theory are the hard scale $E_h\sim E_T \sim E_{\text{cm}}$, the jet scale 
$E_j\sim p_{\perp}$ that describes the width of the jet in momentum space and the scale of 
soft radiation $E_s\sim \Lambda_{\text{QCD}}$. The power counting parameter of SCET $\lambda$ defines the hierarchy between the hard, jet and soft scales. We use the version of SCET, which is sometimes referred to as $\text{SCET}_{\text{I}}$, in which the scales are $E_h\sim \lambda^0$, $E_j\sim \lambda^1$ and 
$E_s\sim \lambda^2$. The degrees of freedom in SCET are 
collinear quarks ($\xi_{n,p}$), collinear gluons ($A_{n,p}$) 
and soft gluons ($A_s$). Collinear particles have momentum in light-cone coordinates $p_c \sim [1,\lambda^2,\vc{\lambda}]$ 
and soft particles $p_s \sim [\lambda^2, \lambda^2, \vc{\lambda}^2]$, where we define our light-cone notation in appendix~\ref{appendix:lightcone}. All other fields, such as hard quarks and gluons,
are integrated out from the QCD Lagrangian. Their effect on the dynamics is contained into the Wilson 
coefficients of the SCET operators, which can be calculated using a standard matching of 
full theory onto effective theory. In order to avoid confusion we note that what we call  soft gluon mode in this paper $p_s^{\mu}\sim \lambda^2$ is sometimes  called ultrasoft, while the soft momentum is defined as $p_s^{\mu}\sim \lambda$, see for example \cite{Bauer:2001yt}. However, below in section~\ref{scetg} when we define
the momentum scaling of the source, one of our choices corresponds to $p_s^{\mu}\sim \lambda$ and we call it a soft source.

The Lagrangian of SCET~\cite{Bauer:2000yr} arises from substituting into the QCD Lagrangian 
$\psi=\sum_{\tilde{p}}\,e^{-i\tilde{p}x}\,\psi_{n,\tilde{p}}$ and integrating out the small 
component $\xi_{\bar{n}}$ of $\psi_{n}$, where $ \xi_n=\frac{\nslash\bnslash}{4}\psi_n,\; \xi_{\bar{n}}=\frac{\bnslash\nslash}{4}\psi_n$ 
and $ \psi_n=\xi_n+\xi_{\bar{n}}$. The result for the collinear-soft Lagrangian is:
\begin{eqnarray}
&& \mathcal{L}_{\text{SCET}}(\xi_n, A_{n}, A_s)=\bar{\xi}_{n}
\left[i n\mcdot{D}+i{\Dslash^{\perp}}\frac{1}{i\bar{n}\mcdot D}\, 
i{\Dslash^{\perp}}\right]\frac{\bnslash}{2}\xi_n+\mathcal{L}_{\text{YM}}(A_n, A_s)\; ,\label{SCETL1}\\
&&\mathcal{L}_{\text{YM}}(A_n, A_s)=\frac{1}{2g^2}\text{tr}
\left\{\left[iD^{\mu}_s+gA^{\mu}_{n,q},iD^{\nu}_s+gA^{\nu}_{n,q'}\right]\right\}^2+\mathcal{L}_{\text{G.F.}}\; ,\label{SCETL2}\\
&&\mathcal{L}_{\text{G.F.}}(R_{\xi})=\frac{1}{\xi}\text{tr}\left\{\left[iD_{s\mu},A^{\mu}_{n,q}\right]\right\}^2,\label{SCETL3}\\
&&\mathcal{L}_{\text{G.F.}}(\text{LCG}(b))=\frac{1}{\xi}\text{tr}\left\{b_{\mu}A^{\mu}_{n,q}\right\}^2\; .\label{SCETL4}
\end{eqnarray}
Here, the covariant derivative $D$ contains both collinear and soft fields: 
$i D=i\partial+g \left(A_{n}+A_s\right)$, while $D_s$ includes only the soft gluons: 
$i D_s=i\partial+g A_s$. Thus, the collinear and soft modes are coupled in the SCET Lagrangian. In the first term of~\eq{SCETL2} the summation over label momenta $q, q'$ is understood implicitly, and in~\eq{SCETL3},~\eq{SCETL4} summation over the label momentum $q$ is understood implicitly.
We have written out explicitly the gauge fixing  terms for the covariant and the light-cone gauges. 
The ghost terms are omitted for brevity.

A key ingredient of the SCET formulation is the BPS transformation~\cite{Bauer:2001yt}. This transformation 
constitutes a  collinear field redefinition which involves soft Wilson lines and removes the interactions 
between soft and collinear fields in the Lagrangian of SCET up to the power corrections. Such decoupling 
is essential in the proof of  factorization theorems  in SCET.
The BPS transformation redefines the collinear quark and gluon fields:
\begin{eqnarray}
   \xi_{n, p}=Y\,\xi_{n,p}^{(0)}\; ,\\
   A^{a,\mu}=\mathcal{Y}^{ab}\,A_{n,p}^{(0)\, \mu, b}\; ,
\end{eqnarray}
where the $Y(x)$ and $\mathcal{Y}^{ab}(x)$ are Wilson lines built out of the soft fields in the 
fundamental and adjoint representations correspondingly:
\begin{eqnarray}
  &&Y(x)=\text{P}\,\text{exp}\left(ig\int_{-\infty}^0 \text{d}s\, n\mcdot A^a_{s}(ns+x)\,T^a\right)\; ,\\
  &&\mathcal{Y}^{ab}(x)=\text{P}\,\text{exp}\left(ig\int_{-\infty}^0 \text{d}s\, n\mcdot A^e_{s}(ns+x)\,\left(-i f^{eab}\right)\right)\; .
\end{eqnarray}

To derive the Lagrangian of SCET in terms of the decoupled collinear fields 
$\xi_{n}^{(0)}, A_{n}^{(0)}$  one needs the following key identities:
\begin{eqnarray}
  &&YT^a Y^{\dagger}=\mathcal{Y}^{ba} T^b \, ,\\
  &&Y^{\dagger}n\mcdot D Y=n\mcdot D_n \, .\label{eqdec}
\end{eqnarray}
In particular, the last equation removes the interactions between the soft gluons and the 
collinear quarks which is contained in the covariant derivative in the left hand side of the~\eq{eqdec}.
As a result, after the BPS transformation we obtain:
\begin{eqnarray}
\mathcal{L}_{\text{SCET}}\left(\xi_n, A_n, A_s\right)=\mathcal{L}_c\left(\xi_n, A_n\right)
+\mathcal{L}_{s}\left(A_s\right)+\mathcal{L}_{cs}\left(\xi_n, A_n, A_s\right)
\xrightarrow[]{\text{BPS}}\mathcal{L}_{c}(\xi_n^{(0)}, A_n^{(0)})+\mathcal{L}_{s}(A_s) \, .
\end{eqnarray}
The couplings between the collinear and soft modes is removed  from the SCET Lagrangian. However, in 
order to preserve the gauge invariance one has to put soft and collinear Wilson lines into the 
external SCET operators. The collinear Wilson line is defined in the following way:
\begin{eqnarray}
W_n(x)=\text{P}\,\text{exp}\left(-ig\int_{0}^{\infty} \text{d}s\, \bar{n}\mcdot A^a_{n}(\bar{n}s+x)\,T^a\right).\label{Wndef}
\end{eqnarray}

\section{Kinematics of  the in-medium jet interactions}
\label{sec:kinematics}

In this section we describe the kinematics of jet interactions in QCD matter. Our goal is to identify the typical
momentum exchanges between the energetic incident partons and the medium, which in turn will help us construct 
an effective theory for these interactions. 

The typical jet transverse energies that are phenomenologically interesting at RHIC are in the range  
$E_T \sim 10$~GeV -  $50\;\text{GeV}$.  At the LHC this range is extended to $E_T \sim$~several hundred GeV. On the other hand, 
the mass of the particles in the medium is on the order of or larger than 1~GeV. In cold nuclear matter this is the mass of the
nucleon $m_N = 0.94$~GeV  and in the quark-gluon plasma this is the mass of  dressed partons 
 $ \propto \mu = g T \sqrt{1+N_f/6}$. Binding effects inside nuclear matter can significantly increase the effective mass of the recoiling particles. 
 One of the main goals of this 
section is to study how the elastic scattering cross section depends on this mass and to identify the kinematic
configurations that give a dominant contribution to this cross section.

The cross section for the elastic scattering of two  particles of masses $m_1$, $m_2$ is given by the 
standard expression:
 \begin{eqnarray}
  {d\sigma}= \frac{1}{4\sqrt{(p_1 \cdot p_2)^2 -m_1^2 m_2^2}}  \langle |M|^2  \rangle   
\frac{d^3p_3}{(2\pi)^3 2 E_3}\frac{d^3 p_4}{(2\pi)^3 2 E_4} (2\pi)^4 \delta^4(p_1+p_2-p_3-p_4) \; .
\label{scat}
 \end{eqnarray}
We work in the rest frame of the medium, which for the  cases of a quark-gluon plasma at mid-rapidity 
coincides  with the laboratory 
frame. The same is true for fixed target p+A experiments. In this frame the flux factor conveniently 
reduces to  ${\rm Flux} =  4 m_2 p_{1}$. 
Let us denote by $\theta$ the scattering angle of the incident parton ($\theta = \angle (p_1,p_3)$)  
and by $p_3$ the physical solution arising from the energy constraint
$E_1(p_1,m_1)+m_2 = E_3 (p_3,m_1) + E_4(\vec{p}_4={\vec p}_1-{\vec p}_3,m_2) $. In this paper we are
interested in the case: $m_1<m_2$. In this case there is a single physical solution $p_3$. 
Substituting in Eq.~(\ref{scat}), we obtain:
\begin{equation}
  \frac{d\sigma}{d\Omega}  = \frac{1}{64 \pi^2}  \frac{p_3^2 }{p_1 m_2  
\left[ p_3(E_1+m_2) - p_1 E_3 \cos(\theta)  \right] } 
\langle |M|^2  \rangle  \, . 
\label{scat1}
\end{equation}

We now discuss the  specific channels that may contribute to  $\langle |M|^2  \rangle$. To study jet broadening 
and energy loss, we will be interested in forward scattering where, on average, the direction of jet 
propagation is
not significantly altered per interaction. Such scattering is dominated by the $t$-channel gluon 
exchange. One
can, of course, write down $u$-channel and $s$-channel diagrams  but these describe hard backward parton 
scattering and isotropisation processes rather than transverse momentum broadening and energy loss. For 
simplicity,  we give the specific example of quark-quark scattering:   
\begin{equation}
\langle |M|^2  \rangle =  g^4 \left( \frac{2}{9} \right)_{\rm color}  \frac{2(u^2+s^2) 
+ 4 m_1^2(2t-2 m_2^2 - m_1^2)  + 4 m_2^2(2t-2 m_1^2 - m_2^2) + 8 m_1^2 m_2^2  }{t^2} \, .
\label{mass}
\end{equation}
Here,  one can  conveniently take the $m_1\rightarrow 0,\, m_2\rightarrow 0$ limits.  We note that if the
interaction is of finite range, i.e. we have an exchange gluon of mass $m \sim \mu$, the only change is
$ t \rightarrow t-\mu^2$ in the denominator of Eq.~(\ref{mass}).

\begin{figure}
\centerline{ \includegraphics[width=0.8\textwidth]{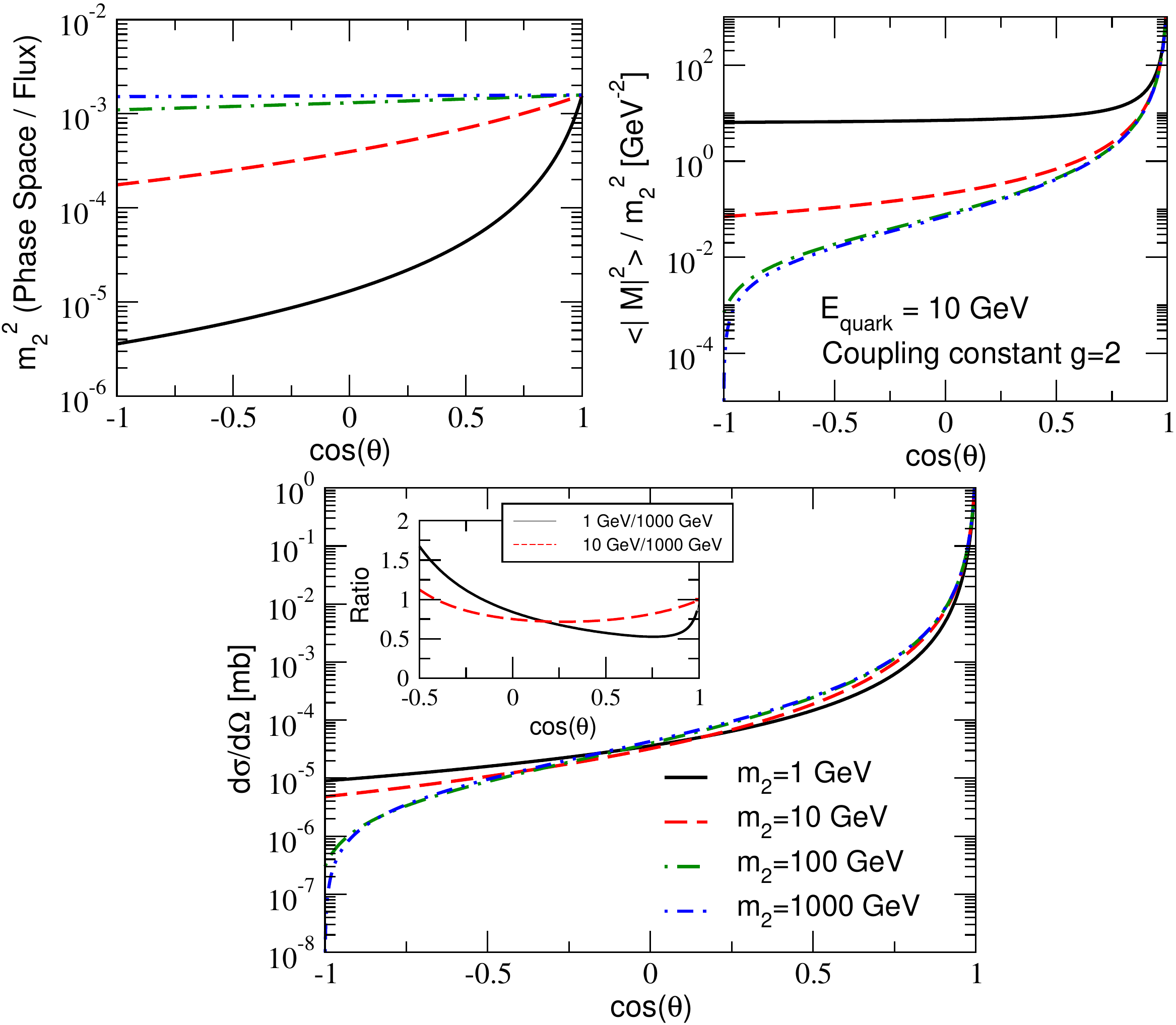}  }
\caption{Kinematic factors and mean squared matrix element for the $t$-channel scattering of interest 
are shown versus 
$\cos(\theta)$ in the top left and top right panels, respectively. These exhibit strong dependence on
the mass $m_2$ of the recoiling particle. The bottom panel shows the differential cross section where
this $m_2$ dependence cancels over most of the $\theta$ range. The insert shows any residual differences
as a ratio of $d\sigma/d\Omega$ for 2 different values of $m_2$. }     
 \label{fig1}
\end{figure}

We can now study the $m_2$ dependence of the average squared matrix element $\langle |M|^2  \rangle$, 
the flux and phase space factors (PS, Flux),  and the differential cross section $d\sigma/d\Omega$. We chose 
a massless incident quark  of energy 10 GeV. The mass of the recoiling particle varies 
from $1$~GeV to $10^3$~GeV and we use $g=2$ for the coupling constant. The top left and right panels 
in figure~\ref{fig1} show 
the differential phase space (normalized by the initial flux factor) and matrix element  versus $\cos(\theta)$ with the leading 
$m_2^2$ dependencies  taken out. While for $\cos(\theta)=1$ the curves come together, the difference
for small values of $m_2$ away from the forward region can be many orders of magnitude. In contrast, 
in the differential cross 
section $d\sigma/d\Omega$ this variation largely cancels everywhere except for the backward 
scattering region. We show this result in the bottom panel of figure~\ref{fig1}. The insert illustrates 
the  remaining subtle mass dependence in the forward scattering region. It is quite remarkable that this
residual variation is less than $\pm 50\%$. At the level of the integrated scattering cross section,
the differences are even much smaller. For incident partons of $E = 10$~GeV at RHIC and $E = 100$~GeV at the LHC  
we obtain:
 $$ \frac{\sigma (m_2 = 1000\; {\rm GeV}) - \sigma (m_2 = 1\; {\rm GeV}) }  
{\sigma (m_2 = 1000\; {\rm GeV})} |_{\rm RHIC}  \approx 13 \%    \, ,  \quad  
\frac{\sigma (m_2 = 1000\; {\rm GeV}) - \sigma (m_2 = 1 GeV) }  
{\sigma (m_2 = 1000\; {\rm GeV}) }|_{\rm LHC}   \approx 2 \%  \; , $$ respectively. Note that the scattering cross 
sections decreases for 
finite and small mass of the recoiling particle. This, in turn, leads to larger mean free paths 
$\lambda=1/\sigma \rho$  in QCD matter  and  
smaller radiative energy loss. This is in contrast to the result of Ref.~\cite{Djordjevic:2009cr}. 
The reason for this difference is that in~\cite{Djordjevic:2009cr} the general term ``dynamical medium'' 
was inaccurately used  to describe a specific hard thermal loop approximation and the reported increase 
arises from the lack of magnetic 
screening. Lattice QCD results and non-perturbative arguments, however, suggest that magnetic screening
effects may be present already at ${\cal O}(g^2T)$. Our results for a general finite-range interaction 
mediated by a massive vector particle allow to precisely quantify the effect of the medium recoil.  
For example, we found that more than 90\% of the cross section comes from configurations where the jet 
is not deflected more than 15\% from its original direction of propagation.

A small difference in the magnitude of the scattering cross section exists between the full calculation
and the often used  analytic approximation  $\sigma = 8\pi \alpha_s^2 / 9 \mu^2 $ to quark-quark scattering. 
We can re-express the exact differential cross section as a function of the transverse momentum transfer
as follows:
\begin{eqnarray}   
&&\frac{d\sigma}{d\Omega} \rightarrow \frac{d\sigma}{d^2{\bf q}_\perp} = 
\frac{C_2(R)C_2(T)}{d_A} \frac{|v({\bf q}_\perp ;E,m_1,m_2 )|^2}{(2\pi)^2} \;. 
\label{transform}
\end{eqnarray}   
In Eq.~(\ref{transform}) $C_2(R)$, $C_2(T)$ are the quadratic Casimirs for the incident and target parton 
representations, respectively, and  $d_A = 8$ is the dimension of the adjoint representation. 
The above expression also defines $v({\bf q};E,m_1,m_2 )$, which now depends on the jet energy and 
the masses of the scattering particles. It is easy to check that such a definition reduces to $v({\bf q};E,0,\infty )=\frac{4\pi\alpha_s}{\vc{q}^2+\mu^2}$, consistent with similar definition in  \cite{Gyulassy:2000er}. The only subtlety is that one allowed value of ${\bf q}_\perp$ 
generally corresponds to two values of $\cos(\theta)$. However, the region $-1 < \cos(\theta) < 0$
contributes $\sim 0.1\%$ to the cross section and we simply ignore the second solution.

Finally we discuss which momentum region for the exchange gluon gives the dominant contribution to the cross section. The momentum transfer in terms of the final and initial jet momentum equals $q=p_3-p_1$. Writing this in laboratory frame in terms of the light-cone components, we get:
\begin{eqnarray}
  &&q^{+}=E_3(1+\cos(\theta))-2E_1\, ,\qquad q^{-}=E_3(1-\cos(\theta))\, , 
\qquad |\vc{q}_{\perp}|=E_3\sin(\theta)\, .\label{leadingregionformula}
\end{eqnarray}
In the formula above we assumed that $m_1=0$ and $m_2$ is arbitrary. As we can clearly see from Figure \ref{fig1} the cross section is dominated in the forward direction. Thus we assign to the leading region for the cross section the following power counting: $\theta\sim \lambda$, $E_3, E_1\sim 1$. We immediately can see from \eq{leadingregionformula} that this power counting corresponds to the following leading momentum region, depending on the mass $m_2$\footnote{Such sensitivity of $q^+$ on the recoil mass arises from the fact that $E_3-E_1=m_2-\sqrt{m_2^2+\vc{q}^2}$, which is energy conservation equation. Thus, for heavy mass $m_2\sim 1$ the recoil energy is negligible $E_3-E_1\sim \lambda^2$ and for the lighter mass $m_2\sim \lambda$ it is comparable to the transverse momentum transfer: $E_3-E_1\sim \lambda$.}:
\begin{eqnarray}
  &&q\sim [\lambda^2, \lambda^2, \vc{\lambda}], \text{  if } m_2\sim \lambda^0, \text{  i.e. static source,} \label{scaling1}\\
  &&q\sim [\lambda^1, \lambda^2, \vc{\lambda}], \text{  if } m_2\sim \lambda^1, \text{  i.e. soft source}.\label{scaling2}
\end{eqnarray}


To summarize, we  investigated in detail the kinematics of jet-medium interactions to determine
the feasibility of an effective theory where these interactions are mediated by Glauber gluons. We showed that for the static source, which 
we use in sections  \ref{collisional}-\ref{bg} below, the cross section is indeed dominated by the Glauber momentum region.
We also calculated the exact dependence of this scattering cross section on the energy and mass of the
incident and recoiling particles. This cross section will be implemented to go beyond the static 
scattering center approximation. We note, however, that this approximation is remarkably good and
within 15\% of the exact result even for low energy jets.

\section{An effective theory for jet propagation in QCD matter: $\text{\SCETG}$}
\label{scetg}

As we have seen from the previous section, the effective theory of jet interactions in matter 
has to contain the Glauber mode, which  carries the exchange momentum  between the incident parton 
and the QCD medium with scaling $q \sim[\lambda^2, \lambda^2, {\bm \lambda}]$. Such  mode is 
absent in SCET and
we have to modify the theory by including it. One possibility was considered 
in~\cite{Idilbi:2008vm} and later on used in~\cite{DEramo:2010ak} 
to study the multiple collisional interactions of jets. There are a few differences between 
our approach and  these references. First of all, 
we write down the Glauber term directly in the momentum space as an effective potential~\cite{StewartTalk}, 
similar to the non-relativistic QCD (NRQCD) potential 
term~\cite{Luke:1999kz}. Secondly, we consider a static source of Glauber gluons, whereas in~\cite{Idilbi:2008vm} 
the source was a massless 
collinear field. We are motivated by the physical picture of nucleons or massive quasi-particles as sources of 
these Glauber gluons in 
nuclei and non-Abelian plasmas, respectively. We  work in the rest frame of nuclear matter and 
also include the collinear gluons into the interaction Lagrangian with Glauber gluons. 

In a different context, namely the Drell-Yan process, it has been shown that the Glauber mode has to be added to SCET 
for the consistency of 
the effective  theory~\cite{Bauer:2010cc}. Having formulated a consistent effective theory $\text{\SCETG}$, it would be 
interesting to revisit the Drell-Yan factorization in this effective theory and 
understand the cancellation of Glauber gluons in inclusive Drell-Yan cross section from the effective theory point 
of view. In traditional QCD this cancellation was derived  in 
Refs.~\cite{Collins:1982wa, Bodwin:1984hc, Collins:1985ue}, however it has not been addressed yet in effective 
theory methods to factorization.

\subsection{The $\text{\SCETG}$ Lagrangian for different sources and gauges}

Consider a quark or a  gluon  propagating in the positive light-cone direction $n$ in  QCD matter. In this subsection
we will derive the effective Lagrangian of $\text{\SCETG}$, which describes the interaction of our propagating 
jet with the source of the Glauber 
gluons. We consider three types of sources of Glauber gluons. The first one is a collinear field propagating in 
the $\bar{n}$ direction 
(considered in Refs \cite{Idilbi:2008vm}, \cite{DEramo:2010ak}).  The second one is a (initially) static nucleon/nucleus 
or a massive
quasi-particle. Interestingly, this source can be adequately described as a heavy quark effective theory 
(HQET) current. Finally, the third type of source 
that we consider is a soft parton of momentum $p\sim [\lambda,\lambda,{\bm \lambda}]$. 
For each of these three sources we consider three gauges: the $R_{\xi}$ gauge and two distinct light-cone gauges 
$A^+=0$ and $A^-=0$. 
In order to verify the symmetry properties of the Glauber interaction term we include the source fields into our 
Lagrangian. However, for practical calculations we integrate out the source fields as well as the Glauber 
gluons and present the resulting Feynman rules.

We start from the vector potential as a function of the QCD current of the source and the gluon propagator. 
Our method 
to derive the scaling for the vector potential $A_G^{\mu}(x)$ created by the Glauber gluons is same as 
in~\cite{Idilbi:2008vm}:
\begin{eqnarray}
  A_G^{\mu,a}(x)=\int d^{4}y \delta^{ab}D^{\mu\nu}(x-y)\, \bar{\psi}(y) gT^b \gamma_{\nu} \psi(y) \, .  \label{Amu0}
\end{eqnarray}
Expanding the propagator and the fermion field in the momentum space we get:
\begin{eqnarray}
  A_G^{\mu,a}(x)=\int \frac{d^3 \vc{p}}{\sqrt{2E_{\vc{p}}}(2\pi)^3}  \frac{d^3 \vc{q}}{\sqrt{2E_{\vc{p}+\vc{q}}}(2\pi)^3}\,
\sum_{s,r}\,a_{\vc{p}+\vc{q}}^{r\dagger} a_{\vc{p}}^s\,\bar{u}(\vc{p}+\vc{q},r) gT^a \gamma_{\nu}u(\vc{p},s)\,\e^{-iq x}\,(-i)\Delta^{\mu\nu}(q)+ \cdots\; ,\quad \label{Amu}
\end{eqnarray}
where we have written down only the part of the vector potential that contributes to particle-particle scattering, 
and the three remaining combinations involving anti-particles are omitted. In each of three gauges under consideration, 
the gluon propagator is equal to:
\begin{eqnarray}
\text{Covariant gauge }  &&\left[\Delta^{\mu\nu}(q)\right]_{R_{\xi}}=
\frac{\left(g^{\mu\nu}-\frac{q^{\mu}q^{\nu}}{\mu^2}(1-\xi)\right)}{q^2-\mu^2}\rightarrow \frac{g^{\mu\nu}}{q^2-\mu^2}\; ,
\label{GP1}\\
\text{Light-cone$(A^+=0)$ gauge} &&\left[\Delta^{\mu\nu}(q)\right]_{A^+}=
\frac{\left(g^{\mu\nu}-\frac{\bar{n}^\mu q^{\nu}+\bar{n}^{\nu}q^{\mu}}{q^+}\right)}{q^2-\mu^2}\rightarrow 
\frac{g^{\mu\nu}-\frac{\bar{n}^{\nu}q^{\mu}}{q^+}}{q^2-\mu^2}\; ,\label{GP2}\\
\text{Light-cone$(A^-=0)$ gauge} &&\left[\Delta^{\mu\nu}(q)\right]_{A^-}=
\frac{\left(g^{\mu\nu}-\frac{{n}^\mu q^{\nu}+{n}^{\nu}q^{\mu}}{q^-}\right)}{q^2-\mu^2}
\rightarrow \frac{g^{\mu\nu}-\frac{{n}^{\nu}q^{\mu}}{q^-}}{q^2-\mu^2}\; ,\label{GP3}
\end{eqnarray}
where the arrow in each line indicates that the term proportional to $q_{\nu}$ vanishes in~\eq{Amu} 
because the external source particles are on-shell: $q^{\nu} \bar{u}(p+q)\gamma_{\nu}u(p)$=0. 
In Eqs.~(\ref{GP1}) - (\ref{GP3}) we also use a finite range interaction mediated by a vector field 
of mass $\mu$.

Using~\eq{Amu} and Eqs.~(\ref{GP1}) - (\ref{GP3}) we obtain the scaling
formula for $A_G^{\mu}(x)$ for all cases.
The scaling of the creation operators is derived from the anti-commutation
relations in momentum space by allowing
the external momentum of the source particles in each of three cases under
consideration  to be correspondingly: collinear, soft\footnote{For the
static source the
external momentum equals $p=mv+k$, however $p_1-p_2=k_1-k_2$,
equals the difference of two soft momenta.} and soft. We
also note that the vector $p$ is collinear, while $q$ has Glauber
scaling for collinear and static sources, and $q$ has scaling given by \eq{scaling2} for the soft source\footnote{Throughout this section we refer to both scalings in \eq{scaling1} and \eq{scaling2} as Glauber gluons, however it should be clear from the context which one is used when.}. As a result we complete in table~\ref{tb1}  all entries for the
scaling of $A_G^{\mu}(x)$.

\begin{table}[!t]
\begin{center}
\begin{tabular}{|c|c||c||c||c||}
	\hline
Gauge& Object&  Collinear source    &   Static source  & Soft source     \\
	\hline \hline
&$p$&  $\left[\lambda^2, 1, {\bm \lambda}\right]$    &   $\left[1, 1, {\bm \lambda}\right]$  & $\left[\lambda, \lambda, {\bm \lambda}\right]$     \\
&$a_{\vc{p}}, a_{\vc{p}}^{\dagger}$& $\lambda^{-1}$ & $\lambda^{-3/2}$ & $\lambda^{-3/2}$   \\
  &$u(p)$& $1$ & $1$ & $\lambda^{1/2}$   \\
   &$\bar{u}(p_2)\gamma_{\nu}u(p_1)$& $\left[\lambda^2, 1, {\bm \lambda}\right] $ & $\left[1, 1, {\bm \lambda}\right]$  & $\left[\lambda, \lambda, {\bm \lambda}\right]$   \\
	\hline \hline
$R_{\xi}$ &$A^{\mu}(x)$& $\left[\lambda^4, \lambda^2, {\bm \lambda^3}\right]$ & $\left[\lambda^2, \lambda^2, {\bm \lambda^3}\right]$ & $\left[\lambda, \lambda, {\bm \lambda}\right]$   \\
	    & $\Gamma_{{\rm qqA_G}}$  & $\Gamma^{\mu}_{1}$ &$\Gamma^{\mu}_{1}$ & $\Gamma^{\mu}_{1}$ \\
		 & $\Gamma_{{\rm ggA_G}}$ & $\Sigma_1^{\mu\nu\lambda}$ & $\Sigma_1^{\mu\nu\lambda}$& $\Sigma_1^{\mu\nu\lambda}$   \\
		 		& $\Gamma_{\text{s}}$  & $\Gamma_1^{\mu}\left(n\leftrightarrow\bar{n}\right)$ &$\Gamma^{\mu}_{3}$& $\Gamma^{\mu}_{4}$   \\
	\hline\hline
$A^+=0$& $A^{\mu}(x)$  & $\left[0,\lambda^2, {\bm \lambda^3}\right]$ & $\left[0, \lambda^2, {\bm \lambda}\right]$ & $\left[0, \lambda, {\bm 1}\right]$\\
		 & $\Gamma_{{\rm qqA_G}}$   &$\Gamma^{\mu}_{1}$ & $\Gamma^{\mu}_1+\Gamma^{\mu}_2$  &  $\Gamma^{\mu}_{1}+\Gamma^{\mu}_2$    \\
		& $\Gamma_{{\rm ggA_G}}$  & $\Sigma_2^{\mu\nu\lambda}$ &$\Sigma_2^{\mu\nu\lambda}$& $\Sigma_2^{\mu\nu\lambda}$   \\
		& $\Gamma_{\text{s}}$  & $\Gamma_2^{\mu}\left(n\leftrightarrow\bar{n}\right)$ &$\Gamma^{\mu}_{3}$& $\Gamma^{\mu}_{4}$   \\
	\hline\hline
  $A^-=0$ &$A^{\mu}(x)$&  $\left[\lambda^2, 0, {\bm \lambda}\right]$ & $\left[\lambda^2, 0, {\bm \lambda}\right]$& $\left[\lambda, 0, {\bm 1}\right]$\\
		   & $\Gamma_{{\rm qqA_G}}$&  $\Gamma^{\mu}_{2}$ & $\Gamma^{\mu}_{2}$  &  $\Gamma^{\mu}_{2}$    \\
		&$\Gamma_{{\rm ggA_G}}$   & $\Sigma_3^{\mu\nu\lambda}$ &$\Sigma_3^{\mu\nu\lambda}$ & $\Sigma_3^{\mu\nu\lambda}$  \\
				& $\Gamma_{\text{s}}$  & $\Gamma_1^{\mu}\left(n\leftrightarrow\bar{n}\right)$ &$\Gamma^{\mu}_{3}$& $\Gamma^{\mu}_{4}$   \\
	\hline
\end{tabular}
\caption{Summary of the scaling behavior of the Glauber gluon source ingredients, the Glauber vector potential and the Feynman rules for the 
newly constructed theory ${\rm SCET_G}$ in the covariant $R_\xi$ gauge and two different light-cone gauges $A^+=0$, $A^-=0$. }
\end{center}
\label{tb1}
\end{table}

Having determined the scaling of the vector potential created by the Glauber field for the cases of interest,
we now derive the effective theory Feynman rules for the coupling of the energetic jet to the Glauber field. 
We read off the Feynman rules from the usual SCET Lagrangian 
Eqs.~(\ref{SCETL1}) - (\ref{SCETL4}) by treating the vector potential created by Glauber gluons in the covariant 
derivative as a background field. All Feynman graphs between collinear and Glauber gluons contained in~\eq{SCETL2}\footnote{Those rules include couplings 
of a Glauber gluon to two and three collinear gluons, and two Glaubers with two collinear gluons. 
We present in the table only the first vertex $\Gamma_{\rm ggA_G}$.} can be found using derived Feynman rules in the background field method from Ref.~\cite{Abbott:1981ke}. 
In addition to these rules, one has to apply the specific power counting of the vector potential, 
which we derived for each source and gauge. 
As a result, we fill in the table the vertices $\Gamma_{\rm qqA_G}, \Gamma_{\rm ggA_G}$ for each source and for each gauge. 
Finally, we include in table~\ref{tb1} the Feynman rules $\Gamma_{\text{s}}$ for the interaction between the source fields and the 
Glauber gluons. This is achieved by noting that we can  view the  jet moving in  the direction of $n$  as a source of 
Glauber gluons for the target fields $\bar{\eta}, \eta$. Thus, the Feynman rules $\Gamma_s$ can be found by using 
our scaling rules for $A_G^{\mu}(x)$ with the collinear source in the $n-$direction\footnote{Note that in the table we have derived the Feynman rules generated by the collinear source moving in the $\bar{n}$ direction. However, since our target is a collinear current in the $n-$direction, its effect on the source can be derived from our table by reversing the $n\leftrightarrow \bar{n}$ in the collinear source column.}. 
Also, note that for the collinear 
source we use SCET current in the $\bar{n}$ direction, for the static source we use the HQET current and, finally, 
for the soft source we use the unexpanded vertex $\gamma_{\mu}$ consistent with the soft quark interaction with
 the background field. As a result we fill in all the elements of table~\ref{tb1}, where we have defined 
$\Gamma_1...\Gamma_4, \Sigma_1...\Sigma_3$ as follows:
\begin{eqnarray}
\Gamma_1^{\mu, a}&=&i g T^a\,n^{\mu}\, \frac{\bnslash}{2}\; ,\\
\Gamma_2^{\mu, a}&=& i g T^a \frac{\gamma_{\perp}^{\mu}\pslash_{\perp}+\pslash'_{\perp} \gamma_{\perp}^{\mu}}{\bar{n}\mcdot p}\frac{\bnslash}{2}\; ,\\
\Gamma_3^{\mu, a}&=&i g T^a\,v^{\mu}\; ,\\
\Gamma_4^{\mu, a}&=&i g T^a\,\gamma^{\mu}\; ,\\
\Sigma_1^{\mu\nu\lambda, abc}&=&g f^{abc}\,n^{\mu}\left[g^{\nu\lambda}\,\bar{n}\mcdot p+\bar{n}^{\nu}\left(p_{\perp}'^{\lambda}-p_{{\perp}}^{\lambda}\right)-\bar{n}^{\lambda}\left(p_{\perp}'^{\nu}-p_{{\perp}}^{\nu}\right)-\frac{1-\frac{1}{\xi}}{2}\left(\bar{n}^{\lambda}p^{\nu}+\bar{n}^{\nu}p'^{\lambda}\right)\right]\; ,\label{eq:sigma1}\\
\Sigma_2^{\mu\nu\lambda,abc}&=&g f^{abc}\,\left[g^{\mu\lambda}_{\perp}\,\left(-\frac{n^{\nu}}{2}p^{+}+p^{\nu}_{\perp}-2p'^{\nu}_{\perp}\right)+g^{\mu\nu}_{\perp}\,\left(-\frac{n^{\lambda}}{2}p^++p'^{\lambda}_{\perp}-2p^{\lambda}_{\perp}\right) \right. \nonumber\\
&& +g^{\nu\lambda}_{\perp}\left(n^{\mu}\,\bar{n}\mcdot p+p_{\perp}^{\mu}+p_{\perp}'^{\mu}\right) \Big]\;, 
\end{eqnarray}
\begin{eqnarray}
\Sigma_3^{\mu\nu\lambda, abc}&=&g f^{abc}\,\left[g^{\mu\lambda}_{\perp}\,\left(\frac{\bar{n}^{\nu}}{2}(p^--2p'^-)+p^{\nu}_{\perp}-2p'^{\nu}_{\perp}\right)+g^{\mu\nu}_{\perp}\,\left(\frac{\bar{n}^{\lambda}}{2}(p'^--2p^-)+p'^{\lambda}_{\perp}-2p^{\lambda}_{\perp}\right)  \right. \nonumber \\  
&& +g^{\nu\lambda}_{\perp}\left(p_{\perp}^{\mu}+p_{\perp}'^{\mu}\right)\Big]\; .
\end{eqnarray}

The derived rules allow us to write down the effective Lagrangian of $\text{\SCETG}$. As a result, we obtain the following 
interaction term between SCET collinear fields and the vector potential $A_G^{\mu}(x)$ of the Glauber gluons:


\begin{eqnarray}
&& \mathcal{L}_{\text{\SCETG}}(\xi_n, A_n, A_G)=\mathcal{L}_{\text{SCET}}(\xi_n, A_n)+
\mathcal{L}_{\text{G}}\left(\xi_n, A_n, A_G\right),\\
&& \mathcal{L}_{\text{G}}\left(\xi_n, A_n, A_G\right)=\sum_{p,p'}\e^{-i(p-p')x}\left(\bar{\xi}_{n,{p'}} 
\Gamma^{\mu,a}_{\rm qqA_G}\frac{\bnslash}{2}\xi_{n,p}-i \Gamma^{\mu\nu\lambda,abc}_{\rm ggA_G}\,
\left({A}^{c}_{n, p'}\right)_{\lambda}\left({A}^{b}_{n, p}\right)_{\nu}\right)\, A_{{\rm G}\, \mu, a}(x)\label{LGdef0}\, .
\end{eqnarray}
Depending on the gauge and the source, the vertices and the vector potential are different and are  
provided in the table above. The Lagrangian of this form for the collinear source in $R_{\xi}$ and 
$A^{-}=0$\footnote{In order to avoid confusion we note that in \cite{Idilbi:2008vm} the source was in the 
$n$ direction while the target jet in the $\bar{n}$ direction, thus our formulas agree with that reference 
if $n\leftrightarrow \bar{n}$ as expected. For example in \cite{Idilbi:2008vm} the light-cone gauge 
$A^+=0$ was considered, while it is analogous to our $A^-=0$ gauge.} gauges was derived in \cite{Idilbi:2008vm} 
and agrees with our expressions for corresponding two entries for $\Gamma_{\rm qqA_G}$ in table~\ref{tb1}.
We also note that for the covariant gauge and $\xi=1$ our Feynman rule for $\Gamma_{\rm ggA_G}(R_{\xi})=
\Sigma_1^{\mu\nu\lambda}$ disagrees with that of~\cite{DEramo:2010ak}. The corresponding Feynman rule from Ref.~\cite{DEramo:2010ak} contains only the first term in \eq{eq:sigma1}. However, note that with such a Feynman rule, the relation \eq{eq:hybridvsRxi} below would be violated, which
would lead to different results for single Born radiative loss (see section~\ref{radiative}) calculated in the covariant and the hybrid gauges\footnote{This gauge choice is defined at the end of this section.}. With our Feynman rule this inconsistency does not happen.

Finally, in order to analyze the invariance of our Glauber exchange terms for quarks and gluons under the gauge symmetries of SCET, we rewrite~\eq{LGdef0} including the source fields 
(see~\eq{Amu0}):
%
\begin{eqnarray}
&& \mathcal{L}_{\text{G}}\left(\xi_n, A_n, \eta\right)=\sum_{p,p', q}\e^{-i(p-p'+q)x}\left(\bar{\xi}_{n,{p'}} 
\Gamma^{\mu,a}_{\rm qqA_G}\frac{\bnslash}{2}\xi_{n,p}-i \Gamma^{\mu\nu\lambda,abc}_{\rm ggA_G}\,
\left({A}^{c}_{n, p'}\right)_{\lambda}\left({A}^{b}_{n, p}\right)_{\nu}\right)\, \bar{\eta}\,
\Gamma^{\delta, a}_{\text{s}}\,\eta\,\Delta_{\mu\delta}(q)\, , \qquad  \label{LGdef1} \label{LG2}
\end{eqnarray}
where all the vertices for the target and the source are provided conveniently in  table~\ref{tb1}. In order to make 
this Lagrangian collinear gauge invariant one needs to dress the quarks and gluons with collinear Wilson 
lines $W_n(x)$, defined in \eq{Wndef}. As a result the Lagrangian that includes the Wilson lines can be obtained as follows:
\begin{eqnarray}
&& \mathcal{L}_{\text{G}}\left(\xi_n, A_n, \eta\right)\rightarrow \mathcal{L}_{\text{G}}
\left(W_n^{\dagger}\xi_n, \mathcal{B}_n(A_n), \eta\right)\equiv 
\mathcal{L}_{\text{G}} \left(\chi_n, \mathcal{B}_n, \eta\right),\label{LGdef}
\end{eqnarray}
where $W_n^{\dagger}\xi_n(\equiv \chi_n)$,  $\mathcal{B}_{n} (A_n)$ are the dressed collinear gauge invariant quark and gluon fields, correspondingly.
In the next subsection we will show that Lagrangian in~\eq{LGdef} is invariant under the soft and collinear gauge transformations of SCET.

The derived Lagrangian of $\text{\SCETG}$ in~\eq{LGdef0} and~\eq{LG2} contains only interaction between a single 
collinear quark or gluon with a single Glauber gluon. However there are additional interactions between the collinear 
particles and Glauber gluons. For example, in the light-cone gauge the first term of~\eq{SCETL1} contains an interaction 
where the two Glauber gluons interact at the same point with a collinear quark line. The same Lagrangian in the same gauge 
contains the interaction where at the same point there are a collinear quark, a collinear gluon and a Glauber gluon. 
While we omit these terms in this section for brevity, their derivation is straightforward and the corresponding 
Feynman rules are listed in the appendix~\ref{Appendix:SCETG}.

While we derived the Feynman rules of $\text{\SCETG}$ for variety of sources and gauges, in the main part of this 
paper we will do calculations in the following cases. For the source, motivated by our study in  
section~\ref{sec:kinematics}, which showed that at RHIC and LHC energies the recoil effect is negligible with 
accuracy better than $15\%$, we will use the (initially) static source. The interested reader will notice that
since the physics results depend on the transverse momentum exchanges (Glauber gluons)  between the projectile and the
target and the jet-medium cross sections, they should not be sensitive to the components of the source in the
$n$ and $\bar{n}$ directions.   

As for the gauge choice, we consider three cases in this paper, which allows us to establish the gauge invariance of 
the broadening and the radiative energy loss results. The first two choices are the covariant and positive 
light-cone gauges. All relevant Feynman rules are derived in this section and summarized in 
 appendix~\ref{Appendix:SCETG}. Note that in both cases the Glauber gluons and collinear gluons are quantized with 
the same gauge-fixing condition. However there is no problem in quantizing the collinear gluons in, say, the light-cone 
gauge, 
while the Glauber gluons are in the covariant gauge. This is our third choice and we call it the hybrid gauge. 
For practical purposes this turns out to be the most convenient gauge choice \cite{Gyulassy:2000er,Vitev:2007ve}. 
Physically, such a choice is 
possible since the scattering part and the medium-induced splitting parts of the calculation factorize.
    From a formal point of view the hybrid gauge 
corresponds to the vector potential created by the source derived in the covariant gauge, and the SCET Lagrangian 
quantized in the light-cone gauge with the corresponding background field vector potential. We discuss in details 
the Feynman rules in this gauge in section~\ref{sec:hybrid} and in appendix~\ref{Appendix:SCETG}.

%
%
%
%

\subsection{Gauge invariance of $\mathcal{L}_{\text{G}}$}\label{App:LGGaugeInvariance}

In this subsection we show that the Glauber Lagrangian  $\mathcal{L}_{\text{G}}$ is invariant under
both collinear and soft gauge transformations in  SCET. 
The collinear gauge symmetry is a simple consequence of dressing the collinear fields with the collinear Wilson lines. The fields transform under the collinear gauge transformation of SCET according to \cite{Bauer:2001yt}: 
\begin{eqnarray}
 &&\xi_n\rightarrow\mathcal{U}_c\,\xi_n\, ,  \qquad\qquad\qquad\qquad\qquad\qquad\chi_n\equiv W_n^{\dagger}\xi_n\rightarrow\mathcal \chi_n\, ,\\
 &&A^{\mu}_n\rightarrow \mathcal{U}_c \, A^{\mu}_n\,\mathcal{U}_c^{\dagger}+\frac{1}{g} 
\mathcal{U}_c\,i{D}^{\mu}\,\mathcal{U}_c^{\dagger}\, , \qquad\qquad\mathcal{B}^{\mu}_n\equiv \frac{1}{g}\left[W_n^{\dagger}\,iD^{\mu}\, W_n\right]\rightarrow \mathcal{B}^{\mu}_n\, ,\\\label{GIBDef}
&&h_v\rightarrow h_v\, ,
\end{eqnarray}
where $W$ is the collinear Wilson line, which is defined in \eq{Wndef}, and the square brackets in the last equation indicate that the derivative operator acts only within the brackets. Note that the massive 
fields $h_v$ do not transform under the collinear transformation, since $p_h \sim [1,1,{\bm \lambda}]$ and 
$p_c\sim [1,\lambda^2,{\bm \lambda}]$. 
As we see from the definition of $\mathcal{L}_G(\mathcal{\chi},\mathcal{B}, h_v)$, since it is explicitly built 
out of the gauge invariant collinear fields, the Lagrangian is invariant under collinear gauge transformation:
\begin{eqnarray}
\mathcal{L}_G(\mathcal{\chi}, \mathcal{B}, h_v)\xrightarrow[]{\text{collinear gauge transformation}} 
\mathcal{L}_G(\mathcal{\chi}, \mathcal{B}, h_v)\, .
\end{eqnarray}

Demonstrating the invariance of the Lagrangian of $\text{\SCETG}$ under the soft gauge transformation is slightly more involved. The soft transformation of SCET looks like~\cite{Bauer:2001yt}:
\begin{eqnarray}
 &&\xi_n\rightarrow V_s\,\xi_n\, ,\qquad\qquad\qquad   \chi_n\equiv W_n^{\dagger} \,\xi_n\rightarrow V_s\,\mathcal \chi_n\, ,\\
 &&A^{\mu}_n\rightarrow V_s \, A^{\mu}_n\,V_s^{\dagger}\, ,\qquad\qquad \mathcal{B}^{\mu}_n\equiv \frac{1}{g}\left[W_n^{\dagger}\,iD^{\mu}\, W_n\right]\rightarrow V_s\,\mathcal{B}^{\mu}_n \,V_s^{\dagger}\, ,\\
 && h_v\rightarrow V_s \,h_v\, ,
\end{eqnarray}
where $V_s=\e^{i\alpha^a(x) T^a}$, such that $\partial_{\mu} V_s(x)=\mathcal{O}(\lambda^2)$. For shorthand notation, define $Y\equiv V_s$. The quark part of the Lagrangian of $\text{\SCETG}$ then transforms into:
\begin{eqnarray}
 && T^c_{ij}\otimes T^c_{kl} \xrightarrow[]{\text{soft gauge transformation}}(Y^{\dagger}T^c Y)_{ij}\otimes (Y^{\dagger}T^cY)_{kl}
= (Y^{\dagger}_{im}T^c_{mn} Y_{nj})\otimes (Y_{kf}^{\dagger}T^c_{fg}Y_{gl}) \nonumber \\
  && = (Y^{\dagger}_{im}Y_{nj})(Y_{kf}^{\dagger}Y_{gl})  \,  
\left(\frac{1}{2}\delta_{mg}\delta_{nf}-\frac{1}{2N} \delta_{mn}\delta_{fg}\right)
=\frac{1}{2}\delta_{il}\delta_{jk}-\frac{1}{2N} \delta_{ij}\delta_{kl}=T^c_{ij}\otimes T^c_{kl}\, . 
\nonumber
\end{eqnarray}
Similarly, for the gluon part of the $\mathcal{L}_G$ in~\eq{LGdef} we have:
\begin{eqnarray}
 && A^{\mu,a}A^{\nu,b} i f^{abc}\otimes T^c\equiv A^{\mu,a}A^{\nu,b}\otimes \left[T^a, T^b\right]
\xrightarrow[]{\text{soft gauge transformation}} \mathcal{Y}^{aa'}A^{\mu,a'}\mathcal{Y}^{bb'}A^{\nu,b'}
\otimes Y^{\dagger}\left[T^{a'}, T^{b'}\right]Y \nonumber\\
&& = A^{\mu,a'}A^{\nu,b'}\otimes Y^{\dagger}\left[\mathcal{Y}^{aa'}T^{a'}, \mathcal{Y}^{bb'}T^{b'}\right]Y
=A^{\mu,a'}A^{\nu,b'}\otimes Y^{\dagger}\left[Y T^{a'}Y^{\dagger}, Y T^{b'} Y^{\dagger}\right]Y
\nonumber\\ && =A^{\mu,a'}A^{\nu,b'}\otimes \left[T^{a'}, T^{b'} \right]=
A^{\mu,a'}A^{\nu,b'} i f^{a'b'c'}\otimes T^{c'}\, .
\end{eqnarray}
As a result, we obtain that the Glauber term in $\text{\SCETG}$ is invariant under the soft 
gauge transformation:
\begin{eqnarray}
\mathcal{L}_G(\mathcal{\chi}, \mathcal{B}, h_v)\xrightarrow[]{\text{soft gauge transformation}} 
\mathcal{L}_G(\mathcal{\chi}, \mathcal{B}, h_v)\, .
\end{eqnarray}

\section{Jet broadening}
\label{collisional}

In this section we derive the modification to the transverse momentum distribution of jets from
an elastic in-medium scattering to first order in opacity. We consider a quark jet or a gluon jet 
interacting with an initially static fermionic center for definitiveness. The necessary effective 
theory $\text{\SCETG}$ Feynman rules were derived in the section~\ref{scetg} and we use 
the covariant gauge for our calculation. The amplitudes that we consider for the quark and gluon 
cases have the following form, respectively:
\begin{eqnarray}
  && A^{\text{(q)}}=\bra{J}T\,\bar{\chi}_{n}(x_0)\,\e^{i\,\int d^4 x\left(\mathcal{L}^{\text{QCD}}+\mathcal{L}^{\text{\SCETG}}\right)}\ket{{\bf{p}}}\, ,\label{CollAquark}\\
  && A^{\text{(g)}}=\bra{J_{\mu, a}}T\,\mathcal{B}_{n}^{\mu, a}(x_0)\,\e^{i\,\int d^4 x\left(\mathcal{L}^{\text{QCD}}+\mathcal{L}^{\text{\SCETG}}\right)}\ket{{\bf{p}}}\, ,\label{CollAgluon}
\end{eqnarray}
where $\chi_n$ and $\mathcal{B}_{n}$ are the gauge invariant quark and Gluon fields in SCET, and $J$ is the 
underlying hard process that creates the quark or gluon jet. The $\mathcal{L}^{\text{QCD}}$ term simply generates the hard QCD process in question, which we take into account by effective Feynman rule $\bra{J}\,\bar{\chi}_n(x_0)\,\ket{{\bf{p}}}=\bar{\chi}_{n,p}\,iJ(p)\, \e^{i p x_0}$. In the next two subsections we calculate 
these amplitudes using the Lagrangian of $\text{\SCETG}$ and combining the single Glauber 
gluon exchange diagram with the contact limit of the two Glauber gluon exchange diagrams. 
The corresponding amplitudes are called single and double Born 
diagrams in the literature~\cite{Gyulassy:2002yv,Qiu:2003pm}.
Each interaction in medium can be considered located at a certain point $\vc{x}$, and an integral over 
the Glauber gluon momentum is introduced. To keep formulas compact, it is convenient to use the 
following shorthand notation:
\begin{eqnarray}
&&d\Phi_i=\frac{d^4 q_i}{(2\pi)^4}\, \e^{i q_i \delta x_i}\, v(q_i)\; ,\label{dphi}\\
&&d\vc{\Phi}_{i\perp}=\frac{d^2 \vc{q}_{i\perp}}{(2\pi)^2}\, 
\e^{-i \vc{q}_{i\perp} \delta \vc{x}_{i\perp}}\, \tilde{v}(\vc{q}_{i\perp})\; ,
\end{eqnarray}
where $i$ is the index of the corresponding scattering center, $v(q)=2\pi\delta(q^0)\tilde{{v}}(\vc{q}_{\perp})$, $\tilde{v}(\vc{q}_{\perp})$ is defined by Eq.~(\ref{transform}) and $\delta x_i=x_i-x_0$. The delta function in $v(q)$ makes the $dq^+$ integral trivial, while the $q^-$ integral still needs to be evaluated. The following identity arises once one works out the $d^4q$ in the light-cone coordinates:
\begin{eqnarray}
d\Phi_i=d\vc{\Phi}_{i\perp}\,\frac{dq^-_i}{2\pi}\,\e^{i q^-\delta z_i} \; , 
\label{jacobianLCC}
\end{eqnarray}
where $\delta z_i=\delta \vc{x}^3_i$.
To also shorten the color notation we substitute the color generator $T^a$ by simply  $a$ and specifying the representation it belongs to. For example:
\begin{eqnarray}
&&(a)_R(a)_{T_i}=T^a(R)\, T^a(i),\\
&&[a,b]_R=i f^{abc}T^c(R).
\end{eqnarray}
We use  $R$ to denote the color representation of the incident energetic quark or gluon, and we use $T$ 
to specify the color representation of the scattering center itself. For example, $d_T=3$ since we  
consider a fermionic scattering center, while $d_R=d_F=3$ for a quark jet and $d_R=d_A=8$ for a gluon jet.

For the quark (or gluon) propagator with the collinear plus Glauber momentum $p-q$, where $p$  
has scaling $[1,\lambda^2, {\bm \lambda}]$ and $q$ is the Glauber gluon with the momentum scaling 
$[\lambda^2,\lambda^2,{\bm \lambda}]$, we define:
\begin{eqnarray}
\Delta_g(p,q)\equiv \frac{1}{p^--q^--\frac{(\vc{p}_{\perp}-\vc{q}_{\perp})^2-i\eps}{p^+}}\; .\label{GlauberProp}
\end{eqnarray}
It is convenient to define the following quantity, since it directly appears in the expression in~\eq{GlauberProp}:
\begin{eqnarray}
\Omega(p,q)=p^--\frac{(\vc{p}_{\perp}-\vc{q}_{\perp})^2-i\eps}{p^+} \, \label{poleofGlauberProp}.
\end{eqnarray}
Note that the quantity in the~\eq{poleofGlauberProp} depends only on $\vc{q}_{\perp}$ and $p$ and always leads to a $q^-$ pole in the propagator~\eq{GlauberProp} which is in the positive imaginary plane.

\subsection{Quark jet}
\begin{figure}[t!]
\begin{center}
\epsfig{file=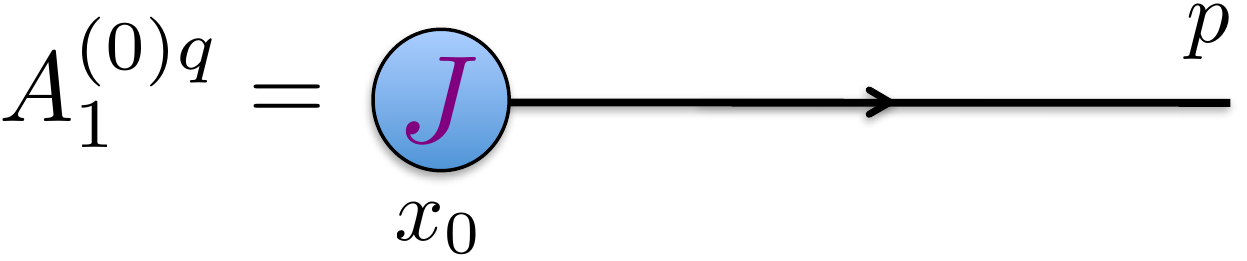, width=5.5cm}\\ \qquad\epsfig{file= 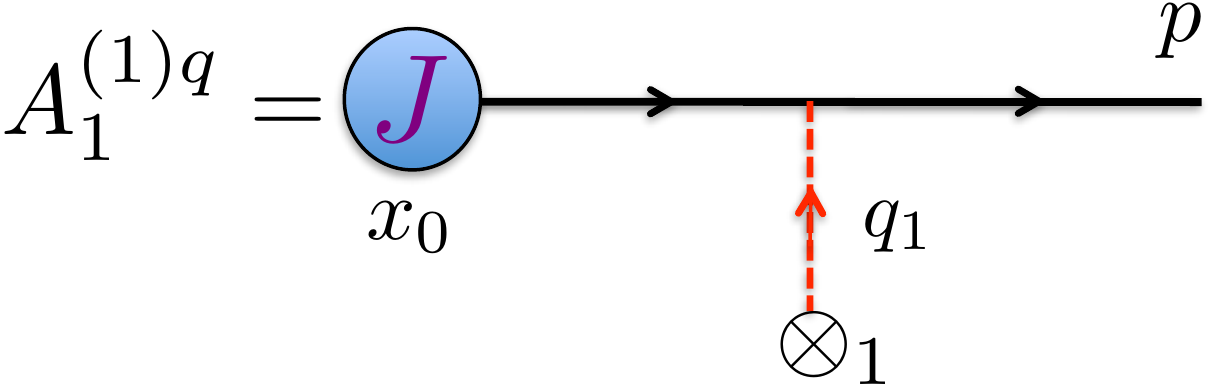, width=5.5cm} \epsfig{file= 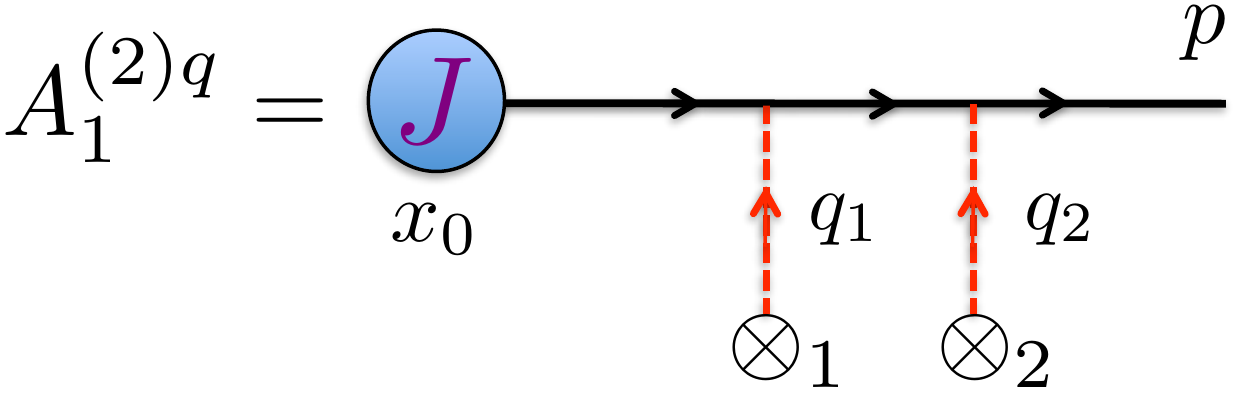, width=5.5cm}
\caption{\label{fig:BroadA} Three lowest order diagrams contributing to a collisional in-medium quark
interaction. Top: tree level, bottom left: single Born diagram, and bottom right: two Glauber gluons exchange diagram. 
The notation for the scattering centers is the following: $\otimes_{1}=[x_1,q_1, (b_1)_i], \otimes_{2}=[x_2,q_2, (b_2)_j]$. The contact limit of the amplitude $A_1^{(2)q}$ is also called double Born amplitude. }
\end{center}
\end{figure}

We start from the first diagram in figure~\ref{fig:BroadA}, which represents the lowest order diagram for the 
matrix element in~\eq{CollAquark}:
\begin{eqnarray}
  A^{(0)q}_1=\bar{\chi}_{n,p}\, i J(p)\,\e^{i p x_0}\, . \label{A0q}
\end{eqnarray}

For a $k$ Glauber gluon exchange diagram one has a factor of $\frac{\nslash}{2}$ for each quark propagator 
and a factor of $\frac{\bnslash}{2}$ for each vertex. These factors can be commuted through to the left and simplified:
\begin{eqnarray}
   \bar{\chi}_{n,p}\, \left(\frac{\bnslash}{2}\frac{\nslash}{2}\right)^k=  \bar{\chi}_{n,p}\, 
\left(\frac{\bnslash}{2}\frac{\nslash}{2}\right)=  \bar{\chi}_{n,p},
\end{eqnarray}
which means that we can (and will) ignore this factors in the Feynman rules. Squaring the tree level 
result in~\eq{A0q} we get:
\begin{eqnarray}
d\sigma \propto  \frac{1}{d_R d_T}\sum_{\text{spin,color}}|A_1^{(0)q}|^2=\text{Tr}
\left(\frac{\nslash}{2}J(p)\bar{J}(p)\right)  \bar{n}\cdot p\, . 
\label{csec}
\end{eqnarray}
It is instructive to examine here the Lorentz structure of the part of the matrix element squared
which is not explicitly written in Eq.~(\ref{csec}). It can be represented completely generally 
as follows: 
\begin{equation}
J(p)\bar{J}(p) = \sum_{i=1}^{16} a_i(p) \Gamma^i \;. 
\label{decompose}
\end{equation}
Here, $\Gamma^i$ is a basis set of 16 matrices for the Dirac algebra, for example: 
$\left\{ {\bf 1}, \gamma^\mu, \gamma^5, \gamma^\mu \gamma^5, \sigma^{\mu \nu} = 
\frac{i}{2}[ \gamma^\mu, \gamma^\nu ] \right\} $. Note that $\sigma^{i j} = \epsilon^{ijk}\Sigma^k$. 
By direct inspection one sees that only the $a_+(p) \gamma^+$ part contributes to the spin
averaged cross section.   

For a $k$ Glauber gluon exchange amplitude we have\footnote{In our notation  $B^{(k)q}$ , 
$B$ stands for ``Broadening".}:
\begin{eqnarray}
  A^{(k)q}_{\text{coll}}=\bar{\chi}_{n,p}\,\int \prod_{m=1}^k\, d\Phi_m\, B^{(k){q}}\,
iJ\left(p-\sum_{l=1}^k q_l\right)\,\e^{i p x_0} \, ,\label{Akquark}
\end{eqnarray}
where $B^{(k)q}$ is the non-trivial part of the amplitude given by the $\text{\SCETG}$ Lagrangian. 
With the notation defined above and using the Feynman rules derived in section~\ref{scetg} and in 
appendix~\ref{Appendix:SCETG} we get in the $R_{\xi}$ gauge:
\begin{eqnarray}
   && B^{(0)q}_1=1\; ,\\
   && B^{(1)q}_1=(b_1)_R\,(b_1)_{T_i}\, i\, i\Delta_g(p,q_1)\; ,\\
      && B^{(2)q}_1=(b_2 b_1)_R (b_1)_{T_i} (b_2)_{T_j}\, i\, i\Delta_g(p, q_2)\,i \, i\Delta_g(p,q_1+q_2)\; .
\end{eqnarray}
We have absorbed all the factors $\tilde{v}(\vc{q}_{\perp})$ and the phases into the conveniently defined differentials in~\eq{dphi}. 
Next we need to perform the longitudinal integrals $d q^+, dq^-$ in the one  and two Glauber gluons exchange diagrams 
and reduce the integration to only transverse components of $\vc{q}_{i\perp}$. For the single Born diagram 
this is done by using the result for the longitudinal integral $I_1^{(1)}$ from 
appendix~\ref{Appendix:integralsColl}. The result is:
\begin{eqnarray}
   \int \, d\Phi_1 \, B_1^{(1)q}&=& i\,(b)_R (b)_{T_i} \, \int d\vc\Phi^{\perp}_1\, 
\left[\e^{i\omega_1\,\delta z_1}\right]\; .\label{singleqres}
 \end{eqnarray}
Similarly, using the results from  appendix~\ref{Appendix:integralsColl} for the integrals 
$I_1^{(2)}$ and its contact limit $I_1^{(2c)}$, we get the following results for the 
two Glauber gluons exchange diagram and its contact limit:
\begin{eqnarray}
   \int \, d\Phi_1 \,d\Phi_2\, B^{(2)q}_2&=& (b_2 b_1)_R (b_1)_{T_i} (b_2)_{T_j}\, \int d\vc\Phi^{\perp}_1 d\vc{\Phi}^{\perp}_2\, \left[\e^{i\left(\omega_{12}\delta z_1+\omega_2(\delta z_2-\delta z_1)\right)}\right]\; \left(-1\right) ,\label{doubleqres1}\\
   \int \, d\Phi_1 \,d\Phi_2\, B^{(2c)q}_2&=& (b_2 b_1)_R (b_1 b_2)_{T_i} \,\int d\vc\Phi^{\perp}_1 d\vc{\Phi}^{\perp}_2\, \left[\e^{i\left(\omega_{12}\delta z_1\right)}\right]\times \left(-\frac{1}{2}\right)\; .\label{doubleqres2}
\end{eqnarray}
In~\eq{singleqres}-\eq{doubleqres2}, the inverse formation times $\omega_{1,2}$ are defined as follows:
\begin{eqnarray}
   \omega_1=\Omega(p,q_1)\; , \qquad\qquad\omega_2=\Omega(p, q_2)\; , \qquad\qquad\omega_{12}=\Omega(p,q_1+q_2)\; .
\end{eqnarray}
Finally, combining together the three amplitudes, squaring them and identifying  the contribution to
first order in opacity, we get:
\begin{eqnarray}
   \frac{1}{d_R\, d_T}\,\text{Tr}|A^{(0)q}_1+A^{(1)q}_1+A^{(2c)q}_1+...|^2=  \frac{1}{d_R\, d_T}\,\text{Tr}\left(|A^{(0)q}_1|^2+|A^{(1)q}_1|^2+2\,\text{Re} \,\left(A^{(0)q }_1\right)^{\dagger}A^{(2c)q}_1+...\right) \; . \qquad \label{squaringA}
\end{eqnarray}
Here, we have omitted the $\text{Re} \,\left(A^{(0)q }_1\right)^{\dagger}A^{(1)q}_1$ term, since it vanishes because of the color trace: 
$\text{Tr}(T^b(R))=~0$. Squaring the $A_1^{(0)q}$ term has been performed in~\eq{csec} and represents the squared 
matrix element of the production 
of the quark jet from the underlying hard process $J$. The same exact overall factor, the differential
jet distribution $d^2\sigma_{q,g}/d^2\vc{p}_\perp$,  appears in all other terms in~\eq{squaringA}. 
We will drop this factor for brevity but will keep in mind that the results derived below should be 
understood as operators acting on the unperturbed by the medium jet distribution. Thus, the first term 
becomes simply unity:
\begin{eqnarray}
   \frac{1}{d_R\, d_T}\,\text{Tr}|A^{(0)q}_1|^2= 1 \; .
\end{eqnarray}
Squaring the single Born amplitude $A^{(1)q}$ we get:
\begin{eqnarray}
   \frac{1}{d_R\, d_T}\,\text{Tr}|A^{(1)q}_1|^2= \frac{1}{d_R\, d_T}\, 
\sum_{i=1}^N\text{Tr}((b)_R (b')_R)\text{Tr}((b)_{T_i} (b')_{T_i})\,
\int d\vc{\Phi}_{\perp}(\vc{q}_i)\,d\vc{\Phi}_{\perp}(\vc{q}'_i)^*\,
\,\e^{i(\omega_1(q_{i\perp})-\omega_1(q'_{i\perp}))\delta z_i}\, .\nonumber\\
\end{eqnarray}
This expression can be simplified further by turning the sum over the scattering centers into a 
continuous integral. This approximation is valid if $A_{\perp}\gg\mu^{-2}$:
\begin{eqnarray}
   \sum_{i=1}^N\, \e^{-i\vc{p}_{\perp}\delta x_i}\approx N\int \frac{d^2\vc{b}}{A_{\perp}}\,\e^{-i\vc{p}_{\perp}\cdot \vc{b}}= N\frac{(2\pi)^2\,\delta(\vc{p}_{\perp})}{A_{\perp}}\; .
\end{eqnarray}
Using this representation and integrating over $\vc{q}'_{i\perp}$, we obtain a particularly simple expression:
\begin{eqnarray}
   \frac{1}{d_R\, d_T}\,\text{Tr}|A^{(1)q}_1|^2=\frac{N}{A_{\perp}}\frac{C_2(R)C_2(T)}{d_A}\int \frac{d^2\vc{q}_{\perp}}{(2\pi)^2} \,|\tilde{v}(\vc{q}_{\perp})|^2 \, \times \,  e^{-{\bf q}_\perp \cdot \stackrel{\rightarrow}{\nabla}_{{\bf p}_\perp} }  \,  .
\label{drel}
\end{eqnarray}
In  Eq.~(\ref{drel}) $e^{-{\bf q}_\perp \cdot \stackrel{\rightarrow}{\nabla}_{{\bf p}_\perp} }  $ 
indicates the  shift in the transverse momentum of the initial jet distribution.
The contribution of the double Born contact term to the first order in opacity cross section correction 
becomes after a similar averaging procedure, which sets $\vc{q}_{1\perp}+\vc{q}_{2\perp}=0$ in~\eq{doubleqres2}:
\begin{eqnarray}
   \frac{1}{d_R\, d_T}\,\text{Tr}\,\left(A^{(0)q}_1\right)^{\dagger} A^{(2c)q}_1
=\left(-\frac{1}{2}\right)\frac{N}{A_{\perp}}\frac{C_2(R)C_2(T)}{d_A}\int \frac{d^2\vc{q}_{\perp}}{(2\pi)^2} 
\,|\tilde{v}(\vc{q}_{\perp})|^2\, .
\label{virel}
\end{eqnarray}
Note that in Eq.~(\ref{virel}) there is no net transverse momentum transfer to the jet and no momentum shift.
Combining the contributions up to first order in opacity we finally get:
\begin{eqnarray}
&&   \frac{1}{d_R\, d_T}\,\text{Tr}\left(|A^{(0)q}_1|^2+|A^{(1)q}_1|^2+2\,\text{Re} 
\,\left(A^{(0)q }_1\right)^{\dagger}A^{(2c)q}_1\right) \nonumber \\
&& \qquad \qquad \qquad \qquad   =1+\frac{N}{A_{\perp}}\int{d^2\vc{q}_{\perp}} 
\left[\frac{d\sigma_{\text{el}}(R,T)}{d^2\vc{q}_{\perp}} e^{-{\bf q}_\perp \cdot 
\stackrel{\rightarrow}{\nabla}_{{\bf p}_\perp} }   
-\sigma_{\text{el}}\delta^{(2)}(\vc{q}_{\perp})\right] \; ,
\end{eqnarray}
where we expressed the answer through the  elastic scattering, which in the lowest Born approximation equals to:
\begin{eqnarray}
\frac{d\sigma_{\text{el}}(R,T)}{d^2\vc{q}_{\perp}}=\frac{C_2(R)C_2(T)}{d_A} \,
\frac{|\tilde{v}(\vc{q}_{\perp})|^2}{(2\pi)^2} \; .
\end{eqnarray}

\subsection{Gluon jet}
\begin{figure}[t!]
\begin{center}
\epsfig{file=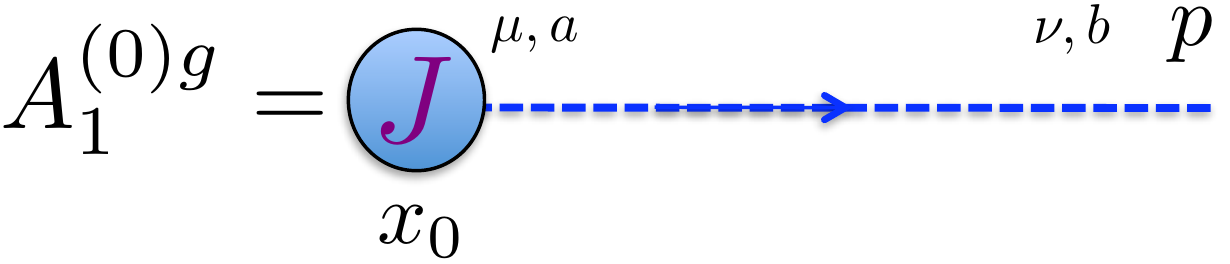, width=5.5cm}\\ \qquad\epsfig{file= 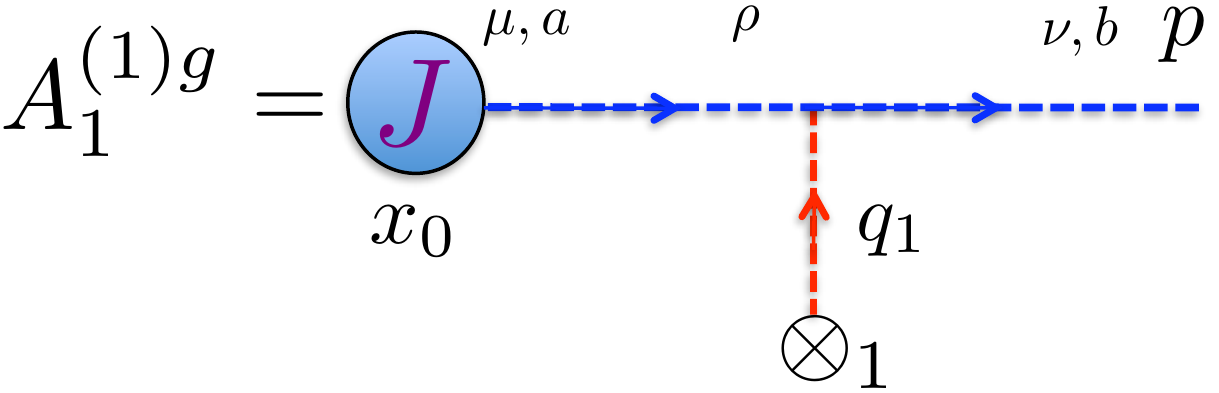, width=5.5cm} \epsfig{file= 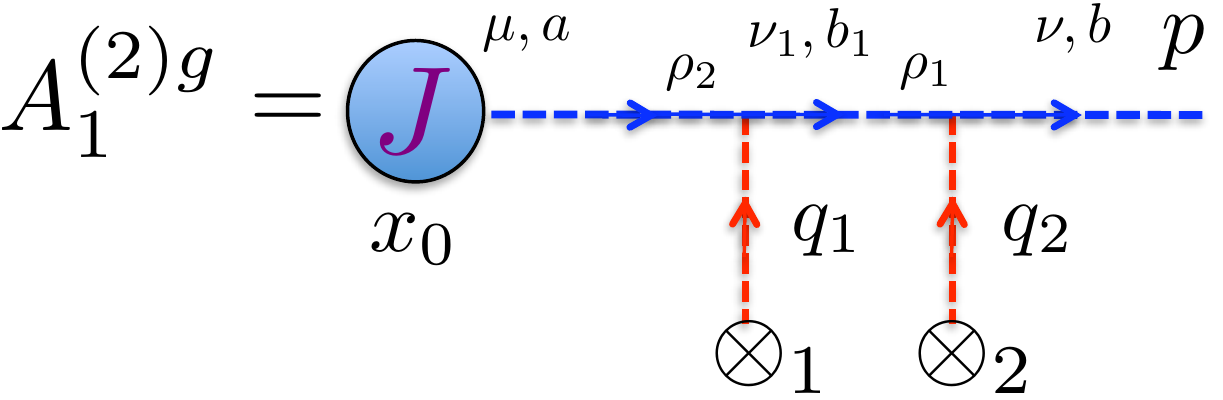, width=5.5cm}
\caption{\label{fig:BroadGA} Three lowest order diagrams contributing to the gluon jet in-medium interaction. 
Top: tree level, bottom left: single Born diagram, and bottom right: two single  Born exchanges diagram. The notation for the scattering centers is the following: $\otimes_{1}=[x_1,q_1, (c_1)_i], \otimes_{2}=[x_2,q_2, (c_2)_j]$.}
\label{fig:gluonsingle}
\end{center}
\end{figure}

We repeat a similar calculation as in the previous subsection for the gluon jet scattering off a 
fermionic scattering center. The first diagram in figure~\ref{fig:gluonsingle} equals to:
\begin{eqnarray}
  A^{(0)g, b}_1= \e^{i p x_0}\,i \,J_{\nu, a}(p)\eps^{\nu}(p)\;\delta^{ab} .
\end{eqnarray}
The square of this matrix element describes the hard scattering production cross section and appears in every term and is dropped for brevity below. For $k$ Glauber gluon exchange with medium amplitude we use the following notation, similarly to the quark case above in~\eq{Akquark}:
\begin{eqnarray}
   A^{(k)g}_{\text{coll}, b} = \eps^{\nu}(p)\,\int \prod_{m=1}^k\, d\Phi_m\, 
B^{(k)g}_{\mu\nu, ab}\,iJ^{\mu, a}\left(p-\sum_{l=1}^k q_l\right)\,\e^{ip x_0}\, .
\label{Akgluon}
\end{eqnarray}
With this notation, using the Feynman rules of $\text{\SCETG}$ in covariant gauge we get for $B^{(k)g}$ amplitudes for $k=0,1,2$: 
\begin{eqnarray}
  (B^{(0)g}_1)^{\mu\nu, ab}&=&g^{\mu\nu}\; \delta^{ab} ,\label{B0g1}\\
    (B^{(1)g}_1)^{\mu\nu, ab}&=&f^{abc_1}\,(c_1)_{T_i}\,\tilde{\Sigma}_1^{\rho\nu}(p-q_1,p)\,  (-i)\Delta(p-q)^{\mu\rho}_{(R_{\xi})}\; ,\label{B1g1}\\
      (B^{(2)g}_1)^{\mu\nu, ab}&=&f^{b_1 b\,c_2}\,(c_2)_{T_j}\,\tilde{\Sigma}_1^{\rho_1\nu}(p-q_2,p)\,  (-i)\Delta(p-q_2)^{\nu_1\rho_1}_{(R_{\xi})}\, 
f^{ab_1c_1}\,(c_1)_{T_i}\,  \nonumber \\
&& \times \,\tilde{\Sigma}_1^{\rho_2\nu_1}(p-q_1-q_2,p-q_2)\,  
(-i)\Delta(p-q_1-q_2)^{\mu\rho_2}_{(R_{\xi})}\; .\label{B2g1}
\end{eqnarray}
In the Equations~\eq{B1g1}-\eq{B2g1}, $\Delta(p)^{\mu\nu}_{(R_{\xi})}$ is a standard gluon propagator in the covariant gauge, and the $\tilde{\Sigma}_1^{\mu\nu}(p,p')$ is the effective Feynman rule for covariant gauge for Glauber gluon coupling to collinear gluons. As one can see from our appendix~\ref{Appendix:SCETG} (also from  section~\ref{scetg}) it reads:
\begin{eqnarray}
   {\tilde\Sigma}_1^{\mu\nu}(p,p')\equiv g^{\mu\nu}\,\bar{n}\mcdot p+\bar{n}^{\mu} \,(p'-p)^{\nu}_{\perp}-\bar{n}^{\nu}\,(p'-p)^{\mu}_{\perp}-\frac{1-\frac{1}{\xi}}{2}\left(\bar{n}^{\nu}p^{\mu}+\bar{n}^{\mu}p'^{\nu}\right)\; .
\end{eqnarray}
To work out the contraction of vertices for single and double Born diagrams, the following identities are useful:
\begin{eqnarray}
 && \eps_{\nu}(p)\,\tilde\Sigma_1^{\rho\nu}(p-q_1, p)\,N^{(R_{\xi})}_{\mu\rho}(p-q_1)=\vc{\eps}_{\perp} \mcdot\left(\vc{p}_{\perp}-\vc{q}_{1\perp}\right)\bar{n}_{\mu}+\eps^{\perp}_{\mu}\,\bar{n}\mcdot p,\label{RxiIdentity} \\
&&\eps_{\nu}(p)\,\tilde\Sigma_1^{\rho_1\nu}(p-q_2, p)\,N^{(R_{\xi})}_{\nu_1\rho_1}(p-q_2)\,\tilde\Sigma_1^{\rho_2\nu_1}(p-q_1-q_2, p-q_2)\,N^{(R_{\xi})}_{\mu\rho_2}(p-q_1-q_2)\nonumber\\
&&\qquad\qquad \qquad =\bar{n}\mcdot p\,\,\vc{\eps}_{\perp} \mcdot\left(\vc{p}_{\perp}-\vc{q}_{1\perp}-\vc{q}_{2\perp}\right)\bar{n}_{\mu}+(\bar{n}\mcdot p)^2\,\eps^{\perp}_{\mu}
\label{RxiIdentity2} \; , \quad
\end{eqnarray}
 where we defined $N^{(R_{\xi})}$ to be the numerator of the covariant gauge gluon propagator:
\begin{eqnarray}
 N^{(R_{\xi})}_{\mu\nu}(p)=g_{\mu\nu}-\frac{p_{\mu}p_{\nu}}{p^2}(1-\xi)\; .
\end{eqnarray}
Note that even though we work in the covariant gauge, we are free to choose any polarization vector for the on-shell final state gluon. We assume for the external scattered gluon as well as its source the physical polarizations:
 \begin{eqnarray}
\bar{n}\mcdot \eps(p)&=&0\, ,\qquad p\mcdot \eps(p)=0\, ,\label{polarization1}\\
\bar{n}\mcdot J(p-q)&=&0\, , \qquad (p-q)\mcdot J(p-q)=0\, .\label{polarization2}
 \end{eqnarray}
 Equation~(\ref{polarization1}) was the only assumption in deriving Eqs.~(\ref{RxiIdentity}, \ref{RxiIdentity2})  above. If in addition one assumes Eqs.~(\ref{polarization2}), then the first terms can be dropped in each of the Eqs.~(\ref{RxiIdentity}, \ref{RxiIdentity2}) and only the transverse term survives. Under these conditions and using results for the same longitudinal integral as in the quark case, which 
are summarized in the appendix~\ref{Appendix:integralsColl},  we get for the single Born amplitude\footnote{Note that in our notation the covariant gauge gluon propagator of collinear plus Glauber momentum can be written as $\Delta(p-q)^{\mu\nu}_{(R_{\xi})}=N^{\mu\nu}_{(R_{\xi})}(p-q)\,\Delta_{g}(p,q)\,\frac{1}{\bar{n}\mcdot p}$. Using this identity along with Eqs.(\ref{RxiIdentity},\ref{RxiIdentity2}) leads to the same results for the broadening of gluon jet as for the quark jet up to the color and Dirac structure.}:
\begin{eqnarray}
   &&\int d\Phi_1 \,\left(B^{(1)g}_1\right)^{\mu\nu, ab}= -\,f^{abc_1} (c_1)_{T_i}\left(g^{\mu\nu}_{\perp}\right) \int d\vc{\Phi}_{1\perp}\, \left[\e^{i\omega_1 \delta z_1}\right]\, ,
\label{M1GSCETG}
\end{eqnarray}
which differs from the quark case in~\eq{singleqres} only by a color and Lorentz structure. Similarly for the two Glauber gluons exchange diagram and its contact limit we obtain the following results:
\begin{eqnarray}
   \int d\Phi_1 d\Phi_2\,\left(B^{(2)g}_1\right)^{\mu\nu, ab}&=& f^{a b_1 c_1}f^{b b_1 c_2} (c_2)_{T_j}(c_1)_{T_i}  \left(g^{\mu\nu}_{\perp}\right)\, \int d\vc\Phi^{\perp}_1 d\vc{\Phi}^{\perp}_2 \left[\e^{i\left(\omega_{12}\delta z_1+\omega_2(\delta z_2-\delta z_1)\right)}\right]\left(-1\right) \, , \qquad \\
   \int d\Phi_1 d\Phi_2\,\left(B^{(2c)g}_1\right)^{\mu\nu, ab}&=& f^{a b_1 c_1}f^{b b_1 c_2} (c_2)_{T_i}(c_1)_{T_i}  \left(g^{\mu\nu}_{\perp}\right)\, \int d\vc\Phi^{\perp}_1 d\vc{\Phi}^{\perp}_2 \left[\e^{i\left(\omega_{12}\delta z_1\right)}\right]\times\left(-\frac{1}{2}\right)\, .
\end{eqnarray}
The final result for the cross section to lowest order in opacity coincides with the one obtained in the last subsection for the quark case, with the substitution of the gluon-quark cross section instead of quark-quark elastic scattering cross section.
Combining the contributions up to first order in opacity we finally get:
\begin{eqnarray}
&& \frac{1}{d_R\, d_T}\,\text{Tr}\left(|A^{(0)g}_1|^2+|A^{(1)g}_1|^2+2\,\text{Re} \,\left(A^{(0)g}_1\right)^{\dagger}A^{(2c)g}_1\right) \nonumber \\
&& \qquad \qquad \qquad \qquad   =1+\frac{N}{A_{\perp}}\int{d^2\vc{q}_{\perp}} 
\left[\frac{d\sigma_{\text{el}}(R,T)}{d^2\vc{q}_{\perp}} e^{-{\bf q}_\perp \cdot 
\stackrel{\rightarrow}{\nabla}_{{\bf p}_\perp} }   
-\sigma_{\text{el}}\delta^{(2)}(\vc{q}_{\perp})\right] \; .
\end{eqnarray}
Here,
\begin{eqnarray}
\frac{d\sigma^{(g)}_{\text{el}}(R,T)}{d^2\vc{q}_{\perp}}=\frac{C_2(R)C_2(T)}{d_A} \,
\frac{|\tilde{v}(\vc{q}_{\perp})|^2}{(2\pi)^2}\, .
\end{eqnarray}
In this case  the $R$ stands for adjoint representation (gluon jet) and $T$ stands for fundamental representation (fermionic static center).

\section{Medium-induced bremsstrahlung }
\label{radiative}

In this section we use the Feynman rules of \text{\SCETG} to derive the 
 probability for an energetic quark to emit a gluon, induced by the jet interactions in QCD matter. 
This is equivalent to evaluating the differential distribution of the number of emitted gluons.
We first present this calculation in the vacuum using SCET and later in the medium using the new \text{\SCETG} Lagrangian. 
In each case we consider the covariant gauge and the initially static source. We also focus on final-state (FS)
radiation. In the literature, such a calculation is 
typically done in the soft (emitted) gluon approximation. However, in SCET and \text{\SCETG} dynamics, the leading 
interaction describes the collinear gluon emission, which will allow us to go easily beyond the conventional $\omega \ll E$
limit. We will perform this new calculation below in section~\ref{bg}, while in this section we will focus 
on taking the soft gluon limit and comparing to the previously derived results for radiative energy loss in 
QCD matter.
\begin{eqnarray}
A_{\text{brem}}=\bra{J}T\,\bar{\chi}_{n}(x_0)\,\e^{i\,\int d^4 x\left(\mathcal{L}^{\text{QCD}}+\mathcal{L}^{\text{\SCETG}}\right)}\ket{{\bf{p}},{\bf{k}}}\, .
\label{CollAquarkGluonEmission}
\end{eqnarray}
To study  gluon emission, we start from the  matrix element,
Eq.~(\ref{CollAquarkGluonEmission}),
where $J$ is the  underlying hard process that creates the quark jet,
$\bar{\chi}_n$ is the gauge invariant
quark field,  and ${{p}}, {{k}}$ are the momenta of the final state quark
and of the emitted gluon,
correspondingly. Since in this section we consider only the case of the
initial quark jet, we omit the
quark index in the amplitudes below for brevity. The matrix element in
Eq.~(\ref{CollAquarkGluonEmission})  gets contributions from
$0,1,2,...$ Glauber gluon exchanges between the collinear quark and/or
gluon and the sources in the medium.
The first three correspond to vacuum emission, single Born amplitude and two single Born exchanges amplitude, 
respectively, 
and are calculated in the subsections below. To simplify the notation we write the $n-$Glauber insertion 
amplitude in the following 
form\footnote{In our notation $R^{\mu}$ stands for ``Radiation".}:
\begin{eqnarray}
 A^{(n), a}&=&g\,\bar{\chi}_{n,p}\left(\prod_{l=1}^n\int d\Phi_l\right)\,R(q_1,...,q_n)^{(n)\mu, a}
\,i J\left(k+p-\sum_{k=1}^{n}q_k\right)\,\e^{i\left(k+p\right)x_0}\,\eps_{\mu}(k)\; .
\label{BremAn}
\end{eqnarray}

\subsection{Obtaining the Altarelli-Parisi splitting function in SCET}
\begin{figure}[t!]
\begin{center}
\epsfig{file=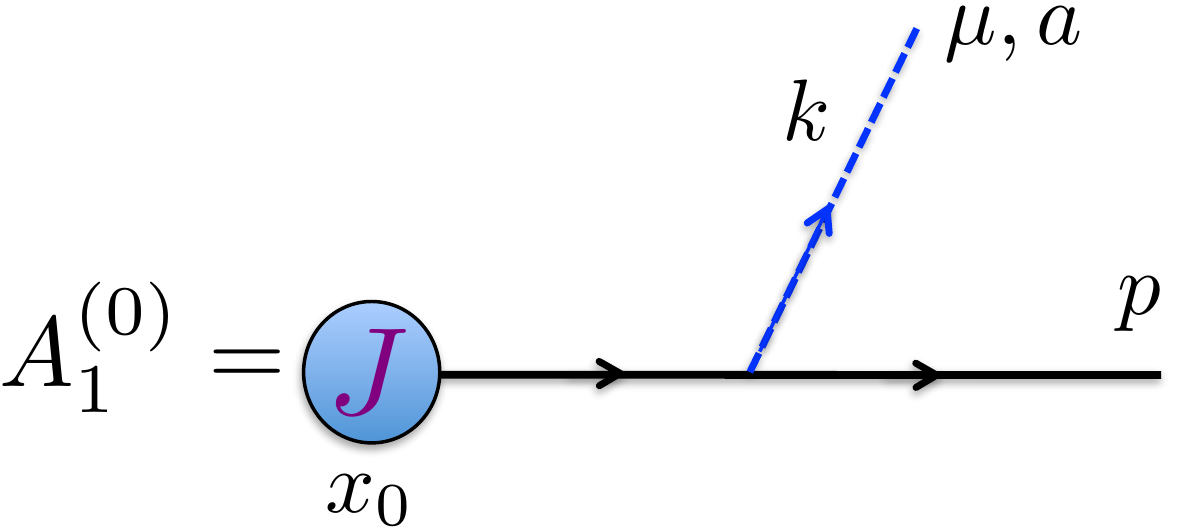, width=5.5cm} \qquad \qquad \qquad\epsfig{file=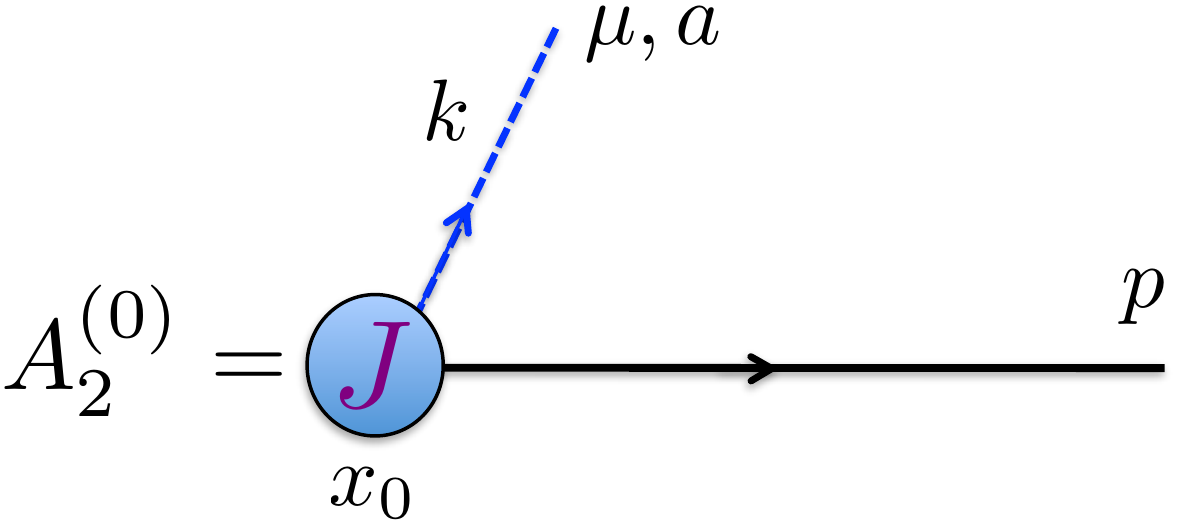, width=5.5cm}
\caption{\label{fig:BremA0} Diagrams in SCET describing splitting of a collinear quark into a collinear 
quark and a collinear gluon.}
\end{center}
\end{figure}

A large $Q^2$ process is accompanied by bremsstrahlung even in the absence of in-medium interactions. 
Knowledge of the corresponding amplitudes is also essential for the evaluation of the interference effects between 
the different sources or radiation for jet production in the QCD medium. 

Calculation of the  vacuum diagrams in figure~\ref{fig:BremA0} leads to the Altarelli-Parisi splitting function for the 
$q\rightarrow qg$ process. This calculation has been performed in Ref.~\cite{Baumgart:2010qf} in the light-cone gauge. 
We perform the same calculation here in the covariant gauge. We also demonstrate how in the small $x=k^+/p^+$ limit
the relevant radiation piece can be identified at the amplitude level.

The calculation of the  relevant radiative matrix element   can be performed using  different set of fields in 
SCET (see~\cite{Bauer:2008qu}): gauge dependent fields $\xi_n$, $A_n$ or gauge independent ones  $\chi_n$ and 
$\mathcal{B}_n$. In the former case the Feynman rules contain Wilson line emissions, 
from the gauge invariant quark field $\chi_n=W_n^\dagger \xi_n$. In the latter case these diagrams are absent, since 
we do calculations directly with $\chi_n$. However, the difference in the second case is that the SCET Lagrangian is 
modified and contains explicit collinear Wilson lines $W_n$ which changes the Feynman rules. The equivalence of different 
formulations was shown in~\cite{Bauer:2008qu} on the example of the 
quark jet function at one loop. We choose the first 
case, i.e. use fields $\xi_n, A_n$ to do the calculation and include the Wilson line graphs, while keeping the SCET 
Lagrangian free from  any collinear Wilson lines. In order to avoid confusion, we note that for the external 
quark spinor 
we use $\bar{\chi}_n$ as before for the broadening, which is simply a matter of notation.

We first recall that for the jet production amplitude we have: 
\begin{eqnarray} 
    A_{J q}&=&\bar{\chi}_{n,p}\, iJ\,\e^{ipx_0} \; ,  \qquad \frac{1}{d_Rd_T} \sum_{\rm spin,color}  d\sigma \propto  |  A_{J q} |^2 =  
\text{Tr}\left(\frac{\nslash}{2}\,\bar{n}\mcdot p\,J(p)\bar{J}(p)\right) \; . \label{SCET0} 
\end{eqnarray}
One notices from Eq.~(\ref{SCET0}) that the contribution to the cross section comes from the part
 $J(p)\bar{J}(p)  \propto \gamma^+ $.
For the case of gluon bremsstrahlung,  the first  amplitude in figure~\ref{fig:BremA0} reads:
\begin{eqnarray}
   R^{(0)\mu, a}_{1}&=&i\,T^a \left(n^{\mu}+\frac{\gamma^{\mu}_{\perp}(\pslash_{\perp}+\kslash_{\perp})}{\bar{n}\mcdot (p+k)}+\frac{\pslash_{\perp}\gamma^{\mu}_{\perp}}{\bar{n}\mcdot p}-\frac{\pslash_{\perp}(\pslash+\kslash)_{\perp}}{\bar{n}\mcdot p  \,\,\bar{n}\mcdot (p+k)}\bar{n}^{\mu}\right) i\frac{\bar{n}(p+k)}{(p+k)^2}\label{SCETBrem}\;  .
\end{eqnarray}
Even though we work in the covariant gauge, we are going to use physical polarization vectors for the emitted real gluon. We choose the following gluon polarization vectors, which are the only possible solutions that satisfy the two conditions
$\bar{n}\mcdot \eps=0,  \,k\mcdot \eps(k)=0$:
\begin{equation}
   \eps^{\mu}_{i}(k) =\left(0, \frac{2\vc{\eps}_{i\perp}\mcdot {\bf{k}}_{\perp}}{k^+}, {\vc{\eps}}_{i\perp}\right), \, \quad i=1,2 \;.\label{polvectors}
\end{equation}
For such polarization vectors the second diagram in figure~\ref{fig:BremA0} vanishes, since it is proportional to $\bar{n}^{\mu}$, and also the last term in~\eq{SCETBrem} vanishes once dotted with the polarization vector. By contracting the polarization vectors from~\eq{polvectors} with the SCET amplitude in~\eq{SCETBrem}, we get\footnote{Note that in~\eq{R01eq} all indices are contravariant, while for example terms in~\eq{SCETBrem} contain contractions between covariant and contravariant vectors, like $\pslash_{\perp}\equiv p^{\mu}\gamma_{\perp\mu}=-\vc{p}^i_{\perp}\,\gamma_{\perp}^i$.}:
\begin{eqnarray}
   && R^{(0)\mu, a}_{1}\eps_{\mu}=-T^a\,\left[\frac{2{\vc{A}}^i_{\perp}}{{\vc{A}}_{\perp}^2}+\frac{x}{\vc{A}_{\perp}^2}\,
\vc{A}_{\perp}^j\,\gamma^i_{\perp}\,\gamma^j_{\perp}\right]\, \vc{\eps}_{\perp}^i\; ,\label{R01eq}\\
   &&\vc{A}_{\perp}\equiv \vc{k}_{\perp}(1-x)-\vc{p}_{\perp} x\; ,
\end{eqnarray}
where $x$ is the fraction of energy taken by the emitted gluon $k^+=x p_0^+, \;  p^+=(1-x)p_0^+$. To derive 
\eq{R01eq} we also assumed that the external quark and gluon are on-shell.
Squaring this amplitude, averaging over the colors of the radiating quark  we obtain:
\begin{eqnarray}
   && \frac{1}{d_R} |A_{J q\rightarrow  q g}|^2 =\frac{g^2}{d_R}\,\text{Tr}\left(\frac{\nslash}{2}\,\bar{n}\mcdot p\,J\bar{J}\,\left[\gamma^0\left(R^{(0)}_{1}\mcdot\eps\right)^{\dagger} \gamma^0\left(R^{(0)}_{1}\mcdot\eps\right)\right]\right)\; ,\\
   &&\gamma^0\left(R^{(0)}_{1}\mcdot\eps\right)^{\dagger} \gamma^0\left(R^{(0)}_{1}\mcdot\eps\right) = 4  C_F \left(1 - x + \frac{x^2}{2} \right)\frac{1}{\vc{A}_{\perp}^2} \,\left(\mathbb{I}\right)_{\text{color}} \left(\mathbb{I}\right)_{\text{Dirac}}\; .
\label{factorizationEq}
\label{ZeroRad} 
\end{eqnarray}
The fact that the expression in~\eq{factorizationEq} is a Dirac scalar means that the cross section of 
splitting factorizes into the cross section of jet production times the Altarelli-Parisi kernel, which has a 
nice probabilistic interpretation. Setting the light-cone direction along the initial quark before splitting, 
i.e. $\vc{p}_{\perp}=-\vc{k}_{\perp}$, we reproduce the  Altarelli-Parisi kernel for $q\rightarrow qg$ splitting:
\begin{eqnarray}
   && \frac{1}{d_R} |A_{J q\rightarrow  q g}|^2 = (1-x)
\text{Tr}\left(\frac{\nslash}{2}p_0^+J(0)\bar{J}(0)\right) 
\times  4 g^2 C_F \left(1 - x + \frac{x^2}{2} \right)\frac{1}{\vc{k}_{\perp}^2} = 
 |    A_{J q} |^2  \times |M^{\rm rad}_0  |^2  \; .
\label{ZeroRad1} 
\end{eqnarray}
We note that the $1-x$ factor is associated with the reduction of the cross section to have 
a high momentum final-state quark when a gluon is emitted.
From Eq.~(\ref{ZeroRad1}) we can identify the radiative correction and supplying the one body
phase space, written as $dk^+ d^2\vc{k}_{\perp}/(2\pi)^3 2k^+$, we obtain:
\begin{equation}
\frac{dN^g}{ dxd^2\vc{k}_{\perp} }  = C_F \frac{\alpha_s}{\pi^2} 
\frac{\left(1 - x + \frac{x^2}{2} \right)}{x}\frac{1}{\vc{k}_{\perp}^2} \; .
\label{APfin}
\end{equation}
One easily recognizes in Eq.~(\ref{APfin}) the usual Altarelli-Parisi splitting probability for quarks 
(up to $x\rightarrow 1-x$).
Taking the small-$x$ limit one can conveniently reproduce the boundary condition~\cite{Gyulassy:2000er}
associated with hard jet production that is subsequently used in the reaction operator approach
to parton energy loss~\cite{Gyulassy:2000er,Vitev:2007ve}.

We finally point out that in the soft gluon limit $x\ll 1$ the amplitude itself, 
Eq.~(\ref{R01eq}), reduces as follows: 
\begin{eqnarray}
   && A_{J q\rightarrow q g} \approx \,\e^{ipx_0}\,\bar{\chi}_{n,p} J 
\left(-g \frac{2\vc{\eps}_{i\perp}\mcdot \vc{k}_{\perp}}{\vc{k}_{\perp}^2}\right) T^a 
= A_{J q} \times M_0^{\rm rad} \; ,
\end{eqnarray}
and, just like in quantum electrodynamics in the soft photon limit, a radiation matrix 
element  can be identified at the amplitude level:
\begin{eqnarray}
   && M_0^{\rm rad} = -g \frac{2\vc{\eps}_{i\perp}\mcdot \vc{k}_{\perp}}{\vc{k}_{\perp}^2}\, T^a \; .
\end{eqnarray}

\subsection{Single Born amplitudes  in $\text{\SCETG}$}\label{singleborn}
\begin{figure}[t!]
\begin{center}
\epsfig{file=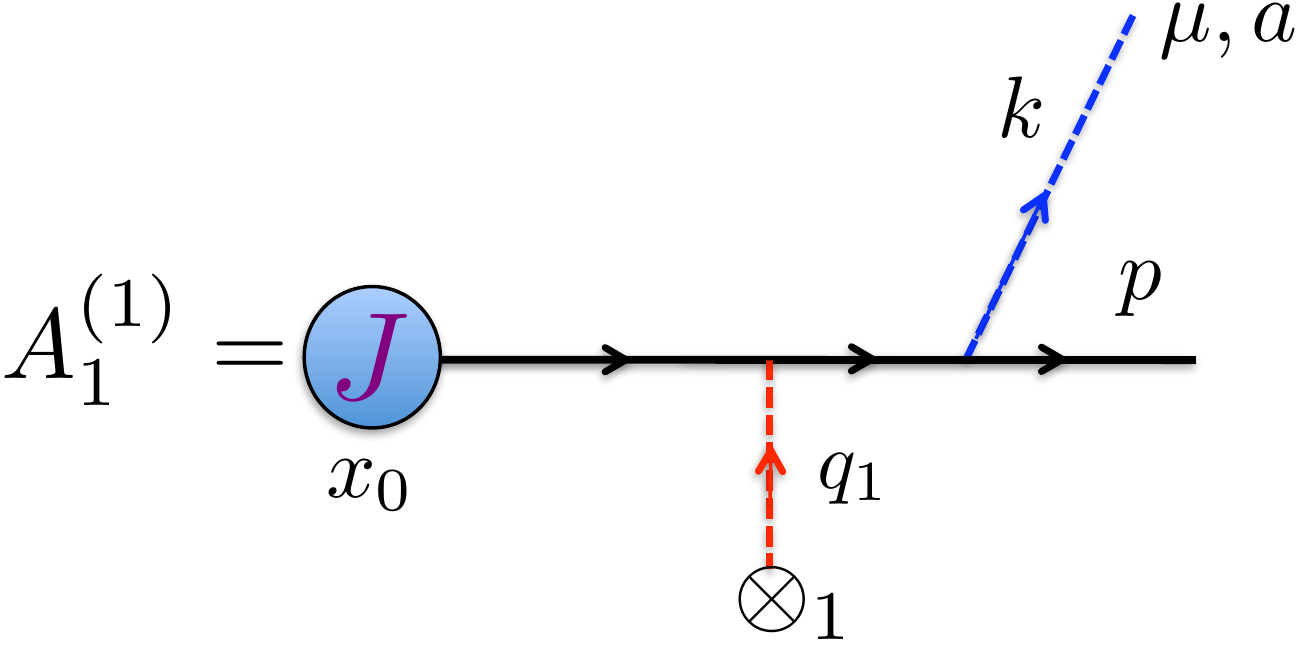, width=5.5cm}\epsfig{file=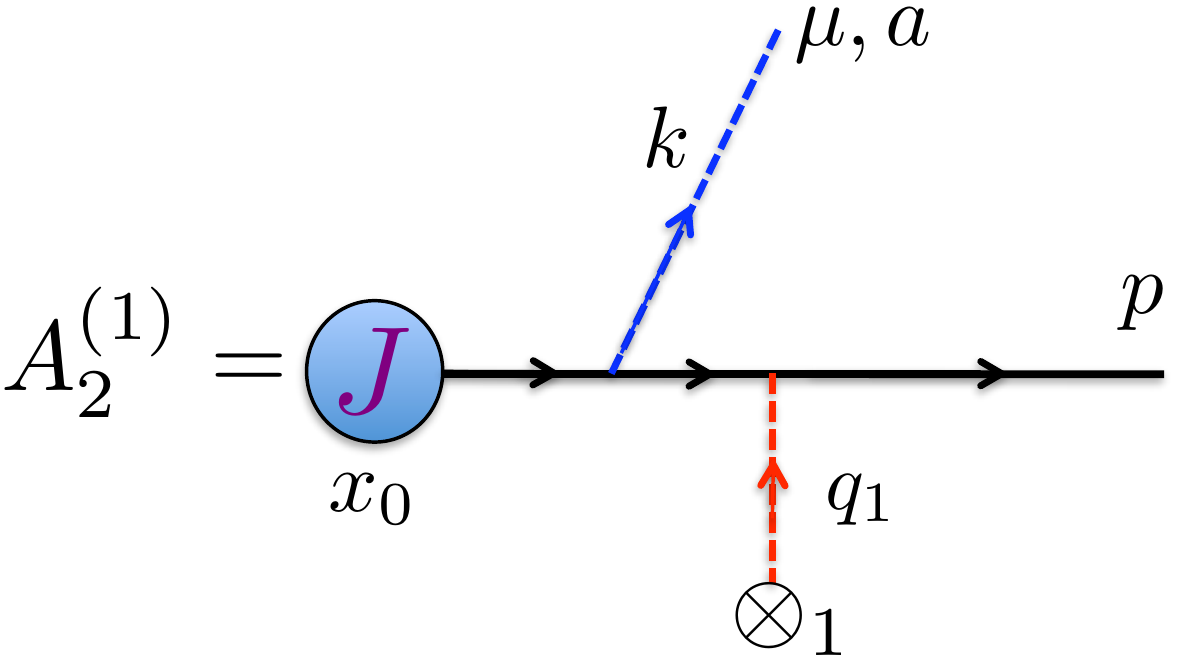, width=5.5cm}\epsfig{file=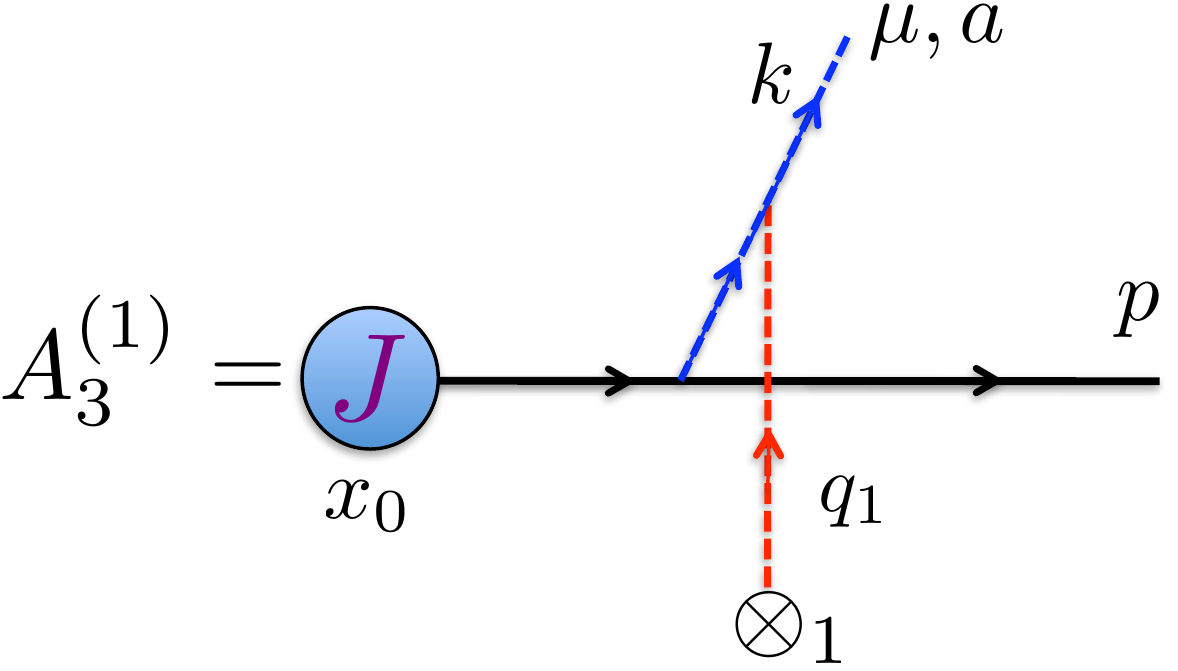, width=5.5cm}\\
\epsfig{file=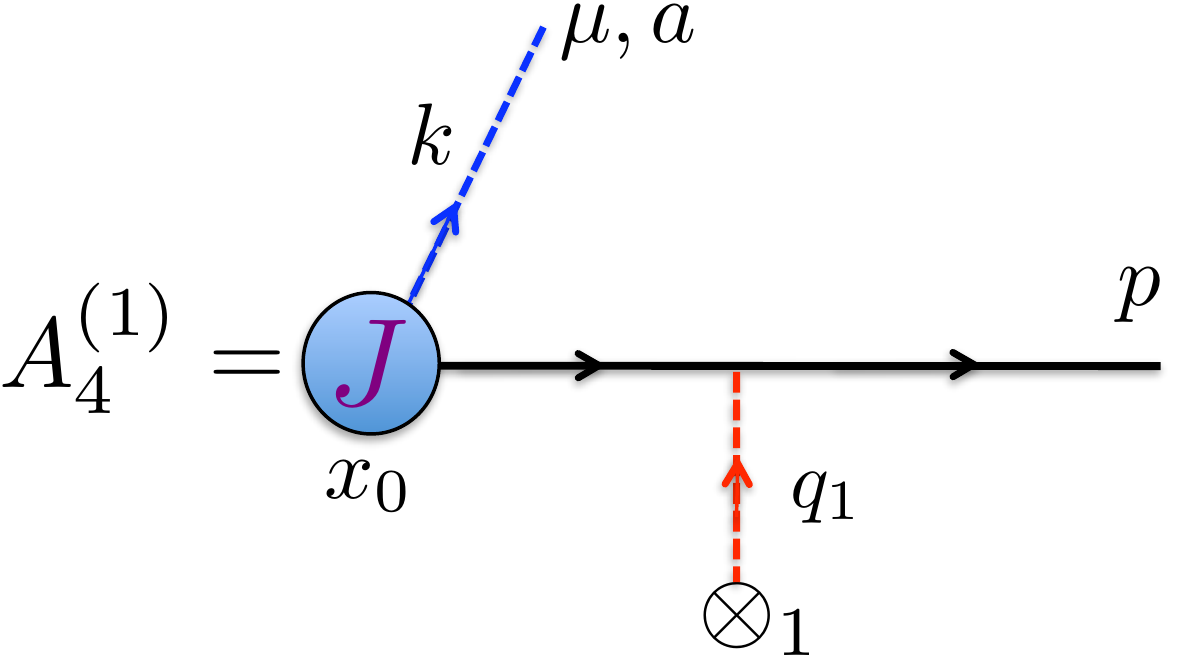, width=5.5cm}\qquad\qquad\epsfig{file=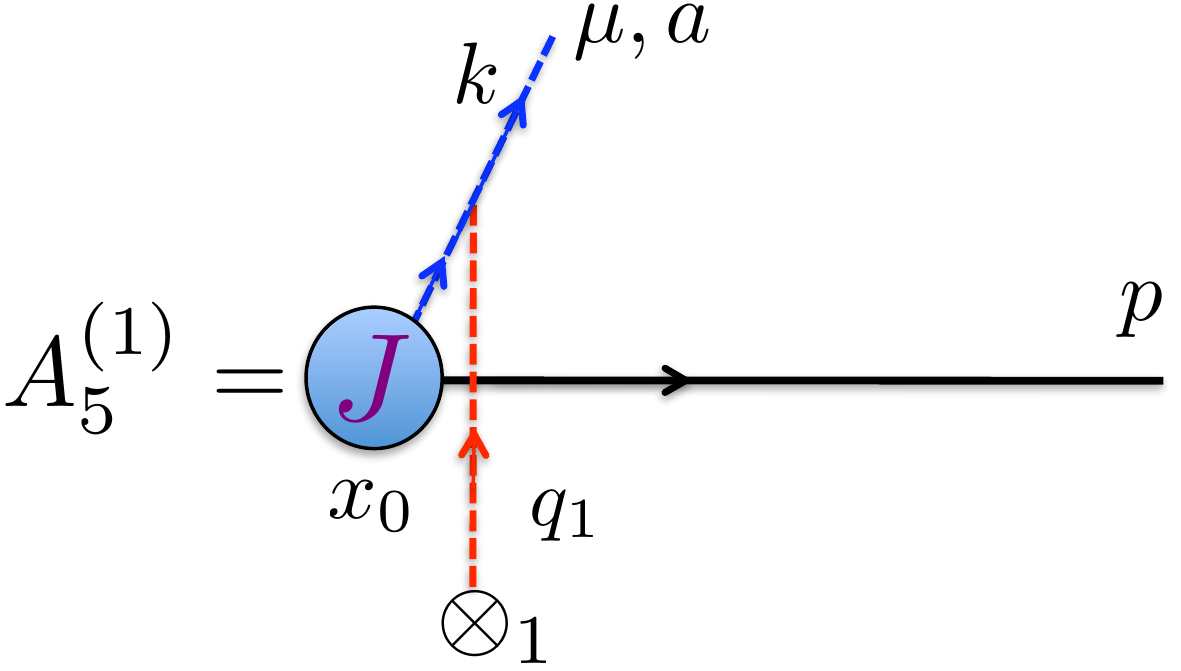, width=5.5cm}
\caption{\label{fig:BremA1} Single Born diagrams contributing to the medium-induced gluon
bremsstrahlung Eq.~({\protect\ref{BremAn})}. The notation for the scattering centers is the following: $\otimes_{1}=[x_1,q_1, (b_1)_i]$.}
\end{center}
\end{figure}

In this subsection we derive the single Born amplitudes in the fully covariant gauge for the scattering off the 
initially static source. In figure~\ref{fig:BremA1} we list all the relevant single Born diagrams. 
In this  gauge two additional diagrams $A_4, A_5$ appear when the collinear gluon appears from 
the Wilson line $W^{\dagger}$. Feynman rule for such collinear gluon vertex is well known: 
$$\Gamma_W^{\alpha, a}(k)=g T^a_r \frac{\bar{n}^{\alpha}}{k^++i \epsilon} \; , $$ 
where $k$ is the outgoing gluon momentum from the Wilson line.
Although we use the $R_{\xi}$ gauge in this subsection, just as in the vacuum case above, we use the physical 
gluon polarization vector for the emitted gluon ~\eq{polvectors}. With the Feynman rules of \text{\SCETG}, 
derived in section~\ref{scetg}, and the notation of~\eq{BremAn}, we get the following expressions for the 
amplitudes in figure~\ref{fig:BremA1}:
\begin{eqnarray}
   R^{(1){\mu, a}}_1&=& i(a)_R\left(n^{\mu}+\frac{\gamma^{\mu}_{\perp} 
(\pslash_{\perp}+\kslash_{\perp}) }{p^++k^+}+\frac{ \pslash_{\perp} \gamma^{\mu}_{\perp}}{p^+}\right) 
\frac{i(p^++k^{+})}{(p+k)^2+i \eps}i \,\,(b_1)_R\,(b_1)_{T_i}\,i\Delta_g(p+k,q_1)\; , \label{eq:BremA11} \\
   R^{(1){\mu, a}}_2&=&i(b_1)_R (b_1)_{T_i} i \Delta_g(p,q_1) i (a)_R\left(n^{\mu}+\frac{ \gamma^{\mu}_{\perp}(\pslash_{\perp}+\kslash_{\perp}-\qslash_{1\perp}) }{p^+}+\frac{ (\pslash_{\perp}-\qslash_{1\perp}) \gamma^{\mu}_{\perp}}{p^+}\right)i\Delta_g(p+k,q_1)\, , \qquad \label{eq:BremA12}\\
     R^{(1){\mu, a}}_3&=& i (c_1)_R\left(n^{\rho_1}+\frac{\gamma^{\rho_1}_{\perp} (\pslash_{\perp}+\kslash_{\perp}-\qslash_{1\perp}) }{p^+}+\frac{ \pslash_{\perp} \gamma^{\rho_1}_{\perp}}{p^+}\right)  \nonumber \\ 
&& \times\, \frac{(-i)\Delta_g(k,q_1)}{\bar{n}\mcdot k}\,N^{(R_{\xi})}_{\rho_1\rho_2}(k-q_1)\, i\Delta_g(p+k,q_1)\,f^{c_1 a b_1} (b_1)_{T_i}\,\tilde\Sigma_1^{\rho_2\mu}(k-q_1,k),\label{eq:BremA13}\\
     R^{(1){\mu, a}}_4&=&0\; ,\label{eq:BremA14}\\
     R^{(1){\mu, a}}_5&=&0\; . \label{eq:BremA15}
\end{eqnarray}
Note that in the collinear gluon vertices in diagrams $A_1, A_2$ we omitted the last term 
proportional to $\bar{n}^{\mu}$ because after contraction with the polarization vector of our choice 
this term vanishes, since $\eps^+=0$. For exactly the same reason diagram $A_4$ vanishes. However, the 
reason why we ignored $\bar{n}^{\rho_1}$ in diagram $A_3$ and why $A_5$ vanishes, is slightly more involved. 
The point is that both $A_3$ and $A_5$ have a common factor given by~\eq{RxiIdentity} with 
$p \rightarrow k$. Since from this identity it is obvious that $\bar{n}^{\rho_1}$ times this combination vanishes, we are allowed to omit this term in $A_3$. For the same reason $A_5$=0.

In order to reduce the integral $d\Phi_1 R^{(1)}_i$ to the $d\vc{\Phi}_{1\perp}$ integral we use the identity in 
\eq{jacobianLCC}. Also, substituting~\eq{RxiIdentity} into the expression for { \bf $R_3^{(1)}$} makes it obvious that 
the entire dependence on $q_1^-$ appears through the propagators $\Delta_g(p,q)$. Using the form of this 
propagator from~\eq{GlauberProp} we define the relevant longitudinal integrals $I_1^{(1)}, I_2^{(1)}, I_3^{(1)}$. 
We evaluate these  integrals in appendix~\ref{appendix:integralsRad}. Thus, using 
\eq{jacobianLCC} and the expressions for  $I_1^{(1)}, I_2^{(1)}, I_3^{(1)}$ from appendix~\ref{appendix:integralsRad},
 we get after taking limit $x\ll1$ in $R^{(1)}_{1,2,3}$:
\begin{eqnarray}
      &&\int d\Phi_1 R^{(1)\mu, a}_1\eps_{\mu}(k)\approx  (-i)\,(ab_1)_R\,(b_1)_{T_i} 
\frac{2\vc{k}_{\perp}\mcdot\vc{\eps}_{\perp}}{\vc{k}_{\perp}^2}    \int  d\vc{\Phi}_{1\perp}\, \e^{i\omega_0\delta z_1}\; ,\\
      &&\int d\Phi_1 R^{(1)\mu, a}_2\eps_{\mu}(k)\approx (-i)\, (b_1a)_R\,(b_1)_{T_i}\,  \frac{2\vc{k}_{\perp}\mcdot\vc{\eps}_{\perp}}{\vc{k}_{\perp}^2} \int  d\vc{\Phi}_{1\perp}\, \left[1-\e^{i\omega_0\delta z_1}\right]\; ,\\
      &&\int d\Phi_1 R^{(1)\mu, a}_3\eps_{\mu}(k)\approx (-i)  \left[a,b_1\right]_R\,(b_1)_{T_i} \int  d\vc{\Phi}_{1\perp}\,\frac{2\left(\vc{k}_{\perp}-\vc{q}_{1\perp}\right)\mcdot\vc{\eps}_{\perp}}{(\vc{k}_{\perp}-\vc{q}_{1\perp})^2} 
\, {\e^{i\omega_0\delta z_1}}\left[\e^{-i\omega_1\delta z_1}-1\right]\; .
\end{eqnarray}
The two inverse formation times $\omega_0$ and $\omega_1$ are defined according to:
\begin{eqnarray}
&&\omega_0=\frac{\vc{k}_{\perp}^2}{xp_0^+}\; ,\qquad
\qquad\omega_1=\frac{(\vc{k}_{\perp}-\vc{q}_{1\perp})^2}{x p_0^+} \; .\label{omega01brem}
\end{eqnarray}

\subsection{Double Born amplitudes  in $\text{\SCETG}$}
\begin{figure}[t!]
\begin{center}
\epsfig{file=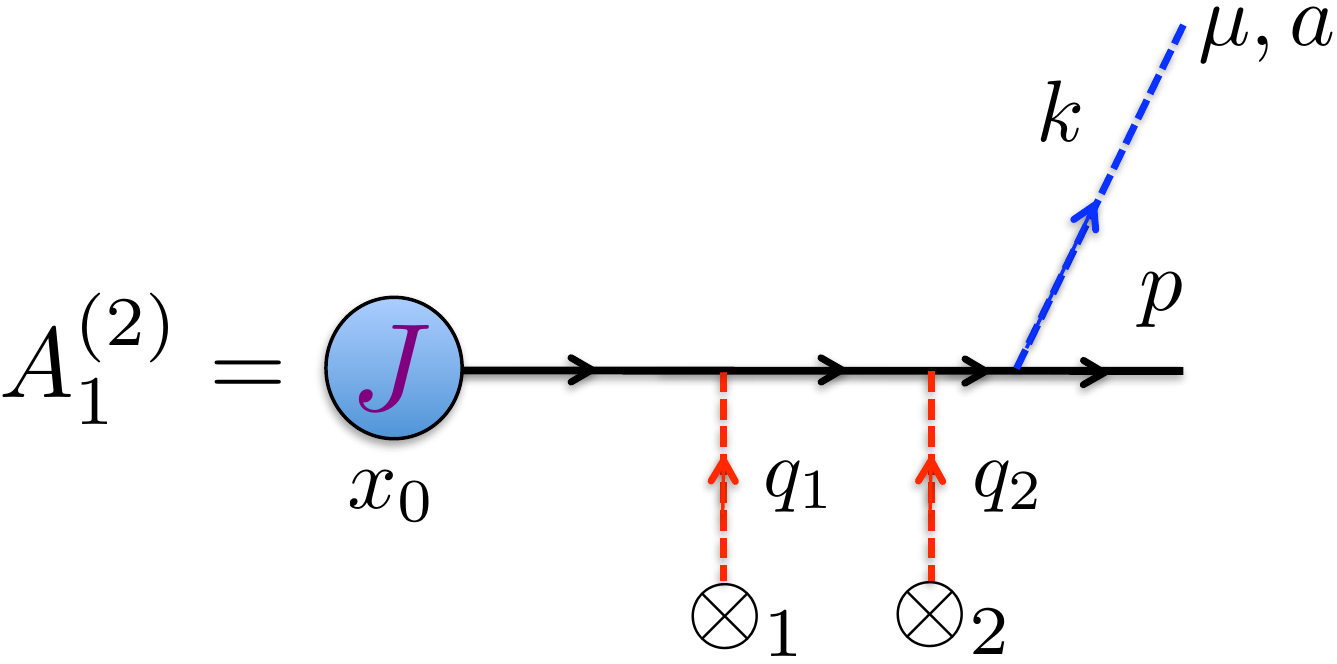, width=5.5cm}\epsfig{file=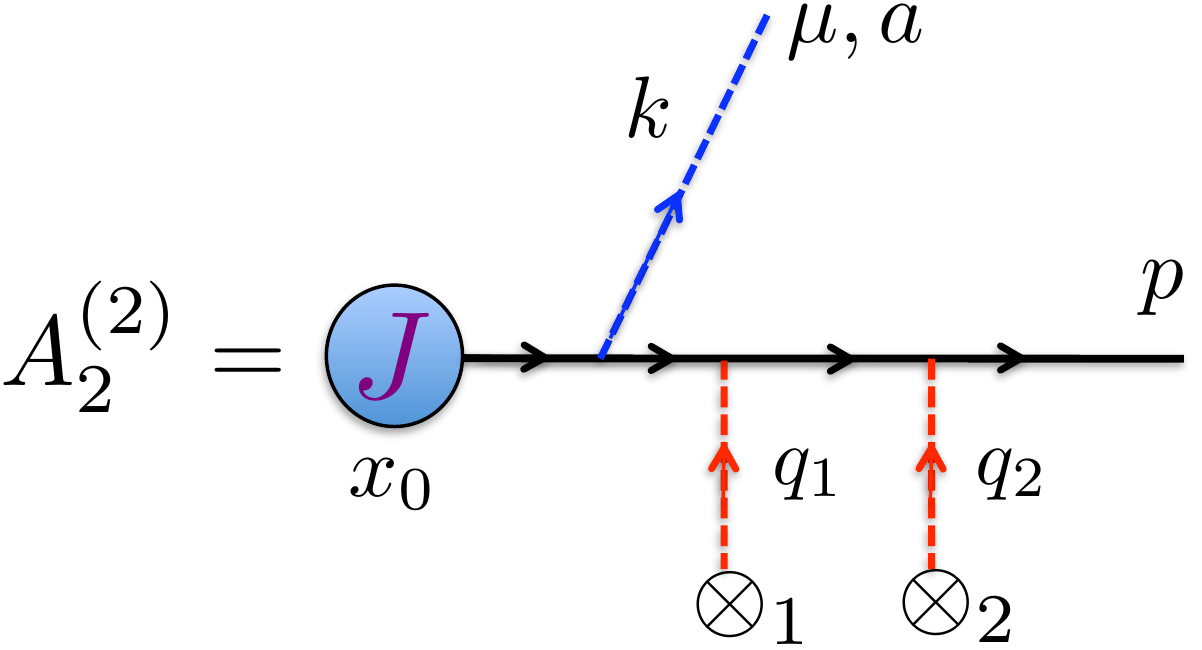, width=5.5cm}\epsfig{file=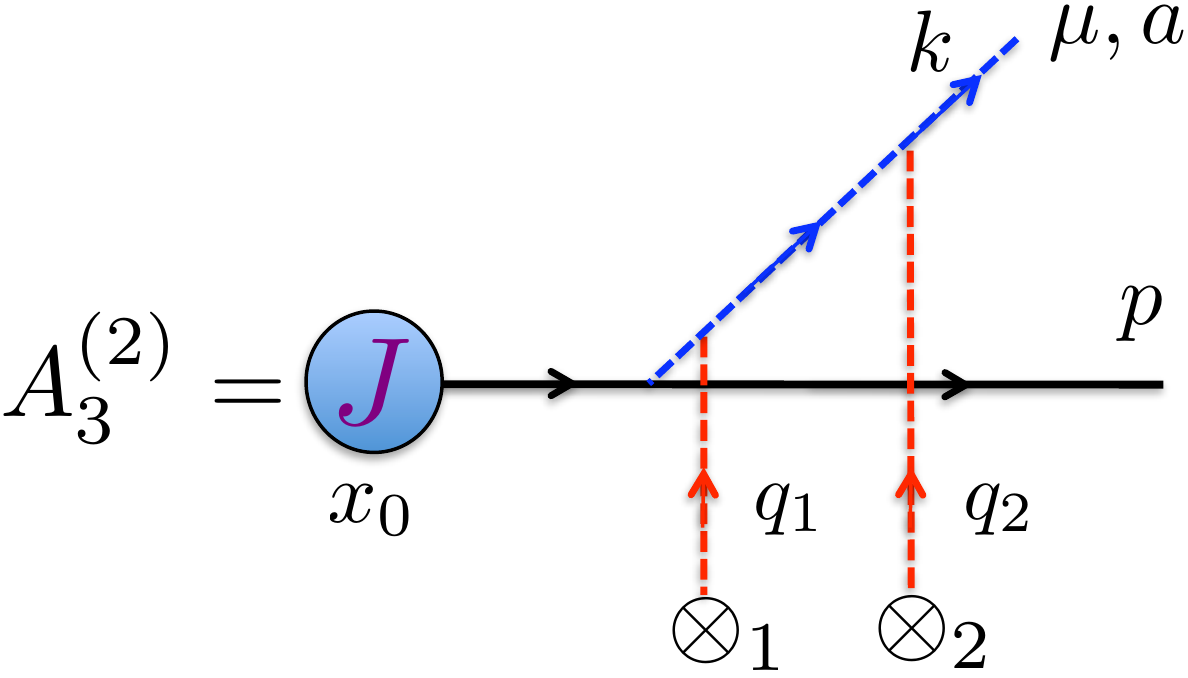, width=5.5cm}\\
\epsfig{file=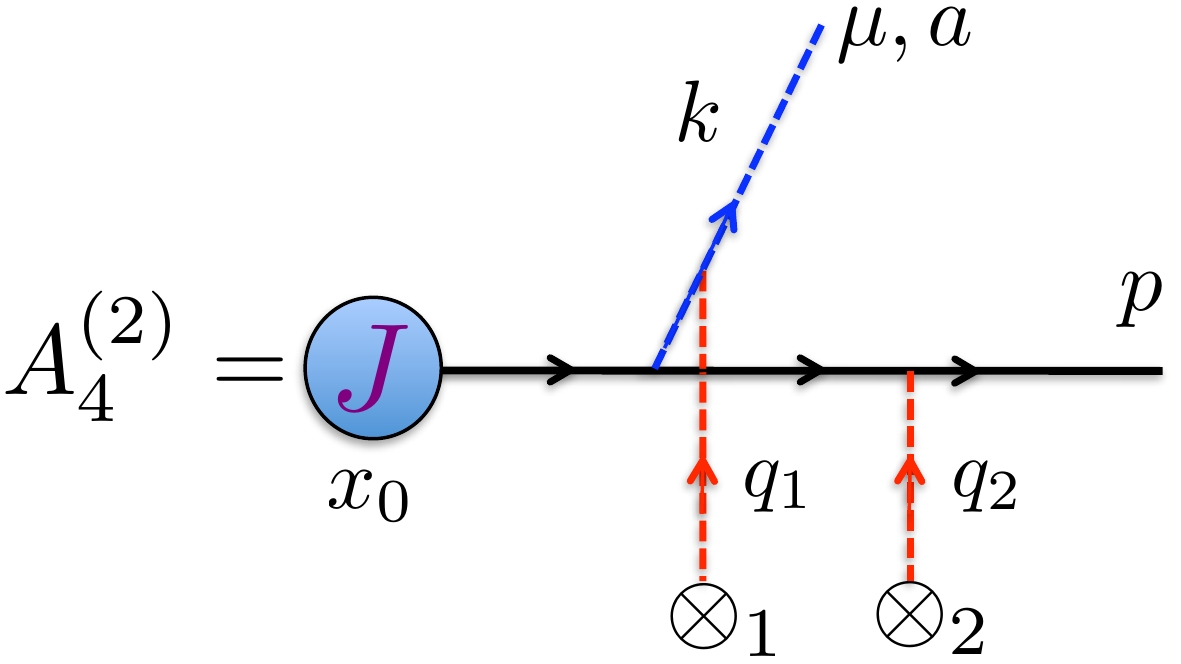, width=5.5cm}\epsfig{file=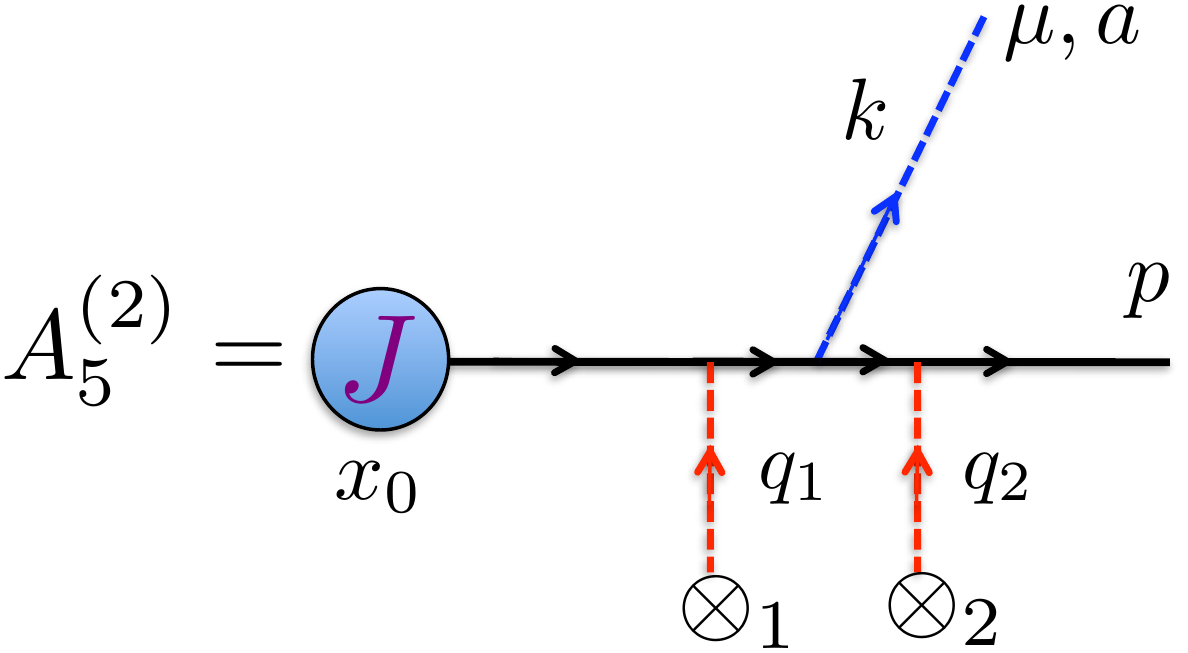, width=5.5cm}\epsfig{file=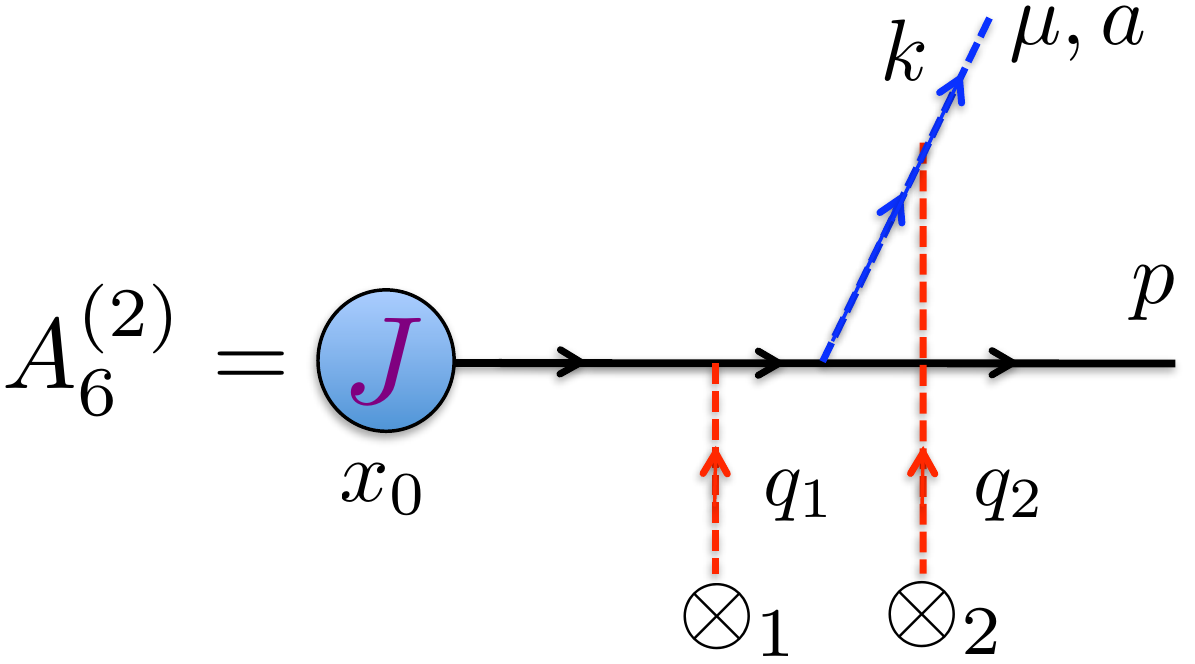, width=5.5cm}\\
\epsfig{file=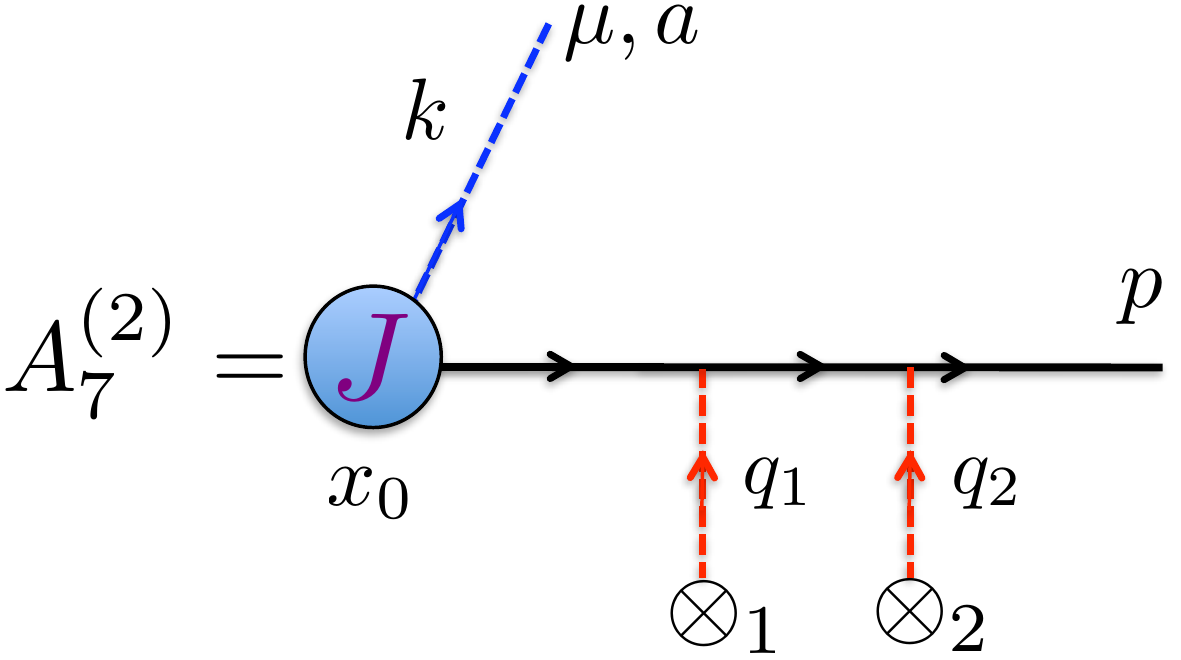, width=5.5cm}\epsfig{file=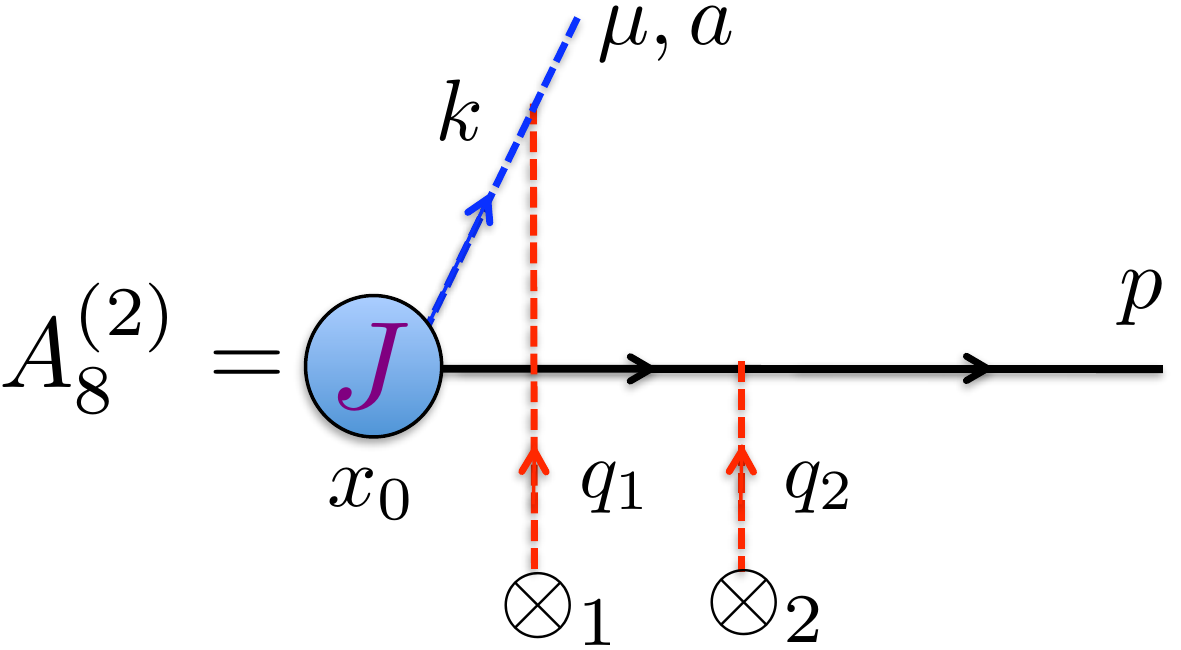, width=5.5cm}\epsfig{file=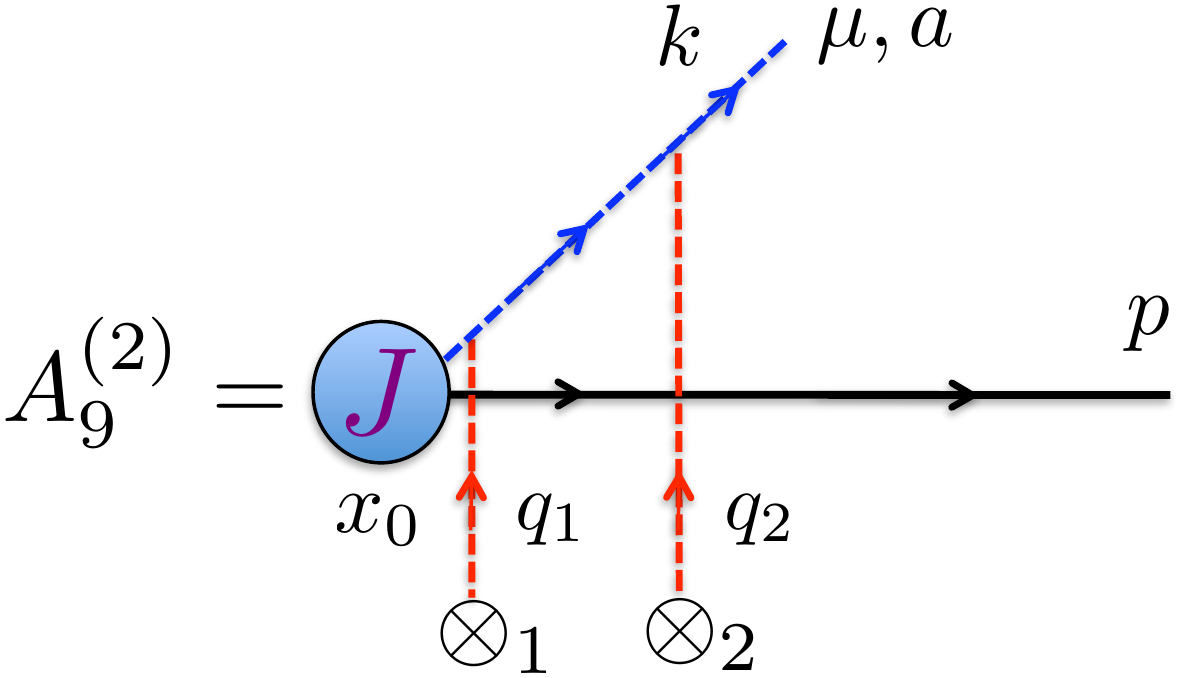, width=5.5cm}
\caption{\label{fig:BremA2} Two single Born exchange diagrams (double Born diagrams in the contact limit) 
contributing to matrix element in Eq.~({\protect\ref{BremAn})}. The notation for the scattering centers is the following: $\otimes_{1}=[x_1,q_1, (b_1)_i], \otimes_{2}=[x_2,q_2, (b_2)_j]$.}
\end{center}
\end{figure}

In this section we calculate all diagrams in figure~\ref{fig:BremA2} in the $R_{\xi}$ gauge for the initially 
static source. Expressions for all of these diagrams are obtained directly from the Feynman rules of $\text{\SCETG}$ 
in the covariant gauge, and are straightforward, though lengthy. However, a nice compact relations can be found 
for these diagrams by relating them to previously calculated amplitudes $R^{(0)}, \, R^{(1)}$ and 
$B^{(1)q}, \,B^{(2)q}, \,B^{(1)g}, \,B^{(2)g}$, in the notation of~\eq{Akquark},~\eq{Akgluon} and~\eq{BremAn}. The result 
is as follows:
\begin{eqnarray}
&&\left(R_1^{(2)}\right)^{\mu, a}=R^{(1)\mu, a}_1(p,k,q_2)\, B^{(1)q}_1(p+k-q_2,q_1)\; , \label{M12Def}\\
  &&\left(R_2^{(2)}\right)^{\mu, a}=B^{(1)q}_1(p,q_2)\,R^{(1)\mu, a}_2(p-q_2,k,q_1)\; , \label{M22Def}\\
     &&\left(R_3^{(2)}\right)^{\mu, a}= R_1^{(0)\rho_1, b}(p,k-q_1-q_2)\, \left(B_{1}^{(2)g}\right)^{\mu\rho_1, b a}(k,q_1,q_2)\; ,  \label{M32Def}\\
      &&\left(R_4^{(2)}\right)^{\mu, a}= B^{(1)q}_1(p,q_2)\,R^{(1)\mu, a}_3(p-q_2,k,q_1)\; , \label{M42Def} \\
     &&\left(R_5^{(2)}\right)^{\mu, a}=R^{(1)\mu, a}_2(p,k,q_2)\, B^{(1)q}_1(p+k-q_2,q_1)\; ,  \label{M52Def}\\ 
      &&\left(R_6^{(2)}\right)^{\mu, a}= R^{(1)\mu, a}_3(p,k,q_2)\, B^{(1)q}_1(p+k-q_2,q_1)\; ,\label{M62Def} \\
      &&\left(R_{7,8,9}^{(2)}\right)^{\mu, a }=0\; .\label{M7892Def} 
\end{eqnarray}
Next, we integrate over the longitudinal momenta and reduce the integrals to transverse ones. This is analogous 
to the procedure for the single Born diagrams. The only poles in $q_i^-$ come from propagators of collinear momenta 
plus Glauber momenta (given by~\eq{GlauberProp}). We summarize the corresponding integrals in 
appendix~\ref{appendix:integralsRad}. Since the amplitudes $A_{7,8,9}$ vanish for physical polarization of the 
radiated gluon, we do not even consider the corresponding longitudinal integrals. Finally, using our results for 
the corresponding longitudinal integrals $I^{(2)}_{1}\, , \cdots   I^{(2)}_{6}$ 
from appendix~\ref{appendix:integralsRad} and taking the
$x \ll 1$ approximation, we get the following expressions:
\begin{eqnarray}
\int d\Phi_1 d\Phi_2\,R_1^{(2)\mu, a}\eps_{\mu}(k)&=&\,\left(a b_2 b_1\right)_R\,(b_1)_{T_i}(b_2)_{T_j}\,\int d \vc{\Phi}_{1\perp} d \vc{\Phi}_{2\perp}\frac{2\vc{k}_{\perp}\mcdot \vc{\eps}_{\perp}}{\vc{k}_{\perp}^2}\e^{i\omega_0 \delta z_2}\; ,\\
\int d\Phi_1 d\Phi_2\,R_2^{(2)\mu, a}\eps_{\mu}(k)&=&\,\left(b_2 b_1 a\right)_R\,(b_1)_{T_i}(b_2)_{T_j}\,\int d \vc{\Phi}_{1\perp} d \vc{\Phi}_{2\perp}\frac{2\vc{k}_{\perp}\mcdot \vc{\eps}_{\perp}}{\vc{k}_{\perp}^2}\left(1-\e^{i\omega_0 \delta z_1}\right)\; ,\\
\int d\Phi_1 d\Phi_2\,R_3^{(2)\mu, a}\eps_{\mu}(k)&=&\,\left[\left[a, b_2\right], b_1\right]_R\,(b_1)_{T_i}(b_2)_{T_j}\int d \vc{\Phi}_{1\perp} d \vc{\Phi}_{2\perp} \nonumber \\
&& \times \,\frac{2(\vc{k}_{\perp}-\vc{q}_{1\perp}-\vc{q}_{2\perp})\mcdot \vc{\eps}_{\perp}}{(\vc{k}_{\perp}-\vc{q}_{1\perp}-\vc{q}_{2\perp})^2}\left(\e^{i(\omega_0-\omega_{12})  \delta z_1}-\e^{i\omega_{0}  \delta z_1}\right)\e^{i(\delta z_2-\delta z_1)(\omega_0-\omega_1)}\; ,\\
\int d\Phi_1 d\Phi_2\,R_4^{(2)\mu, a }\eps_{\mu}(k)&=&\left(b_2\left[a, b_1 \right]\right)_R\,(b_1)_{T_i}(b_2)_{T_j}\,\int d \vc{\Phi}_{1\perp} d \vc{\Phi}_{2\perp}\nonumber\\
&& \times \frac{2(\vc{k}_{\perp}-\vc{q}_{1\perp})\mcdot \vc{\eps}_{\perp}}{(\vc{k}_{\perp}-\vc{q}_{1\perp})^2}\left(\e^{i(\omega_{0}-\omega_1)\delta z_1}-\e^{i\omega_0 \delta z_1}\right)\; ,\\
\int d\Phi_1 d\Phi_2\,R_5^{(2)\mu, a}\eps_{\mu}(k)&=&\,\left(b_2 a b_1 \right)_R\,(b_1)_{T_i}(b_2)_{T_j}\,\int d \vc{\Phi}_{1\perp} d \vc{\Phi}_{2\perp}\frac{2\vc{k}_{\perp}\mcdot \vc{\eps}_{\perp}}{\vc{k}_{\perp}^2}\left(1-\e^{i\omega_0 (\delta z_2-\delta z_1)}\right)\e^{i\omega_0 \delta z_1}\; ,\\
\int d\Phi_1 d\Phi_2\,R_6^{(2)\mu, a}\eps_{\mu}(k)&=&\left(\left[a, b_2 \right] b_1\right)_R\,(b_1)_{T_i}(b_2)_{T_j}\,\int d \vc{\Phi}_{1\perp} d \vc{\Phi}_{2\perp}\nonumber\\
&&\times\frac{2(\vc{k}_{\perp}-\vc{q}_{2\perp})\mcdot \vc{\eps}_{\perp}}{(\vc{k}_{\perp}-\vc{q}_{2\perp})^2}\left(\e^{-i\omega_2 (\delta z_2-\delta z_1)}-1\right)\e^{i\omega_0 \delta z_2}\; , \qquad
\end{eqnarray}
where $\omega_0$ and $\omega_1$ are defined in~\eq{omega01brem} above and $\omega_2, \omega_{12}$ are equal to:
\begin{eqnarray}
&&\omega_2=\frac{(\vc{k}_{\perp}-\vc{q}_{2\perp})^2}{xp_0^+}\; ,\qquad
\qquad\omega_{12}=\frac{(\vc{k}_{\perp}-\vc{q}_{1\perp}-\vc{q}_{2\perp})^2}{x p_0^+} \; .
\end{eqnarray}
In order to understand the lowest opacity contribution to the induced bremsstrahlung, one needs to combine the 
single Born diagrams computed in the 
previous section with the contact double Born limit of the two single Born exchange diagrams. The contact limit 
of two single  Born exchange longitudinal 
integrals is derived in appendix~\ref{appendix:integralsRad}. Using these results we get the following contact 
limits (or double Born amplitudes):
\begin{eqnarray}
R^{(2c)}_{1,2,3}&=&\frac{1}{2}R^{(2)}_{1,2,3}(\delta z_2=\delta z_1)\; ,\\
R^{(2c)}_{4}&=&R^{(2)}_{4}(\delta z_2=\delta z_1)\; ,\\
R^{(2c)}_{5,6}&=&0\; .
\end{eqnarray}
All results in this and the previous subsections, derived in the framework of {\rm \SCETG}, agree with the 
soft gluon approximation  previously derived in the literature~\cite{Gyulassy:2000er,Vitev:2007ve}.  
For example, to first order in opacity and without 
explicitly showing the integral over the position of the scattering center we find: 
\begin{equation}
k^+ \frac{dN^g(FS)}{dk^+ d^2 {\bf k}_\perp } = \left( \frac{N}{A_{\perp}} \right)
\frac{C_F \alpha_s}{\pi^2}
\int{d^2\vc{q}_{\perp}} 
\left[\frac{d\sigma_{\text{el}}(R,T)}{d^2\vc{q}_{\perp}} \right] 
\left(\frac{2\vc{k}_\perp\mcdot \vc{q}_\perp}{\vc{k}_\perp^2 (\vc{k}_\perp-\vc{q}_\perp )^2}\right)
\left(1-\cos\left[\frac{(\vc{k}_\perp-\vc{q}_\perp )^2}{k^+} \delta z\right]\right) \, .     \quad      
\end{equation}
Note that in the equation above the superscript ``g'' stands for the radiated gluons and should
not be confused with Glauber gluons, denoted by ``G''.
However, we can go beyond that and calculate the finite-$x$ corrections to single and double Born diagrams, 
similarly to the full Altarelli-Parisi splitting kernel, and not just its soft gluon limit. In section~\ref{bg} below we 
derive analytical formulas for these finite-$x$ corrections to radiative energy loss at first order in opacity.

\section{Gauge invariance of the jet broadening and the medium-induced bremsstrahlung results}
\label{invariant}
In this section we demonstrate that the single and double Born amplitudes calculated in the previous two 
sections are gauge invariant. As it is known on the example of SCET, the gauge structure of effective 
theory is more rich than that of a full theory. This is a simple consequence of having multiple modes 
for the gauge field. In our calculation we deal with two types of gluons: collinear and Glauber. Thus, 
we can gauge fix these two modes completely independently without changing any physical result. Since Glauber 
mode is an off-shell mode, it is integrated out from the theory and is presented in the form of the 
potential term in~\eq{LGdef1}. Thus, the only gauge freedom for Glauber gluons is the choice of the 
propagator $\Delta_{\mu\nu}(q)$ in our effective potential, which in principle can be arbitrary. The collinear 
gluon field on the other hand is a truly propagating degree of freedom, with the corresponding kinetic 
term contained in the SCET Lagrangian. For each collinear gluon one could choose a certain gauge-fixing term.

In the previous two sections we considered the fully covariant gauge, in the sense that both collinear 
gluons are quantized in the covariant gauge, and also for the Glauber Lagrangian we choose covariant 
gluon propagator $\Delta_{\mu\nu}(q)_{R_{\xi}}$. Below we consider two alternative gauge choices and 
demonstrate there equivalence to the previous results. First, we consider a hybrid gauge where the 
collinear gluons are in the positive light-cone gauge and the Glauber potential is in the 
covariant gauge. Second, we choose both the collinear gluons and the Glauber potential term in the 
positive light-cone gauge $A^+_{c,g}=0$.

The equivalence of all considered cases can be formulated in the following way. All the diagrams 
in consideration are some combination of elastic scattering amplitudes with a real gluon emission amplitude. 
Since each of these two processes is gauge-independent, the resulting amplitudes, when contracted with the 
physical external gluon polarization vectors, are the same even when 
each of the gauges are chosen independently. This is equivalent to the statement that different modes 
for the gauge field in the effective theory can be gauge-fixed independently.

\subsection{Hybrid gauge $A_c^+=0,\;  R_{\xi}(A_g)$}\label{sec:hybrid}
We start from the hybrid gauge because it is simpler from a practical point of view.  In this case we 
consider the light-cone gauge for the collinear gluons, while the Glauber propagator in the potential is 
taken to be in the covariant gauge. The Feynman rules for this gauge are contained in figure~\ref{fig:SCETGFR}. 
Note that the collinear Wilson line $W_n=1$ is absent in this case. For example, for the single Born 
radiative energy loss case we only have the first three diagrams compared to the covariant gauge case in 
figure~\ref{fig:BremA1}, while the remaining two are absent. Since the collinear gluons are in the light-cone 
gauge, the transverse gauge link at infinity could in principle add new Feynman rules, where Glauber gluons 
arise from the transverse Wilson line $T_n$ (see appendix~\ref{appendix:TLine} for a brief review of this subject).
However, in this case it doesn't generate such a Feynman rule for the following reason. The diagram for which 
the $T_n$ Wilson line generates a Glauber gluon that couples to the source vanishes in this hybrid gauge 
because the propagator of Glauber gluon doted with the transverse gauge field and the source term vanishes: 
$A_{\perp}^i g^{i\mu} v_{\mu}=0$. However as we will see from the next subsection this is not the case in the 
fully positive light-cone gauge, where this interaction plays an important role. This is another reason 
why the hybrid gauge is so convenient.

For brevity, we consider explicitly only single Born diagrams for the  radiative energy loss in detail and 
then quote results for the remaining cases. We have only the first three diagrams in figure~\ref{fig:BremA1} for 
this case. Using the Feynman rules from the appendix~\ref{Appendix:SCETG} we obtain:
\begin{eqnarray}
   \left(R_1^{\mu, a}\right)_{\text{hyb}}&=& i (a)_R\left(n^{\mu}+\frac{\gamma^{\mu}_{\perp} (\pslash_{\perp}+\kslash_{\perp}) }{p^++k^+}+\frac{ \pslash_{\perp} \gamma^{\mu}_{\perp}}{p^+}\right) \frac{i(p^++k^{+})}{(p+k)^2+i \eps}i \,\,(b_1)_R\,(b_1)_{T_i}\,i\Delta_g(p+k,q_1)\; ,\\
   \left(R_2^{\mu, a}\right)_{\text{hyb}}&=&i(b_1)_R\,(b_1)_{T_i}\, i\, \Delta_g(p,q_1) i  (a)_R\left(n^{\mu}+\frac{ \gamma^{\mu}_{\perp}(\pslash_{\perp}-\qslash_{1\perp}) }{p^+}+\frac{ \pslash_{\perp} \gamma^{\mu}_{\perp}}{p^+}\right)i\Delta_g(p+k,q_1)\; ,\\
   \left(R_3^{\mu, a}\right)_{\text{hyb}}&=& i  (c_1)_R\left(n^{\rho_1}+\frac{\gamma^{\rho_1}_{\perp} (\pslash_{\perp}+\kslash_{\perp}-\qslash_{1\perp}) }{p^+}+\frac{ \pslash_{\perp} \gamma^{\rho_1}_{\perp}}{p^+}\right) \frac{(-i)\Delta_g(k,q_1)}{\bar{n}\mcdot k}\,N^{(\text{hyb})}_{\rho_1\rho_2}(k-q_1)\,{i\Delta_g(p+k,q_1)} \nonumber \\
   &&\times \, g^{\mu\rho_2}_{\perp}\,f^{c_1a b_1}(b_1)_{T_i}\bar{n}\mcdot k \; .\label{M13FormulasHybrid}
\end{eqnarray}
While first two amplitudes immediately match those in the covariant gauge given in~\eq{eq:BremA11}, 
\eq{eq:BremA12}, in order to work out the third amplitude, the numerator $N^{\text{hyb}}_{\mu\nu}$ of the positive light-cone gauge gluon propagator should be inserted. The following identity is straightforward to 
verify using the Feynman rules at hand:
  \begin{eqnarray}
 \eps_{\nu}(k)\,g^{\rho\nu}_{\perp}\,\bar{n}\mcdot k\,N^{(\text{hyb})}_{\mu\rho}(k-q_1)= \eps_{\nu}(k)\,\tilde\Sigma_1^{\rho\nu}(k-q_1, k)\,N^{(R_{\xi})}_{\mu\rho}(k-q_1).\label{eq:hybridvsRxi}
 \end{eqnarray}
 Thus, we arrive to the result that  $ \left(R_{1,2,3}^{\mu, a}\right)_{\text{hyb}}\equiv  
\left(R_{1,2,3}^{\mu, a}\right)_{R_{\xi}}$. The remaining two diagrams in the figure~\ref{fig:BremA1} are zero in 
the covariant 
gauge\footnote{This is true only for the same choice of physical polarization vectors  in the covariant, 
light-cone and hybrid gauge calculations. If one chooses arbitrary polarization vectors, once 
squared and summed over polarizations the result is independent on them. However, in this case it would be 
impossible to verify the gauge invariance with the hybrid or positive light-cone gauge, 
at the amplitude level.} and are simply absent in the hybrid gauge. 
 
 Similarly, proceeding with the remaining single and double Born amplitudes for the broadening and radiation we 
verify that the hybrid gauge results match the covariant gauge results at the amplitude level, as they should:
 \begin{eqnarray}
\left(B^{(1)q,g}\right)_{\text{hyb}}&=&\left(B^{(1)q,g}\right)_{R_{\xi}}\; ,\qquad \left(B^{(2c)q,g}\right)_{\text{hyb}}=\left(B^{(2c)q,g}\right)_{R_{\xi}}\; ,\nonumber\\
\left(R^{(1)}\right)_{\text{hyb}}&=&\left(R^{(1)}\right)_{R_{\xi}}\; ,\qquad\left(R^{(2c)}\right)_{\text{hyb}}=\left(R^{(2c)}\right)_{R_{\xi}}\; .\nonumber\\
\end{eqnarray}

\subsection{Positive light-cone gauge $A_{c,g}^+=0$}
We consider the positive light-cone gauge in this subsection. The collinear gluons are treated in the 
light-cone gauge as well as the  Glauber gluons. Similarly to the previous subsection, we consider in details 
the single Born diagrams for radiative energy loss and later on quote the result in all other cases. Clearly,
the collinear Wilson line vanishes again in this gauge $W_n=1$. However, the transverse gauge link 
(see appendix \ref{appendix:TLine}) gives a new non-trivial Feynman diagram. In figure~\ref{fig:BremA1LCG} we 
show the two new diagrams that appear in our gauge choice. The first one directly follows from the collinear part 
of the Lagrangian of SCET, once the background field with the light-cone vector potential scaling is added. 
It is summarized in figure~\ref{fig:SCETGFR} of appendix~\ref{Appendix:SCETG}. The second diagram in 
figure~\ref{fig:BremA1LCG} arises from the $T_n$ Wilson line emitting a Glauber gluon which interacts with the 
source. The corresponding Feynman rule is derived in the appendix~\ref{appendix:TLine} and is summarized in 
figure~\ref{fig:TWLine}. As one can see from this Feynman rule, it depends on the light-cone prescription. 
So does the light-cone (Glauber) gluon propagator. We will see in this section that these two dependences cancel 
non-trivially and the final answer in this case is identical to the covariant gauge calculation above for all 
light-cone prescriptions. Similar cancellation was found in \cite{Idilbi:2010im} by introducing the $T$ 
Wilson line to SCET and calculating the jet function at one-loop in the light-cone gauge.

Using the Feynman rules of this gauge we  evaluate the first three diagrams in figure~\ref{fig:BremA1}, which are present in this gauge as well:
\begin{eqnarray}
\left(R^{(1)\mu, a}_{1,2,3}\right)_{A^+}=\left(R^{(1)\mu, a}_{1,2,3}\right)_{R_{\xi}}+ \Delta {R_{1,2,3}^{(1)\mu, a}}\; ,
\end{eqnarray}
where the correction terms $\Delta {R_i^{(1)\mu, a}}$ sum up to the following expression:
\begin{figure}[t!]
\begin{center}
\epsfig{file=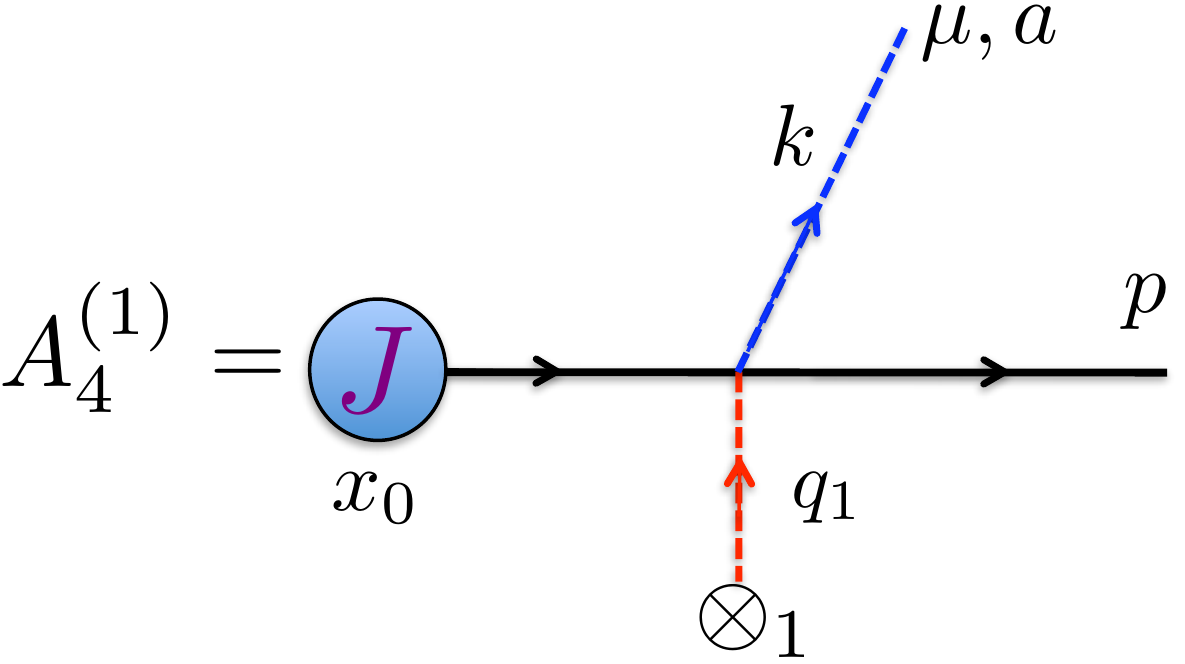, width=5.5cm} \qquad\epsfig{file= 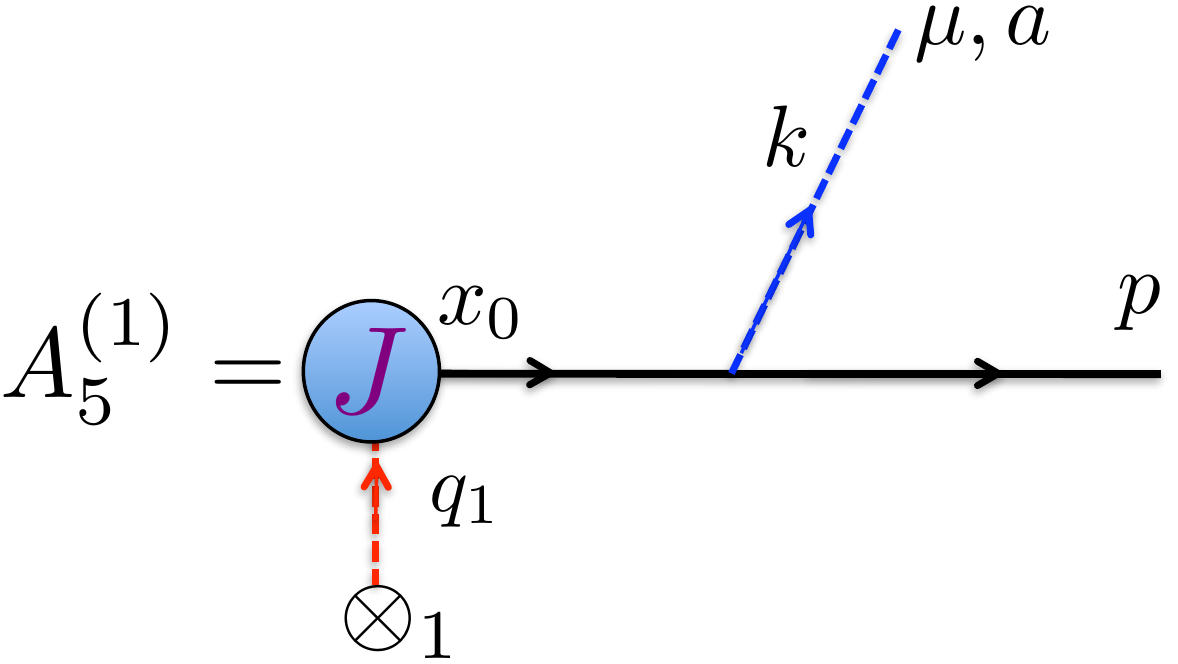, width=5.5cm}
\caption{\label{fig:BremA1LCG} Two additional diagrams that appear in the light-cone gauge for the single Born 
amplitude for radiative energy loss. Left: term arising from the $\text{\SCETG}$ Lagrangian in the light-cone gauge. 
Right: $T_n$ Wilson line contribution to the matrix element. The notation for the scattering centers is the following: $\otimes_{1}=[x_1,q_1, (b_1)_i]$.}
\end{center}
\end{figure}
\begin{eqnarray}
&&\int d\Phi_1 \left(\Delta {R_1^{(1)\mu, a}}+\Delta {R_2^{(1)\mu, a}}+\Delta {R_3^{(1)\mu, a}}\right)=\int d\vc{\Phi}_{1\perp}\Bigg[ (b_1a)_R\, (b_1)_{T_i}\left(-\qslash_{1\perp}\gamma^{\mu}_{\perp}\frac{p^++k^+}{p^+}\,\Delta I\right)\nonumber\\
&& \qquad \qquad \qquad  + (ab_1)_R\, (b_1)_{T_i}\left(-\gamma_{\perp}^{\mu}\qslash_{1\perp}\,\Delta I+\frac{p^++k^+}{(p+k)^2}\Delta_{A^+}\left(\frac{2\vc{k}_{\perp}^{\mu}}{k^+}+\frac{\gamma_{\perp}^{\mu}\left(\pslash_{\perp}+\kslash_{\perp}\right)}{p^++k^+}+\frac{\pslash_{\perp}\gamma_{\perp}^{\mu}}{p^+}\right)\right)\Bigg] \; . \qquad \;
\label{LCGEq1234}
\end{eqnarray}
The two longitudinal integrals $\Delta I$ and $\Delta_{A^{+}}$ are defined
in the following way\footnote{Note that $q_1^+\equiv -q_1^{-}$ in this
equations, set by the $\delta(q^0_1)$ in $v(q_1)$.}:
\begin{eqnarray}
 &&\Delta I=\int \frac{dq^-_1}{2\pi}\e^{i q^-_1 \Delta
z_1}\frac{1}{(p+k-q_1)^2}\frac{1}{\left[q^+_1\right]}\; ,\\
 &&  \Delta_{A^{+}}=\int \frac{d q_1^-}{2\pi}\e^{iq_1^-\delta z_1}\,
\frac{1}{\left[q_1^{+}\right]}\; .
\end{eqnarray}
The second integral is summarized in the table~\ref{tb2} and vanishes in the $-i\eps$ light-cone prescription. 
However, the first integral is clearly non-zero in any of the prescriptions, since its value comes from the 
poles of the propagator denominator. Luckily the first diagram in figure~\ref{fig:BremA1LCG} completely 
cancels this term:
\begin{eqnarray}
&&\int d\Phi_1 \,\left(R_{4}^{(1)}\right)^{\mu, a}_{A^+}=\int d\vc{\Phi}_{1\perp}\Bigg[  (b_1a)_R\, (b_1)_{T_i}\,\qslash_{1\perp}\gamma^{\mu}_{\perp}\frac{p^++k^+}{p^+}\,\Delta I+ (ab_1)_R\, (b_1)_{T_i}\,\gamma_{\perp}^{\mu}\qslash_{1\perp}\,\Delta I\Bigg]\; .
\end{eqnarray}
\begin{table}[!t]
\begin{center}
\begin{tabular}{||c||c||c|c|c|}
       \hline
Prescription& $\frac{1}{[k^+]}$&  $i\mcdot\Delta_{A^+}$      \\
       \hline \hline
$+i\eps$& $\frac{1}{k^++i\eps}$&  $1$ \\
$-i\eps$& $\frac{1}{k^+-i\eps}$&  $0$ \\
PV& $\frac{1}{2}\left(\frac{1}{k^++i\eps}+\frac{1}{k^+-i\eps}\right)$&
$\frac{1}{2}$  \\
ML& $\frac{1}{k^++i\eps \text{sign}(k^-)}$&  $\frac{1}{2}$       \\
       \hline
\end{tabular}
\caption{Prescription dependent integral appearing in the light-cone gauge.}
\end{center}
\label{tb2}
\end{table}
When the amplitude $R_{4}^{(1)}$ is added to the first three diagrams, the $\Delta I$ integral vanishes as expected, 
while the $\Delta_{A^+}$ integrals stays. It vanishes only in the $-i\eps$ light-cone prescription, exactly in 
which the transverse gauge link vanishes according to Feynman rule derived in appendix~\ref{appendix:TLine}. 
In all other prescriptions, the contribution from the transverse Wilson line does not vanish and we include it 
by calculating the second amplitude in figure~\ref{fig:BremA1LCG}. We explicitly see that by including 
these Feynman diagram the prescription dependence cancels in all three remaining cases: $+i\eps$, PV, ML, 
which happens through pushing the $q^-$ pole into the negative complex plane which leads to the zero integral. 
Thus, by including the $T$-Wilson line into the calculation we get the same result in light-cone gauge as in 
the covariant gauge for all light-cone prescriptions. To see this explicitly we apply Feynman rule for 
the single Glauber gluon exchange from  quark $T-$ Wilson line, using appendix~\ref{appendix:TLine} 
(see figure~\ref{fig:TWLine}) and obtain:
\begin{eqnarray}
&&\int d\Phi_1 \,\left(R_{5}^{(1)}\right)^{\mu, a}_{A^+}=(ab_1)_R\, (b_1)_{i}\,\int d\vc{\Phi}_{1\perp}\,i\,\Bigg[n^{\mu}+\frac{(\pslash_{\perp}+\kslash_{\perp})\gamma_{\perp}^{\mu}}{p^++k^+}+\frac{\gamma_{\perp}^{\mu}\pslash_{\perp}}{p^+}\Bigg]\,i\frac{(p^++k^+)}{(p+k)^2}\,\Delta_{T}\; ,
\end{eqnarray}
where the longitudinal integral $\Delta_T$ is defined according to:
\begin{eqnarray} 
 \Delta_{T}=C_{\infty}^{\text{Pres}}\,\int \frac{d q_1^-}{2\pi}\e^{iq_1^-\delta z_1}\, \left(\frac{1}{q_1^{+}+i\eps}-\frac{1}{q_1^{+}-i\eps}\right)\; .
\end{eqnarray}
Using  table~\ref{tb3} for the prescription dependence of $C_{\infty}^{\text{Pres}}$ we verify that 
it vanishes in the $-i\eps$ prescription\footnote{Note that in the $-i\eps$ prescription the integral 
$\Delta_{A^+}$ vanishes as well.} and in all other prescriptions makes the $\Delta_T$ integral cancel 
exactly the contribution from $\Delta_{A^+}$ integral above. Thus, in the presence of a transverse gauge link 
we verify the gauge invariance for all prescriptions:
\begin{eqnarray}
&&\int d\Phi_1 \left(\Delta {R_1^{(1)\mu, a}}+\Delta {R_2^{(1)\mu, a}}+\Delta {R_3^{(1)\mu, a}}+ 
{R_4^{(1)\mu, a}}+ {R_5^{(1)\mu, a}}\right)=0\; .
\end{eqnarray}
Calculations in all other cases, like jet broadening and radiative energy loss in 
single and double Born amplitudes, are similar to case considered above. We quote the results:
\begin{eqnarray}
\left(B^{(1)q,g}\right)_{A^+}&=&\left(B^{(1)q,g}\right)_{R_{\xi}}\; ,
\qquad \left(B^{(2c)}_{q,g}\right)_{A^+}=\left(B^{(2c)}_{q,g}\right)_{R_{\xi}}\; ,\nonumber\\
\left(R^{(1)}\right)_{A^+}&=&\left(R^{(1)}\right)_{R_{\xi}}\; ,
\qquad\left(R^{(2c)}\right)_{A^+}=\left(R^{(2c)}\right)_{R_{\xi}}\; .\nonumber\\
\end{eqnarray}
Thus, for all light-cone prescriptions we get an unambiguous result, same as in the covariant gauge. 
If one ignores the transverse gauge link, one should employ one particular prescription, $-i\eps$ in this case, 
in order to recover the correct result. For practical purposes this gauge choice is the most difficult 
from all three that we considered in this paper.

\section{Identifying the in-medium jet evolution kernels}\label{sec:ROP}

In the preceding sections we calculated the amplitudes for collisional and radiative
jet interaction at a particular position  $x_i$  and demonstrated the gauge invariance 
of the end  results. It is clear that a direct calculation of all diagrams for any
number of scattering positions $x_{i_1},\, x_{i_2}, \cdots $ is not possible. From the results
at hand, however, we can deduce the effect of an in-medium interaction at the amplitude 
and cross section levels in momentum space and employ the resulting kernels to describe
a number of collisional and radiative processes in cold and hot nuclear matter as 
solutions to algebraic recurrence relations with suitably chosen initial 
conditions~\cite{Gyulassy:2000er,Gyulassy:2002yv,Qiu:2003pm,Vitev:2007ve,Vitev:2008vk,Sharma:2009hn}.

In the absence of long-range color correlations in the target, the relevant interactions at position
$i_n$ to lowest order in $\alpha_s$ that can build up to a jet-medium cross sections (a total of two
Glauber exchanges in the forward cut diagram) can 
be expresses as follows: 
\begin{eqnarray} 
\label{basit0}
{\cal A}_{i_1\cdots i_{n-1},0}({\alpha}) &\equiv& 
\htI  \vAim({\alpha}) \;\;,  \\
{\cal A}_{i_1\cdots i_{n-1},1}({\alpha}) &\equiv& 
\htD  \vAim({\alpha})  \;\;,   \\
{\cal A}_{i_1\cdots i_{n-1},2} ({\alpha}) &\equiv&  
\htV  \vAim({\alpha})  \;\;.  
\label{basit}
\end{eqnarray}  
Here, $\alpha$ represents a set of relevant quantum numbers, such as longitudinal momentum, transverse 
momentum and color. The kinematic and color structure of the scattering is contained in
the unit $(\,\htI\,)$, direct $(\,\htD \,)$ and virtual $(\, \htV \,)$ 
operators, which evolve the amplitude ${\cal A}_{i_1\cdots i_{n-1}}({\alpha})$ of the 
propagating system. The unit operator indicates that there are no Glauber gluon exchanges 
between the projectile and the target at position $x_{i_n}$ ($i_n=0$).
 The direct operator indicates a single Glauber gluon exchange 
between the projectile and the target at position $x_{i_n}$ ($i_n=1$). The virtual operator
indicates two Glauber gluon exchanges  between the projectile and the target at position 
$x_{i_n}$ ($i_n=2$), i.e. in the contact limit. The set of amplitude indices in 
Eqs.~(\ref{basit0})-(\ref{basit}), thus, 
encodes the complete history of jet interactions in QCD matter. Repeating the basic operator steps in  
Eqs.~(\ref{basit0})-(\ref{basit})  any amplitude that includes parton scatterings inside 
the medium  can be iteratively derived from the unperturbed jet production amplitude
\beq
\vAi({\alpha})
=\prod_{m=1}^n
\left[ \delta_{0,i_m} + \delta_{1,i_m} \htD_m + \delta_{2,i_m} 
\htV_m
\right] J_0({\alpha})
\;  . \eeq{atens}
Time ordering is implicit in the above formula. The amplitudes $\bar{\cal A}^{i_1\cdots i_n}(p,c)$  
are the complementary amplitudes to  $\vAi({\alpha})$   given by 
\beqar
\vAbi({\alpha})&\equiv& J_0^\dagger
({\alpha}) \prod_{m=1}^n
\left[ \delta_{0,i_m} \hat{V}_m^\dagger + 
\delta_{1,i_m} \hat{D}_m^\dagger  + \delta_{2,i_m}  \right] 
\; .\eeqar{atens2}
The differential jet or radiative gluon distribution, depending on the problem,  
can be  expresses as a sum over the interactions in the medium  
$dN({\alpha})/dPS = \sum_{n=0}^\infty dN^{(n)}({\alpha})/dPS$. Each contribution 
with a  fixed number of interactions $n$ can, in turn  be  represented as 
\beq
dN^{(n)}(\alpha)/dPS \propto \vAbi(\alpha)\vAi({\alpha}) \propto 
\tr \sum_{i_1=0}^2\cdots \sum_{i_n=0}^2 \bar{\cal A}^{i_1\cdots i_n}(\alpha)    
A_{i_1\cdots i_n}(\alpha)  \; .
\eeq{pnten}
The trace is over any uncontracted color and spin/polarization indices.
Using Eqs.~(\ref{atens}), (\ref{atens2}) we obtain a simple recursion identity 
which relates 
$dN^{(n)}/dPS$ to $dN^{(n-1)}/dPS$ through the reaction operator $\htR$ 
\beq
dN^{(n)}/dPS \propto \;\bar{\cal A}^{i_1\cdots i_{n-1}}
\htR_n {\cal A}_{i_1\cdots i_{n-1}}
\;, \qquad \htR_n = \hat{D}_n^\dagger
\hat{D}_n+\hat{V}_n+\hat{V}_n^\dagger
\; \; . \eeq{rop}
Thus, the most important step in investigating the effects of the medium on jet propagation
is the identification of the reaction operator $\htR$. We note that in problems that 
involve long-range coherence effects, such as radiative gluon re-interactions in the QCD matter,  
the recurrence relations may be inhomogeneous. 
 
\begin{figure}[h!]
\begin{center}
\epsfig{file=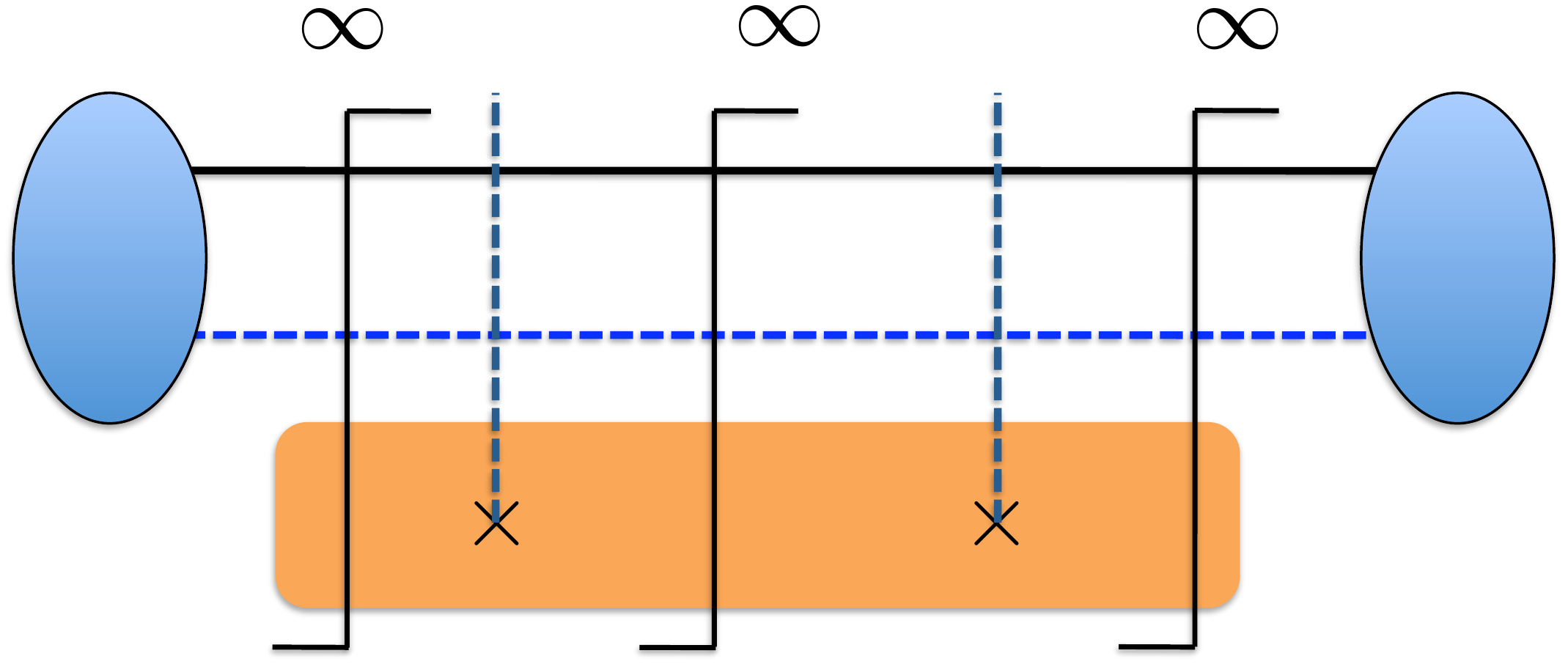, width=12cm}
\caption{\label{fig:reaction} Three contributions to the cross section, associated with  
the application of the reaction operator, correspond to the three $t=\infty$ cuts.} 
\end{center}
\end{figure}

\subsection{Reaction operator for collisional interactions}

With the direct and virtual interactions of a jet calculated in section~\ref{collisional}, we can 
identify the form of the reaction operator for collisional interactions acting on a jet of
momentum $p$
\beqar
\htR_n  & = &  
\int_{z_{n-1}}^L dz_n \rho(z_n)  \int d^2 {\bf q_{\perp\,n}}  \left[ \frac{d\sigma_{el} }
{d^2{\bf q}_{\perp\, n}} \, 
e^{-i {\bf q}_{\perp\,n} \cdot {\bf \hat{b}}^\dagger}
\cdot e^{i {\bf q}_{\perp\, n} \cdot {\bf \hat{b} } }- \sigma_{el} 
\delta^2({\bf q}_{\perp n})\, \right] \,  \; .  
\eeqar{dit}
Here,  $L$ is the thickness of the target, ${\bf \hat{b}} = i \stackrel{\rightarrow}{\nabla}_{{\bf p}_\perp}$ 
is the  2-dimensional impact parameter operator conjugate to the transverse momentum
${\bf p}$  acting to the right  and  $ {\bf \hat{b}}^\dagger = 
- i \stackrel{\leftarrow}{\nabla}_{{\bf p}_\perp} $ is its Hermitian conjugate acting to the left.  
In this paper we explicitly proved the gauge invariance of this reaction operator. 
Our results are not surprising since $\htR$ is expressed in terms of physical quantities,
such as cross sections and densities and kinematic modification to the initial jet
distribution.

The contribution to the jet  transverse momentum distribution from $n$ medium 
interactions  can be written as: 
\beqar
\frac{dN^{(n)}({\bf p}_\perp)}{d^2{\bf p}_\perp} 
& = &   
  \prod_{i=1}^n  \int_{z_{i-1}}^L \frac{d z_i}{\lambda} \int d^2 {\bf q}_{\perp \, i}   
 \left[ \frac{1}{\sigma_{el}(z_i)} \frac{d\sigma_{el}(z_i)}{d^2{\bf q}_{\perp\,i}}
\left(   e^{-{\bf q}_{\perp \, i} \cdot \stackrel{\rightarrow}{\nabla}_{{\bf p}_\perp} }\right)
 - \delta^2({\bf q}_{\perp i})  \, \right]  \;  \frac{dN^{(0)} ({\bf p}_\perp)}{d^2{\bf p}_\perp}  \;,
\eeqar{ropit}
where $\lambda$ is the mean free path of the jet in the medium. Summing over all fixed-$n$ contributions that Eq.~(\ref{ropit}) specifies, we obtain 
the most general result for the broadening of jets that 
propagate and interact in strongly-interacting matter.

Simplifications of the sum over the contributions given by Eq.~(\ref{ropit}) 
can be  obtained for special  cases. For example, 
for a uniform density the integrals along the path of the jet can be performed 
immediately. With the notation $\chi=L/\lambda$, the result reads:
\beqar
\frac{dN({\bf p}_\perp)}{d^2{\bf p}_\perp} &=& \sum_{n=0}^\infty 
\frac{dN^{(n)}({\bf p}_\perp)}{d^2{\bf p}_\perp} 
 =  \sum_{n=0}^\infty e^{-\chi} \frac{\chi^n}{n!}  
\int \prod_{i=1}^n d^2 {\bf q}_i \frac{1}{\sigma_{el}} 
 \frac{d \sigma_{el} }{d^2 {\bf q}_{\perp \, i}}  \;  
dN^{(0)}({\bf p}_\perp - {\bf q}_{\perp \, 1} - \cdots - {\bf q}_{\perp \, n})  \;. 
\label{glau}
\eeqar{glaberf}
This is the partonic version of the Glauber multiple collision series where the jet 
interaction in the medium are described by a Poisson distribution and the momentum 
distribution is modified by the normalized differential scattering cross 
section \cite{Gyulassy:2002yv,Qiu:2003pm}.

A final simplification can be achieved if one recognizes that the individual 
jet-medium interactions are of finite  range $r=1/\mu$, such as the one 
discussed in sections~\ref{sec:kinematics} and~\ref{scetg}. Let us take the initial jet distribution to be in the 
positive light-cone direction,  ${dN({\bf p}_\perp)}/{d^2{\bf p}_\perp} = \delta^2({\bf p}_\perp) $. The Gaussian
approximation can be best understood in impact parameter space where the  Fourier 
transform  of the  normalized scattering cross section reads
\beq
\frac{d \tilde{\sigma}_{el} }{d^2 {\bf q}_\perp}({\bf b}) = 
\int \frac{d^2 {\bf q}_\perp}{(2 \pi)^2} \; e^{-i{\bf q}_\perp \cdot {\bf b}}
\frac{1}{\pi} \frac{\mu^2}{({\bf q}_\perp^2+\mu^2)^2} = 
\frac{\mu \, b}{4 \pi^2} K_1(\mu\, b)   \approx 
\frac{1}{4 \pi^2} \left(1- \frac{\xi \, \mu^2\, b^2}{2}
+ {\cal O}(b^3)  \right) \; .
\eeq{ftdc} 
Here $b=|{\bf b}|$ and in the quadratic term in Eq.~(\ref{ftdc}) 
the $\log 2/(1.08\, \mu\, b) $ multiplicative factor has been 
absorbed into an approximately $b$-independent constant $\xi$.
Fourier transforming back to momentum space we obtain:
\beq
\frac{dN({\bf p}_\perp)}{d^2{\bf p}_\perp} = \int d^2{\bf b} \;
 e^{i{\bf p}_\perp \cdot {\bf b} } 
\frac{1}{(2\pi)^2}{e^{-\frac{\chi\, \mu^2 \, \xi \, b^2}{2}}  } 
= \frac{1}{2\pi} \frac{e^{-\frac{p^2_{\perp}}{2\, \chi  \, \mu^2 \, \xi} }}
{\chi\, \mu^2\, \xi } \; .
\eeq{gauss}
The resulting distribution is of two dimensional Gaussian form and has a width of
$2 \chi\, \mu^2\, \xi$,  i.e. $\left\langle {\bf p}_\perp^2  \right\rangle = 2 \chi\, \mu^2\, \xi$.   
It should be noted that the Gaussian approximation Eq.~(\ref{gauss}) is only 
applicable for small transverse momenta. It misses the Rutherford scattering 
power law $1/{\bf p}_\perp^4$ behavior for ${\bf p}_\perp^2 \geq  2 \chi\, \mu^2$. The general 
solution the problem of jet multiple scattering is given by Eq.~(\ref{ropit})~\cite{Gyulassy:2002yv}.

\subsection{Reaction operator for medium-induced gluon bremsstrahlung}

The derivation of the reaction operator for radiative processes is significantly more 
complicated in comparison to the one for collisional interactions. The first step is to 
reorganize the amplitudes obtained in section~\ref{radiative} and to identify the contributions that
can be interpreted as an interaction of the parent parton, an interaction of the already 
radiated gluon ($\htS_n$) and, finally, a genuinely new source of radiation associated with this
particular scattering ($\htB_n$).   

We introduce the following notation for the radiative gluon inverse formation times:
\beq
\omega_0=\frac{{\bf k}_\perp^{2}}{2\omega}\; ,\;
\omega_i=\frac{({\bf k}_\perp-{\bf q}_{\perp\, i})^2}{k^+}\; ,\;
\omega_{(ij)}=\frac{({\bf k}_\perp- {\bf q}_{\perp\,i}-{\bf q}_{\perp\,j})^2}{k^+}
\;,\; \omega_{(i\cdots j)}=\frac{({\bf k}_{\perp}-
\sum_{m=i}^j {\bf q}_{\perp m})^2}{k^+} \;  ,
\eeq{shorth}
and transverse momentum propagators:
\beqar
{\bf H}&=&\frac{{\bf k}_\perp }{ {\bf k}_\perp^2 }\; , \qquad \qquad
{\bf C}_{(i_1i_2 \cdots i_m)}=  \frac{ {\bf k}_\perp - {\bf q}_{\perp\, i_1} - 
{\bf q}_{\perp\, i_2} -  \cdots -{\bf q}_{\perp \, i_m} } 
{ ({\bf k}_\perp - {\bf q}_{\perp \, i_1} - 
{\bf q}_{\perp\, i_2} - \cdots -{\bf q}_{\perp\, i_m}   )^2 } \;, 
\nonumber \\[1.ex]
{\bf B}_i &= &{\bf H} - {\bf C}_i \; , \qquad
{\bf B}_{(i_1 i_2 \cdots i_m )(j_1j_2 \cdots j_n)} = 
{\bf C}_{(i_1 i_2 \cdots i_m)} - {\bf C}_{(j_1 j_2 \cdots j_n)}\; .
\eeqar{hbgcdef}
Further, we recall that each double Born exchange yields a factor $-1/2$ 
and the numbers of these exchanges in the amplitude and its complementary are: 
\beqar
N_v=N_v(\vAim)=\sum_{m=1}^{n-1} \delta_{2,i_m}\;,  \qquad
\bar{N}_v=N_v(\vAbim)=\sum_{m=1}^{n-1} \delta_{0,i_m}  \; .
\eeqar{kv}
Finally,  the color  matrices associated with the collisional 
interactions of the jet are denoted by
\beq
T_{el}({\cal A}_{i_1\cdots i_{n-1}})\equiv (a_{n-1})^{i_{n-1}}\cdots (a_1)^{i_1}
\;, \qquad  T_{el}^\dagger(\bar{\cal A}^{i_1\cdots i_{n-1}})
\equiv (a_1)^{2-i_1} \cdots (a_{n-1})^{2-i_{n-1}} \;   . 
\eeq{tel}

We first examine the single Glauber exchange and the direct operator reads:
\beqar
 \htD_n   &\equiv&   (a_n + \htS_n + \htB_n)  \nonumber \\
&=&a_n  + e^{i(\omega_0-\omega_n)z_n } e^{i{\bf q}_{n} \cdot \hat{\bf b}}  
\times if^{ca_n d}
 - \left(-\half \,\right )^{N_v(\vAim)} {\bf B}_n \, 
e^{i \omega_0 z_n} [c,a_n] T_{el}(\vAim) \;.
\eeqar{didamit}  
We note that the common factor to all diagrams 
$2ig {\bm \epsilon} \cdot (\cdots )$, where the transverse polarization 
vector ${\bm \epsilon}$ contracts with the 2D propagators in Eq.~(\ref{hbgcdef}),
 is not explicitly shown for brevity. It is understood in Eq.~(\ref{didamit})
that   $if^{ca_n d}$ rotates the color of the radiated gluon $d$:  
$if^{ca_n d} d = [c,a_n]$. Next, we identify the virtual operator:
\beqar
\htV_n  &\equiv&  -\half(C_A+C_R) - a_n \htS_n- a_n\htB_n
= -a_n \htD_n - \half(C_A-C_R) \nonumber \\
&= & - \frac{C_R+C_A}{2} 
-  e^{i(\omega_0-\omega_n)z_n }  e^{i{\bf q}_{n} \cdot \hat{\bf b}}  a_n if^{ca_n d} 
- \left(-\half\,\right)^{N_v(\vAim) }
\frac{C_A}{2} \, {\bf B}_n \,
e^{i \omega_0 z_n} c a_{n-1}^{i_{n-1}}\cdots a_1^{i_1} \;.
\eeqar{vidamit} 
Substituting Eqs.~(\ref{didamit}) and (\ref{vidamit}) in the definition of the 
reaction operator we find: 
\beqar
\htR_n &=&  (\htD_{n}-a_{n})^\dagger  (\htD_{n}-a_{n}) - C_A =
(\htS_{n}+\htB_{n})^\dagger  (\htS_{n}+\htB_{n}) -C_A \nonumber \\
&=& C_A \left(e^{-{\bf q}_{n} \stackrel{\leftarrow}{{\nabla}_{{\bf k}}}}
e^{-{\bf q}_{n} \stackrel{\rightarrow}{{\nabla}_{{\bf k}}}} -1 \right)
-2 C_A\, {\bf B}_n\cdot \left( {\bf Re}\; e^{-i\omega_nz_n} 
e^{i{\bf q_n}\cdot\hat{\bf b}} {\bf I}_{n-1} \right)
 + \delta_{n,1} C_A C_R |{\bf B}_1|^2   \; .  
\eeqar{np1}
The diagonal Bertsch-Gunion term contributes only for the first
scattering $n=1$~\cite{Gyulassy:2000er,Vitev:2007ve}.  The off-diagonal terms depend on the
current ${\bf I}_{n}$, which in turn obeys a recurrence relation itself:
\beqar
{\bf I }_n &=&C_A(e^{i(\omega_0-\omega_n)z_n}
 e^{i{\bf q}_{n}\cdot\hat{\bf b}} -1) {\bf I}_{n-1}
-\delta_{n,1} C_A C_R {\bf B}_{1} e^{i\omega_0 z_{1}}
\; . 
\eeqar{bfin}
To summarize, the medium-induced bremsstrahlung depends
sensitively  on the boundary conditions - both at the amplitude 
and cross sections levels.

Finally, we give one example for the complete solution to the final-state  
medium-induced bremsstrahlung for a jet produced in a large $Q^2$ process.
The boundary condition is represented by the  amplitude: 
$J_0 =  - 2 i g \, \frac{ { \bm \epsilon} \cdot
{\bf k}_\perp }{  {\bf k}_\perp ^2 } \,  e^{i \omega_0 z_0 } \, c \;, $ associated 
with the real hard bremsstrahlung for $x\ll 1$. Rewriting the $z_n$ position 
integrals as integrals over the separation between the scattering centers 
$\Delta z_n = z_n - z_{n-1}$ and including the integrals over the momentum 
distribution of the jet-medium scattering cross section we obtain~\cite{Gyulassy:2000er}:  
\beqar
&& k^+ \frac{dN^g}{dk^+ d^2 {\bf k}_\perp } = \frac{C_R \alpha_s}{\pi^2}
\sum_{n=1}^{\infty}  \left[ \prod_{i = 1}^n  \int
\frac{d \Delta z_i}{\lambda_g(z_i)}  \right] 
\left[ \prod_{j=1}^n \int d^2 {\bf q}_{\perp\,j} \left( \frac{1}{\sigma_{el}(z_j)} 
\frac{d \sigma_{el}(z_j) }{d^2 {\bf q}_{\perp \, j}}   
-  \delta^2 ({\bf q}_{\perp\, j}) \right)    \right] \nonumber  \\
&&  \times \;  \left[ -2\,{\bf C}_{(1, \cdots ,n)} \cdot 
\sum_{m=1}^n {\bf B}_{(m+1, \cdots ,n)(m, \cdots, n)} 
\left( \cos \left (
\, \sum_{k=2}^m \omega_{(k,\cdots,n)} \Delta z_k \right)
-   \cos \left (\, \sum_{k=1}^m \omega_{(k,\cdots,n)} 
\Delta z_k \right) \right)\; \right]  \;.   \qquad 
\eeqar{full-final}
To use unified notation above, we have to specify $\sum_2^1~\equiv 0$ and  
${\bf B}_{(n+1, n)}~\equiv {\bf B}_n$. In the case of final-state interactions, 
$z_0 \approx 0$ is the point of the initial hard scatter and $z_L = L$ is 
the extent of the medium.  The path ordering  of the 
interaction points, $z_L > z_{j+1} > z_j > z_0$, leads 
to the  constraint 
$\sum_{i=1}^n \Delta z_i  \leq  z_L $. One implementation of this
condition would be $\Delta z_i \in [\, 0,z_L -\sum_{j=1}^{i-1} 
\Delta z_j \, ]$  and it is implicit in Eq.~(\ref{full-final}). 
The complete solution to the problem of medium induced bremsstrahlung 
in the $x\ll 1$ limit for three different boundary conditions can be found
in Ref.~\cite{Vitev:2007ve}.  With the proof of the gauge invariance of the
reaction operator for  collisional and radiative processes, we have
also proven the gauge invariance of these results. As we pointed before,
when the end result is expresses in terms of physics quantities, such as 
scattering cross sections, mean free paths and formation times, gauge-invariance
can easily be recognized.

\section{Bremsstrahlung: beyond the soft gluon approximation}
\label{bg}

In this section we calculate the corrections to the medium induced bremsstrahlung at 
finite splitting fraction $x\sim 1$. Effective theory Feynman rules allow us to do so easily. 
The advantage of the effective theory
approach is that we have the appropriate interactions in medium at the Lagrangian level, which allows us to straightforwardly calculate any process of interest. Doing similar calculations, for example calculating radiative energy loss to first order in opacity keeping the 
full $x$ dependance in
the traditional approach would be more difficult, since it will require to do approximations at the level of Feynman diagrams.

\subsection{Incoherent radiation}

We first discuss the simpler case of  gluon emission without the Landau-Pomeranchuk-Migdal
destructive interference effects. Starting from expressions in~\eq{eq:BremA11}~$-$\eq{eq:BremA13} and without making approximation on $x$ we identify 
three pieces of these  matrix elements that are proportional to $\exp[i\Omega_1(z-z_0)]$ 
(see appendix~\ref{appendix:integralsRad} for details) 
as  the ones that contribute to the Bertsch-Gunion amplitude - the QCD analog of the Bethe-Heitler 
radiation in electrodynamics. Squaring the sum of three diagrams and summing over the physical 
polarizations  we  obtain: 
\begin{eqnarray}
 && \frac{1}{d_Rd_T} \left \langle |A_{J q\rightarrow J q g}|^2 \right\rangle_{\rm medium \; {\bf q}_\perp} 
= (1-x) \text{Tr}\left(\frac{\nslash}{2}p_0^+J \bar{J}\right) 
\times  \frac{N}{A_\perp} \int d^2{\bf q}_\perp  \frac{d\sigma_{el}^{g\; {\rm medium}}}{d^2 {\bf q}_\perp} \; 
    | M^{\rm rad}_{BG}  |^2  \; .
\label{BGRad} 
\end{eqnarray} 
It easy to see that in Eq.~(\ref{BGRad})  $N/A_\perp = dz\,\rho(z)$, where $\rho$ is the density of scattering
centers in the medium. Integrating over the path of the quark propagation we find that the differential
spectrum of the incoherent Bertsch-Gunion bremsstrahlung can be written as:
\begin{eqnarray}
   && x \frac{dN^g}{ dxd^2\vc{k}_{\perp} }  = \int \frac{dz}{\lambda_g(z)}  
\int d^2{\bf q}_\perp  \frac{1}{\sigma_{el}} \frac{d\sigma_{el}^{g\; {\rm medium}}}{d^2 {\bf q}_\perp} \; 
   \frac{1}{2(2\pi)^3} | M^{\rm rad}_{BG}  |^2  \; .
\label{BGRad1} 
\end{eqnarray} 
Here, $\lambda_g(z) = 1/\left[\sigma_{el}^{g\; {\rm medium}} \rho(z)\right] $ and the initial jet direction
$( p_0=p+k) || n$. Very generally, one can express the radiative amplitude squared as follows:
\begin{eqnarray}
   && 
   \frac{1}{2(2\pi)^3} | M^{\rm rad}_{BG}  |^2  = C_F\frac{\alpha_s}{\pi^2} \left(1-x+\frac{x^2}{2} \right) 
\frac{ {\bf q}_\perp^2 }{2 {\bf A}_\perp^2 {\bf B}_\perp^2  {\bf C}_\perp^2    }  
\left[ {\bf A}_\perp^2 +(1-x)^2 {\bf B}_\perp^2  -\frac{1}{N_c^2} x^2  {\bf C}_\perp^2  \right] \;. 
\label{BGRad2} 
\end{eqnarray} 
In Eq.~(\ref{BGRad2})
\begin{eqnarray}
  {\bf A}_\perp  & = &  (1-x){\bf k}_{\perp} -x {\bf p}_\perp   =  {\bf k}_\perp\; |_{{\bf p}_\perp= -{\bf k}_\perp} 
\;,  \\
  {\bf B}_\perp  & = &   (1-x){\bf k}_{\perp} -x ({\bf p}_\perp - {\bf q}_\perp  ) = 
{\bf k}_\perp + x {\bf q}_\perp \; |_{{\bf p}_\perp= -{\bf k}_\perp}    \;,  \\
  {\bf C}_\perp  & = &   (1-x)( {\bf k}_{\perp}- {\bf q}_{\perp})  -x {\bf p}_\perp  =  
{\bf k}_\perp - (1-x){\bf q}_\perp \;  |_{{\bf p}_\perp= -{\bf k}_\perp}    \;.  
\end{eqnarray}
By taking the small $ x \ll 1 $ limit above and substituting in Eq.~(\ref{BGRad2}) we obtain the
known result:
\begin{eqnarray} 
   \frac{1}{2(2\pi)^3} | M^{\rm rad}_{BG}  |^2  &=& C_F\frac{\alpha_s}{\pi^2}  
\frac{ {\bf q}_\perp^2 }{ {\bf k}_\perp^2  ({\bf k}_\perp - {\bf q}_\perp)^2 } \;. 
\label{BGRad3} 
\end{eqnarray} 
The general expression for the incoherent  medium-induced gluon bremsstrahlung for a medium of constant density
and  length $L$ without such approximations is:
\begin{eqnarray}
    x \frac{dN^g}{ dxd^2\vc{k}_{\perp} }  & = &    
C_F\frac{\alpha_s}{\pi^2} \left(1-x+\frac{x^2}{2} \right)   \frac{L}{\lambda_g}  
\int d^2{\bf q}_\perp  \frac{1}{\sigma_{el}} \frac{d\sigma_{el}}{d^2 {\bf q}_\perp} \; 
\frac{ \vc{q}_{\perp}^2}{2 \vc{k}_{\perp}^2 \, \left(\vc{k}_{\perp}+x\vc{q}_{\perp}\right)^2 \, 
\left(\vc{k}_{\perp}-(1-x)\vc{q}_{\perp}\right)^2 } \qquad \, \nonumber \\
&& \times  
\left[ \left(\vc{k}_{\perp}^2+(1-x)^2\left(\vc{k}_{\perp}+x\vc{q}_{\perp}\right)^2\right)
 - \frac{1}{N_c^2}\, x^2\left(\vc{k}_{\perp}-(1-x)\vc{q}_{\perp}\right)^2\right] \; .
\label{MExact}
\end{eqnarray}

\begin{figure}[t!]
\begin{center}
\epsfig{file=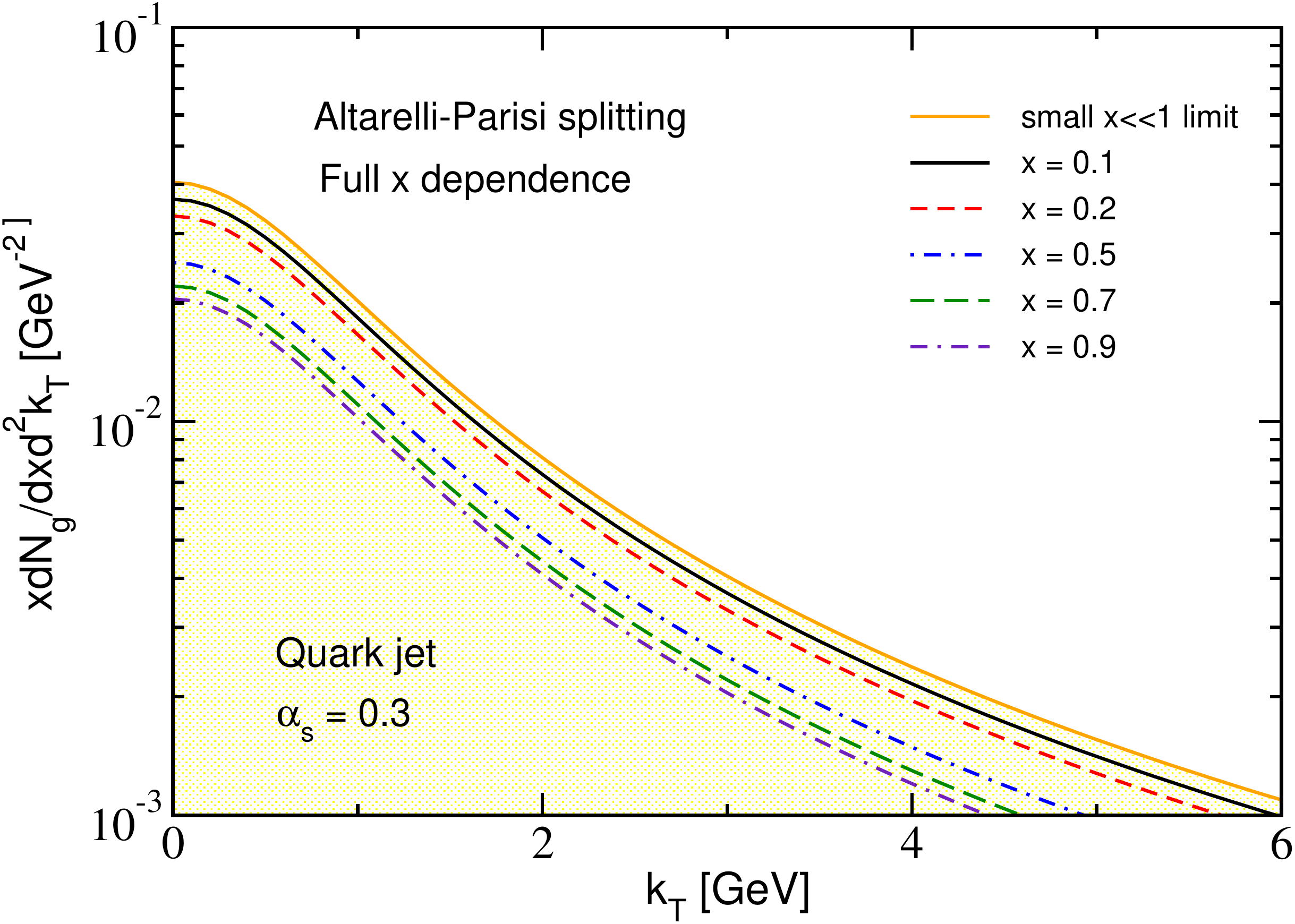, width=8cm} \epsfig{file=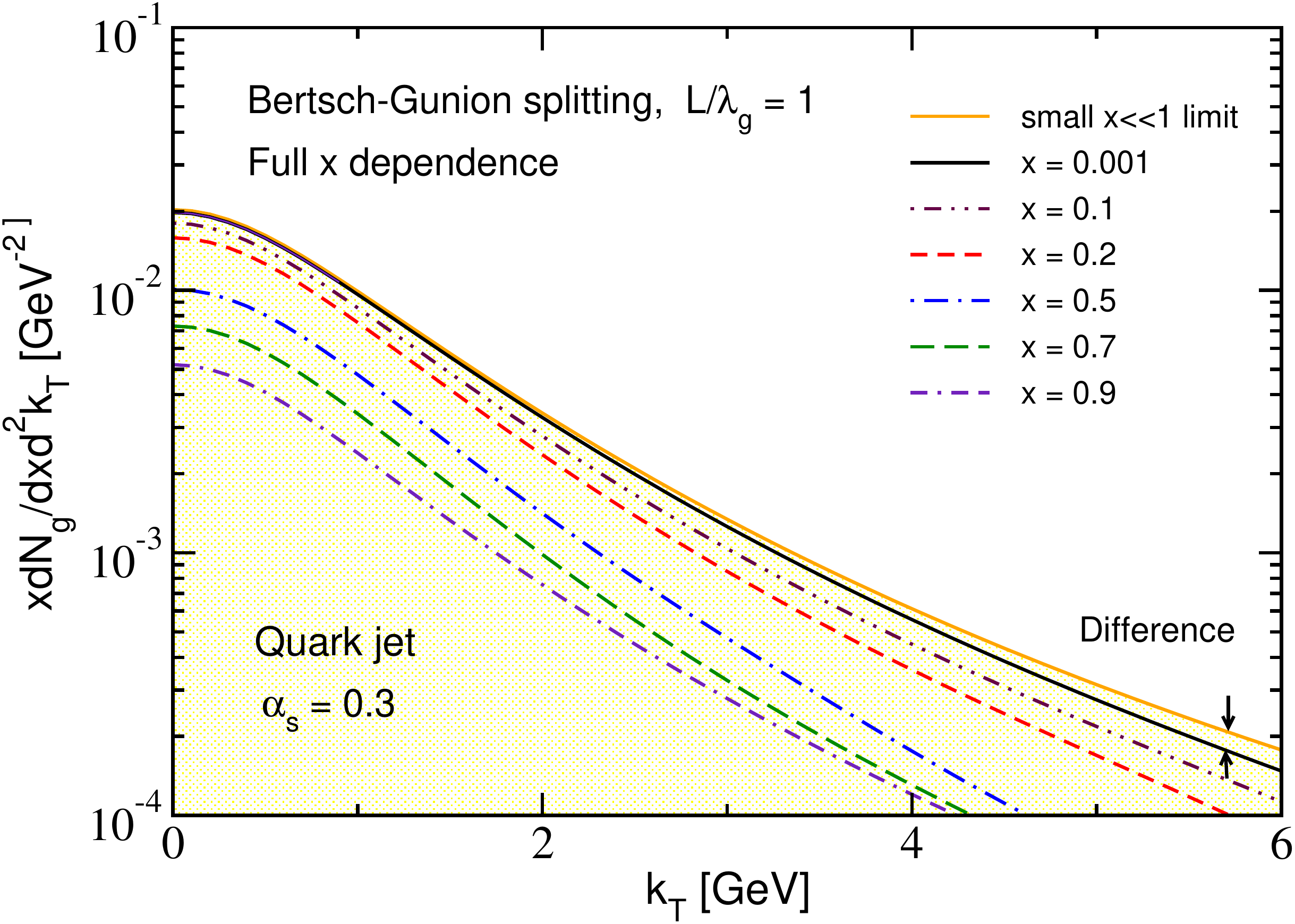, width=8cm}
\caption{\label{fig:BeyondReaction} Left panel: the differential gluon number distribution
versus $k_\perp$ is shown for selected values of $x=k^+/p^+$ for a quark jet that has undergone
large $Q^2$ scattering. A finite in-medium mass $m_{\rm eff}=1$~GeV regulates the collinear
divergence.  Right panel:  the  same spectrum for the incoherent medium-induced 
Bertsch-Gunion radiation. We have used $L/\lambda_g = 1$ and  for the $x\rightarrow 0$ 
also shown an analytic calculation in the massive scattering center limit for comparison 
to the exact calculation with $m_{\rm med.}=1$~GeV. } 
\end{center}
\end{figure}

Finite-$x$ effects are illustrated in figure~\ref{fig:BeyondReaction}. The left panel shows the 
Altarelli-Parisi real gluon differential spectrum Eq.~(\ref{APfin}) for selected values of 
$x = 0.1,\; 0.2, \; 0.5, \; 0.7, \; 0.9$  versus $k_\perp$. We use $x=0.001$ 
to simulate the small $x$-independent limit and represent the result by a solid line.  
The normalization  is fixed by the choice of a quark jet and $\alpha_s=0.3$. A finite 
effective mass  $m_{\rm eff.} = 1$~GeV to simulate medium effects is implemented as  
$k_\perp^2 \rightarrow k_\perp^2 + m_{\rm eff.}^2$ and regulates any collinear divergence. 
The dashed 
lines show that the finite-$x$  corrections can be as large as a factor of 2. The kinematic
bound for the transverse momentum of the emitted gluon in large $Q^2$ jet production
$k_{\perp \; \max} = \sqrt{x(1-x)Q^2}$ is not explicitly shown in the figure.  

The right panel of figure~\ref{fig:BeyondReaction} shows the medium-induced differential gluon spectrum 
for the same values of $x$ in the incoherent Bertsch-Gunion limit. We have chosen 
$L/\lambda_g = 1$ in Eq.~(\ref{MExact}) to facilitate direct comparison to the 
Altarelli-Parisi case. In our 
calculation we used the exact form of the $2 \rightarrow 2$ scattering cross section 
Eqs.~(\ref{scat1}), (\ref{mass}), (\ref{transform}), including the finite mass of the 
medium particles, the medium
recoil and the kinematic bounds on $q_{\perp \; \max}$. Having chosen a quark of energy 
$E=100$~GeV, we note that in the Bertsch-Gunion  case
large-$x$ effects are even more pronounced than in the Altarelli-Parisi case.

It should be noted that if an approximate form for the $q_\perp$ dependence of the
normalized differential scattering cross section is employed:
\begin{equation}
 \frac{1}{\sigma_{el}} \frac{d\sigma_{el}}{d^2 {\bf q}_\perp} =  
\frac{\mu^2}{\pi (q_\perp^2+\mu^2)^2} \;, \qquad  { q}_{\perp\; \max} \rightarrow \infty \;,   
\label{appxs}
\end{equation}
the remaining integral in Eq.~(\ref{MExact}) can be performed analytically.  
This form is motivated by the infinitely massive scattering center approximation. One uses a
Feynman change of variables developed for the calculation of loop diagrams. We here
quote the final result  for the soft gluon approximation $x \ll 1$:   
\begin{eqnarray}
 x \frac{dN^g}{ dxd^2\vc{k}_{\perp} }  &\approx & 
\frac{\alpha_s C_F}{\pi^2}\,\left(\frac{L}{\lambda_g}\, \right) 
\frac{\mu^2}{ \left({\bf{k}}_{\perp}^2\right)^2\, ({\bf{k}}_{\perp}^2+m_{\rm eff.}^2 ) }
 \nonumber \\ 
&& \times \frac{\mu^2(\lambda_1^2-\lambda_2^2)+
\left[\mu^2(\lambda_1+\lambda_2-2)+{\bf{k}_{\perp}^2}(\lambda_1-\lambda_2)^2\right]\, \lambda_1 \lambda_2\,
\ln\left(\frac{\lambda_1(1-\lambda_2)}{\lambda_2(1-\lambda_1)}\right)}{ \lambda_1\lambda_2(\lambda_1-\lambda_2)^3},
\end{eqnarray}
where $\lambda_{1,2}$ are the solutions of the quadratic equation:
\begin{equation}
\lambda\, m_{\rm eff.}^2+(1-\lambda)\,\mu^2+\lambda(1-\lambda)\,{\bf{k}}_{\perp}^2=0\; .
\end{equation}
The analytical formula for the  finite $x$ case is also calculable, though slightly more 
involved. To illustrate the differences between the exact form of the $2\rightarrow 2$ 
scattering cross section and the approximate form given by Eq.~(\ref{appxs}) we show the
analytic small-$x$  gluon distribution versus $k_\perp$ in the right panel of 
figure~\ref{fig:BeyondReaction}. We note that the differences between the exact and approximate 
calculations are very small and appear only at high $k_\perp$ - a part of phase space that 
does not contribute significantly to the medium-induced energy loss.

\subsection{Final-state radiation and the Landau-Pomeranchuk-Migdal effect}
In this subsection we calculate the combination of single and double Born amplitudes to first 
order in opacity by keeping the finite-$x$ corrections. This evaluation proceeds analogously to the 
one for the incoherent Bertsch-Gunion limit. Directly from the Feynman rules of $\text{\SCETG}$, derived 
in this paper, we find the combined squared amplitude from single and double Born diagrams equals to:
\begin{eqnarray}
&& \frac{1}{d_Rd_T} 
\left\langle \text{Tr}\left[A_1 A_1^{\dagger}+A_0 A_2^{(c)\dagger}
+A_0^{\dagger} A_2^{(c)}\right]\right\rangle_{\rm medium \; {\bf q}_\perp}  \nonumber \\
&& \qquad \qquad \qquad 
=\frac{N}{A_{\perp}}\int\,\frac{d^2\vc{q}_{\perp}}{(2\pi)^2}\,|\tilde{v}(\vc{q}_{\perp})|^2\,
\text{Tr}\left(\frac{\nslash}{2}\bar{n}\mcdot p\,J\bar{J}\, \frac{g^2}{d_Rd_T} 
\left[\rho^{\text{SB}}+\rho^{\text{DB}}\right] \right),
\end{eqnarray}
where both $\rho^{\text{SB}}$ and $\rho^{\text{DB}}$ have the following form:
\begin{eqnarray}
\rho=\sum_{i=1}^2\, c_i\,\left(F_i\,\mathbb{I} + G_i\, \Sigma^3\right).
\label{sdecomp}
\end{eqnarray}
In the equation above $F_i, G_i$ are  form-factors, which we give below, and $c_i$ are the color factors: 
$$c_1=C_2(R)^2C(r)d_R\, , \qquad c_2=C_2(R)C(r) \left(C_2(R)-\frac{1}{2}C_A\right)d_R\;.$$ We recall that in our example 
$C_2(R)=C_F$, $d_R=N_c$ and $C(r)=\frac{1}{2}$. The two operators that appear are the identity operator 
in the Dirac indices $\mathbb{I}$ and  $\Sigma^3$, which is the third component of the spin 
operator for the fermion:
\begin{equation}
\Sigma^3= \left [ \begin{array}{ccc} \sigma_3&&0\\
0 &&\sigma_3 \end{array} \right] \, . 
\end{equation}
The first term in Eq.~(\ref{sdecomp}) allows for the factorization of the medium-induced radiative 
corrections and the hard scattering cross section even in the large-$x$ limit. The second term 
in principle does not vanish. If we, however, recall the decomposition of $J(p)\bar{J}(p)$ in the 
Dirac algebra basis, see Eq.~(\ref{decompose}), it is easy to verify that there must be 
a non-zero pseudo-vector contribution for this term not to vanish identically. Therefore,
one might expect further corrections for processes with electroweak boson exchanges. For QCD, 
 medium-induced radiative corrections always factorize.
A simple exercise for the interested reader is to show this on the example of inclusive tree level
jet production. 
  
The form-factors for the single Born diagrams in Eq.~(\ref{sdecomp})  are given by the following 
expressions:
\begin{eqnarray}
F^{\text{SB}}_1&=&\left(1-x+\frac{x^2}{2}\right)\left(|\vc{\beta}_1|^2+|\vc{\beta}_2|^2\right)\, ,\qquad G^{\text{SB}}_1=2\left(x-\frac{x^2}{2}\right)\,\text{Im}\left[\left(\vc{\beta}_1^x\right)^*\vc{\beta}_1^y+\left(\vc{\beta}_2^x\right)^*\vc{\beta}_2^y\right]\, ,\label{FG1}\\
F^{\text{SB}}_2&=&\left(1-x+\frac{x^2}{2}\right)\left(\vc{\beta}_1\mcdot \vc{\beta}_2^{*}+\vc{\beta}^{*}_1\mcdot \vc{\beta}_2\right)\, , \qquad G^{\text{SB}}_2=2\left(x-\frac{x^2}{2}\right)\,\text{Im}\left[\left(\vc{\beta}_2^x\right)^*\vc{\beta}_1^y+\left(\vc{\beta}_1^x\right)^*\vc{\beta}_2^y\right]\, .
\end{eqnarray}
On the other hand, the double Born diagrams contributions can be written as follows:
\begin{eqnarray}
F^{\text{DB}}_1&=&\left(1-x+\frac{x^2}{2}\right)\left(\vc{\alpha}_0\mcdot \vc{\gamma}_1^{*}+\vc{\alpha}^{*}_0\mcdot \vc{\gamma}_1\right)\, ,\qquad G^{\text{DB}}_1=2\left(x-\frac{x^2}{2}\right)\,\text{Im}\left[\left(\vc{\gamma}_1^x\right)^*\vc{\alpha}_0^y+\left(\vc{\alpha}_0^x\right)^*\vc{\gamma}_1^y\right]\, ,\\
F^{\text{DB}}_2&=&\left(1-x+\frac{x^2}{2}\right)\left(\vc{\alpha}_0\mcdot \vc{\gamma}_2^{*}+\vc{\alpha}^{*}_0\mcdot \vc{\gamma}_2\right)\, , \qquad G^{\text{DB}}_2=2\left(x-\frac{x^2}{2}\right)\,\text{Im}\left[\left(\vc{\gamma}_2^x\right)^*\vc{\alpha}_0^y+\left(\vc{\alpha}_0^x\right)^*\vc{\gamma}_2^y\right]\, . \label{FG4}
\end{eqnarray}
In the expressions above the vector $\vc{\alpha}_0$ appears from the vacuum diagram, the vectors 
$\vc{\beta}_{1,2}$ appear from the single Born diagrams and $\vc{\gamma}_{1,2}$ appear from the
 double Born diagrams calculation. They all are given in the equations below:
\begin{eqnarray}
&&\vc{\alpha}_{0}=\frac{2\vc{A}_{\perp}}{\vc{A}_{\perp}^2}\, ,\\
&&\vc{\beta}_{1}=\left(\frac{2\vc{A}_{\perp}}{\vc{A}_{\perp}^2}-\frac{2\vc{C}_{\perp}}{\vc{C}_{\perp}^2}\right)\,\e^{i\Omega_1\delta z}+\frac{2\vc{C}_{\perp}}{\vc{C}_{\perp}^2}\,\e^{i\Omega_3\delta z}\, ,\\
&&\vc{\beta}_{2}=-\left(\frac{2\vc{B}_{\perp}}{\vc{B}_{\perp}^2}-\frac{2\vc{C}_{\perp}}{\vc{C}_{\perp}^2}\right)\,\e^{i\Omega_1\delta z}+\frac{2\vc{B}_{\perp}}{\vc{B}_{\perp}^2}\,\e^{i\Omega_2\delta z}-\frac{2\vc{C}_{\perp}}{\vc{C}_{\perp}^2}\,\e^{i\Omega_3\delta z}\, ,\\
&&\vc{\gamma}_{1}=-\frac{3\vc{A}_{\perp}}{\vc{A}_{\perp}^2}+\left(\frac{2\vc{A}_{\perp}}{\vc{A}_{\perp}^2}-\frac{2\vc{D}_{\perp}}{\vc{D}_{\perp}^2}\right)\,\e^{i\Omega_4\delta z}+\frac{2\vc{D}_{\perp}}{\vc{D}_{\perp}^2}\,\e^{i\Omega_5\delta z} \, ,\\
&&\vc{\gamma}_{2}= \frac{2\vc{A}_{\perp}}{\vc{A}_{\perp}^2}-\left(\frac{2\vc{A}_{\perp}}{\vc{A}_{\perp}^2}-\frac{2\vc{D}_{\perp}}{\vc{D}_{\perp}^2}\right)\,\e^{i\Omega_4\delta z}-\frac{2\vc{D}_{\perp}}{\vc{D}_{\perp}^2}\,\e^{i\Omega_5\delta z} \, ,\\
\end{eqnarray}
where the vectors $\vc{A}_{\perp}, \vc{B}_{\perp}, \vc{C}_{\perp}$ are same as in the previous subsection, 
the phase factors $\Omega_{1,2,3}$ are same as in appendix~\ref{appendix:integralsRad}. 
The remaining propagators and inverse formation times
$\vc{D}_{\perp}, \Omega_{4}, \Omega_5$ appear in the double Born diagrams and are equal to:
\begin{eqnarray}
&&\vc{D}_{\perp}=\vc{A}_{\perp}-\vc{q}_{\perp} 
= \vc{k}_{\perp} - \vc{q}_{\perp} |_{{\bf p}_\perp= -{\bf k}_\perp} \, ,\\[1ex]
&&\Omega_4=\frac{\vc{A}_{\perp}^2}{p_0^+x(1-x)}\, , \qquad
\Omega_5=\frac{\vc{A}_{\perp}^2-\vc{D}_{\perp}^2}{p_0^+x(1-x)}\, .
\end{eqnarray}

First, we should check that our result agrees with previous calculation in the soft gluon 
approximation~\cite{Vitev:2007ve,Gyulassy:2000er}. For this we need to expand all quantities 
to lowest order in $x\ll 1$:
\begin{eqnarray}
&&\vc{A}_{\perp}\approx \vc{k}_{\perp}\, , \quad
\vc{B}_{\perp}\approx \vc{k}_{\perp}\, , \quad 
\vc{C}_{\perp}\approx \vc{k}_{\perp}-\vc{q}_{\perp}\, , \quad
\vc{D}_{\perp}\approx \vc{k}_{\perp}-\vc{q}_{\perp}\, ,\\
&&\Omega_1\approx \omega_0 \, , \quad  \Omega_2\approx 0\, , \quad
\Omega_3\approx \omega_0-\omega_1\, , \quad \Omega_4\approx \omega_0\, , \quad
\Omega_5\approx \omega_0-\omega_1\, ,\\
&&G^{\text{SB}}_i\approx 0\, , \quad 
G^{\text{DB}}_i\approx 0\, .
\end{eqnarray}
Note that in this limit factorization is exact even if $J(p)\bar{J}(p)$ contains a pseudo-vector component.
The final answer depends on the form-factors $F_i$, which we calculate below in the 
soft gluon limit. It is convenient to rewrite the $\vc{\alpha}, \vc{\beta}_{1,2}, \vc{\gamma}_{1,2}$ in terms of standard definitions in the 
literature\footnote{Note that $\vc{B}_1$ is distinct 
from $\vc{B}_{\perp}$ and $\vc{C}_1$ is distinct 
from $\vc{C}_{\perp}$.}~\cite{Vitev:2007ve,Gyulassy:2000er}:
\begin{eqnarray}
   \vc{H}_1=\frac{\vc{k}_{\perp}}{\vc{k}_{\perp}^2}\, ,\qquad
    \vc{C}_1=\frac{\vc{k}_{\perp}-\vc{q}_{\perp}}{(\vc{k}_{\perp}-\vc{q}_{\perp})^2}\, ,\qquad
     \vc{B}_1=\vc{H}_1-\vc{C}_1\, .
\end{eqnarray}
The corresponding $x\rightarrow 0$ limit is particularly simple:
\begin{eqnarray}
&&\vc{\alpha}_{0}=2\vc{H}_1 \, ,\\
&&\vc{\beta}_{1}=2\vc{B}_1\,\e^{i\omega_0\delta z}+2\vc{C}_1\,\e^{i(\omega_0-\omega_1)\delta z}\, , \qquad
\vc{\beta}_{2}=2\vc{H}_1-2\vc{B}_1\,\e^{i\omega_0\delta z}-2\vc{C}_1\,\e^{i(\omega_0-\omega_1)\delta z}\, ,\\
&&\vc{\gamma}_{1}=-3\vc{H}_1+2\vc{B}_1\,\e^{i\omega_0\delta z}+2\vc{C}_1\,\e^{i(\omega_0-\omega_1)\delta z}\, , \qquad 
\vc{\gamma}_{2}=2\vc{H}_1-2\vc{B}_1\,\e^{i\omega_0\delta z}-2\vc{C}_1\,\e^{i(\omega_0-\omega_1)\delta z} \,  .
\end{eqnarray}
Useful relations between these vectors, that help in deriving the expression for the form-factors $F_i$ are the following ones:
\begin{eqnarray}
\vc{\beta}_1+\vc{\beta}_2=2\vc{H}_1\, , \qquad \vc{\gamma}_1+\vc{\gamma}_2=-\vc{H}_1 \, .
\end{eqnarray}
From the definitions in~\eq{FG1}-\eq{FG4} we get:
\begin{eqnarray}
&&F^{\text{SB}}_1 = |\vc{\beta}_1|^2+|2\vc{H}_1-\vc{\beta}_1|^2=2|\vc{\beta}_1|^2+4\vc{H}_1^2-4\,\text{Re}\,\vc{H}_1\mcdot\vc{\beta}_1 \nonumber \\
&&  \qquad \;  = 8\vc{B}_1^2+8\vc{C}_1^2+4\vc{H}_1^2+16\vc{B}_1\mcdot \vc{C}_1\cos(\omega_1\delta z)-8\vc{H}_1\mcdot \vc{B}_1\cos(\omega_0\delta z) 
-8\vc{H}_1\mcdot \vc{C}_1\cos((\omega_0-\omega_1)\delta z)\, , \qquad \quad \\
&&F^{\text{SB}}_2=\vc{\beta}_1\left(2\vc{H}_1-\vc{\beta}_1\right)^*+(2\vc{H}_1-\vc{\beta}_2)^*\vc{\beta}_2=-F_1^{\text{SB}}+4\,\vc{H}_1^2\, ,\\
&&F^{\text{DB}}_1=2\vc{H}_1\mcdot\,2\,\text{Re}\,\vc{\gamma}_1=-12\vc{H}_1^2+8\vc{H}_1\mcdot \,\vc{B}_1\cos(\omega_0\delta z)+8\vc{H}_1\mcdot \,\vc{C}_1\cos((\omega_0-\omega_1)\delta z)\, ,\\
&&F^{\text{DB}}_2=2\vc{H}_1\mcdot\,2\,\text{Re}\,\vc{\gamma}_2=-F_1^{\text{DB}}-4\vc{H}_1^2\, .
\end{eqnarray}
Finally, using these equations we combine the single and double Born form-factors into the sum:
\begin{eqnarray}
&&F^{\text{SB}}_1+F^{\text{DB}}_1= 8\vc{B}_1^2+8\vc{C}_1^2-8\vc{H}_1^2
+16\vc{B}_1\mcdot \vc{C}_1\cos(\omega_1\delta z)=   
-16\vc{B}_1\mcdot \vc{C}_1 \, \left(1-\cos(\omega_1 \delta z)\right)\, ,\\
&&F^{\text{SB}}_2+F^{\text{DB}}_2=-F^{\text{SB}}_1-F^{\text{DB}}_1 \, .
\end{eqnarray}
Thus, in the soft gluon approximation we get:
\begin{eqnarray}
\left(\rho^{\text{SB}}+\rho^{\text{DB}}\right)_{x\ll 1}\approx(c_1-c_2)\, 
\left(-16\vc{B}_1\mcdot \vc{C}_1\right)\left(1-\cos(\omega_1 \delta z)\right) \, .
\end{eqnarray}
Taking into account the phase space factors, the color factors and the final-state 
coherent medium-induced emission contribution above, we find:
\begin{eqnarray}
   && x \frac{dN^g}{ dxd^2\vc{k}_{\perp} }_{|x \ll 1}  = C_F \frac{\alpha_s}{\pi^2}  
\int \frac{d\Delta z}{\lambda_g(z)}  
\int d^2{\bf q}_\perp  \frac{1}{\sigma_{el}} \frac{d\sigma_{el}^{g\; {\rm medium}}}{d^2 {\bf q}_\perp} \; 
 \left(-2\vc{B}_1\mcdot \vc{C}_1\right)\left(1-\cos(\omega_1 \Delta z)\right) \, .           
\label{CohRadSX} 
\end{eqnarray} 
in agreement with Eq.~(70) of \cite{Vitev:2007ve}.

Beyond the soft gluon approximation, the full result for the coherent medium-induced bremsstrahlung
reads:
\begin{eqnarray}
  x \frac{dN^g}{ dxd^2\vc{k}_{\perp} }  &=& C_F \frac{\alpha_s}{\pi^2}
\left(1-x+\frac{x^2}{2} \right) 
\int \frac{d\Delta z}{\lambda_g(z)}  
\int d^2{\bf q}_\perp  \frac{1}{\sigma_{el}} \frac{d\sigma_{el}^{g\; {\rm medium}}}{d^2 {\bf q}_\perp} \; 
 \Bigg[  -\left(\frac{\vc{A}_\perp}{\vc{A}_\perp^2}\right)^2  + 2 \left(\frac{\vc{C}_\perp}{\vc{C}_\perp^2}\right)^2 
- \frac{\vc{A}_\perp}{\vc{A}_\perp^2} \mcdot \frac{\vc{C}_\perp}{\vc{C}_\perp^2}    \nonumber \\
&&  - \frac{\vc{B}_\perp}{\vc{B}_\perp^2} \mcdot \frac{\vc{C}_\perp}{\vc{C}_\perp^2} \big( 1-\cos[(\Omega_1 -\Omega_2)\Delta z] 
+\cos[(\Omega_2 -\Omega_3)\Delta z]  \big)  \nonumber \\  
&& + \frac{\vc{C}_\perp}{\vc{C}_\perp^2} \mcdot \left( \frac{\vc{A}_\perp}{\vc{A}_\perp^2} + \frac{\vc{B}_\perp}{\vc{B}_\perp^2}  
-2 \frac{\vc{C}_\perp}{\vc{C}_\perp^2}   \right) \cos[(\Omega_1 -\Omega_3)\Delta z] 
+ \frac{\vc{A}_\perp}{\vc{A}_\perp^2} \mcdot \left( \frac{\vc{A}_\perp}{\vc{A}_\perp^2} - 
\frac{\vc{D}_\perp}{\vc{D}_\perp^2} \right) 
\cos[\Omega_4\Delta z]  \nonumber \\
&& +\frac{\vc{A}_\perp}{\vc{A}_\perp^2} \mcdot \frac{\vc{D}_\perp}{\vc{D}_\perp^2}\cos[\Omega_5\Delta z]   
+ \left(  \frac{N_c^2-1}{N_c^2} \left(\frac{\vc{B}_\perp}{\vc{B}_\perp^2}\right)^2    
+    \frac{1}{N_c^2} \frac{\vc{A}_\perp}{\vc{A}_\perp^2} \mcdot \frac{\vc{B}_\perp}{\vc{B}_\perp^2}   \right)
\big( 1-\cos[(\Omega_1 -\Omega_2)\Delta z] \big)   \Bigg] \, . \qquad          \; 
\label{CohRadSX1} 
\end{eqnarray} 
We leave the discussion of this new result and phenomenological applications to new RHIC and LHC
experimental data~\cite{Salur:2009vz,Lai:2009zq,Ploskon:2009zd,Aamodt:2010jd,Chatrchyan:2011sx,Aad:2010bu} 
for future work. We note however that in Ref.~\cite{Zhang:2003yn} an evaluation of the medium-induced
energy loss beyond the helicity amplitude approximation found $\sim 18\%$ reduction in the mean
medium-induced energy loss. Our plan for the future is to investigate the large-$x$ radiative correction 
reduction effects at the most differential level.

\section{Conclusions}
\label{conclude}

In summary, we constructed an effective theory SCET$_{\rm G}$ for energetic quark and gluon 
$p\sim [1,\lambda^2, {\bm \lambda}] $  
propagation and interaction in dense QCD matter. This theory is well-suited to calculations 
both in the quark-gluon plasma~\cite{Vitev:2008rz,Vitev:2009rd,Renk:2009hv,Neufeld:2010fj,Sharma:2009hn} 
 and in cold nuclear matter~\cite{Qiu:2003pm,Vitev:2008vk,Sharma:2009hn,Neufeld:2010dz}. 

To construct this theory, we examined the relevant 
$t$-channel parton scattering cross sections and demonstrated that they
are dominated by forward scattering. The corresponding momentum exchanges
$ q \sim [\lambda^2,\lambda^2, {\bm \lambda}] $  
are approximately transverse to the direction of jet propagation and mediated
by Glauber gluons. We also demonstrated that a fully dynamical treatment of the 
scattering centers in QCD matter leads to a small reduction of the scattering cross 
section and smaller medium-induced effects in contrast to early speculations~\cite{Djordjevic:2009cr}. 
The SCET$_{\rm G}$  Lagrangian was shown to be invariant 
under soft and collinear gauge transformations. We derived the Feynman rules for this 
new Lagrangian in the covariant and light-cone gauges. Also we provided a third choice, 
which we call the hybrid gauge, when the
Glauber gluons are quantized in the covariant gauge, while the collinear fields are gauge-fixed with 
the light-cone gauge. This choice provides us with the simplest form of Feynman rules and is useful from
practical point of view.

The new effective theory was used  to evaluate the broadening and medium-induced radiation for
energetic quarks traversing a region of hot/dense QCD matter. SCET$_{\rm G}$ was shown
to recover exactly the known results for the transverse momentum diffusion of particles
in the strongly-interacting medium~\cite{Gyulassy:2002yv,Qiu:2003pm}. 
It allows to study the Molliere multiple  parton
scattering beyond the limitations of the Gaussian approximation - namely, the inability
to describe the Rutherford power-law  tails of transverse momentum distributions.
For the case of inelastic quark interactions we derived the fully differential  
medium-induces gluon bremsstrahlung spectrum for both the incoherent and Landau-Pomeranchuck-Migdal
suppressed cases. In the soft gluon approximation we obtained the kernel for the 
reaction operator approach to medium-induced energy loss~\cite{Gyulassy:2000er,Vitev:2007ve}. 
We also evaluated the large
$x=k^+/E^+$ corrections to the medium-induced bremsstrahlung spectrum.
Gauge invariance of the jet broadening and radiative energy loss results 
was  demonstrated  explicitly for the first time by performing the calculations in 
the  covariant $R_{\xi}$, light-cone and hybrid gauges. On the example of an energetic quark, we 
found that in QCD the process-dependent medium-induced radiative corrections factorize from the 
hard jet production  cross section. This allows us to write down perturbative jet and leading particle 
observables in heavy ion collisions as a convolution of the corresponding observables in the more 
elementary nucleon-nucleon reactions and the medium-induced collisional and radiative  corrections
specific to the process under investigation.  
Our results put jet quenching phenomenology in heavy-ion collisions on more solid theoretical 
grounds.

In the near future it will be a high priority for us to implement the improved theory
of parton propagation and energy loss in matter in perturbative QCD predictions for  
analysis of the energetic particle and jet quenching data from the heavy-ion experiments
at RHIC and at the LHC. With the calculations at hand, we expect to be able to reliably
combine the process-dependent medium-induced radiative corrections with next-to-leading 
order perturbative effects~\cite{Vitev:2010ci}. We plan to extend these calculations 
beyond applications to jet propagation in the quark-gluon plasma~\cite{Vitev:2008rz,Vitev:2009rd,Neufeld:2010fj} 
and also improve the 
accuracy in the evaluation of cold nuclear matter effects~\cite{Neufeld:2010dz}.

The derived Lagrangian of $\text{\SCETG}$ can be used to revisit the factorization of the 
Drell-Yan process in the effective theory and include the spectator interactions into the analysis. 
The importance of Glauber (or Coulomb) gluons for the Drell-Yan factorization has been addressed, and their 
cancellation in the inclusive cross section has been proved in traditional QCD approach 
to factorization~\cite{Collins:1982wa, Bodwin:1984hc, Collins:1985ue},  while similar understanding 
in effective theory method is still missing~\cite{StewartTalk, Bauer:2010cc}.

\acknowledgments
We thank Christian Bauer and Gerry Hale for many illuminating discussions. We thank Christian Bauer for useful comments on the manuscript.
This research is  supported by the US Department of Energy, Office  of Science, under
Contract No. DE-AC52-06NA25396 and in part by the LDRD program at LANL.

\appendix

\section{Light-cone notation}\label{appendix:lightcone}

The momentum of any  particle can be conveniently represented  in light-cone coordinates. 
A light-cone vector $\beta$ is defined by the condition $\beta^2=0$. The two vectors that 
specify the positive and negative light-cone directions are  $n^{\mu}=(1,0,0,1)$ and  
$\bar{n}^{\mu}=(1,0,0,-1)$, respectively. 
An arbitrary four-vector $p$ then can be expanded in the light-cone vectors basis: 
\begin{eqnarray}
p^{\mu}=p^{+}\,\frac{n^{\mu}}{2}+p^{-}\,\frac{\bar{n}^{\mu}}{2}+p^{\mu}_{\perp},
\end{eqnarray}
where the $+,-$  components are defined as follows: $p^{+}\equiv \bar{n}\mcdot p,\,  
p^{-}\equiv {n}\mcdot p$. The four-vector $p^{\mu}$ thus can be written in the light-cone 
basis as $[p^+,p^-,\vc{p}_{\perp}]$, where we will use the square brackets to emphasize that 
the light-cone notation is being used.  For example, the light-cone vectors have the following 
coordinates:  $n = [2,0,\vc{0}], \; \bar{n}=[0,2,\vc{0}]$, corresponding to positive and negative 
light-cone directions. Thus, the plus component is in the $n$ direction
and the minus component is in the $\bar{n}$  direction. 
The degrees of freedom of SCET are collinear quarks and gluons, with the following 
momentum scaling $p_c=[1,\lambda^2,{\bm \lambda}]$ and soft gluons 
with momentum 
$p_s=[\lambda^2,\lambda^2,{\bm \lambda^2}]$. The Glauber modes that we consider in this 
paper  have momentum scaling $p_g=[\lambda^2,\lambda^2,{\bm \lambda}]$.

\section{Kinematics in  the laboratory frame}
In section~\ref{sec:kinematics} we estimated the recoil effect in $2\rightarrow 2$ scattering of 
a projectile particle with mass $m_1$ on an originally at rest target particle with the mass $m_2$. 
An important feature of the laboratory frame is that the final energy $E_3$ of the scattered particle is 
a non-trivial function of the laboratory angle $\theta$. This function can be found from energy 
and momentum conservation laws. The solution is given by:
\begin{eqnarray}
E_{3}(\theta)=m_2 \,\frac{\rho_1\gamma^2\pm\left(\gamma^2-1\right)r\sin\theta\cos\theta}
{\gamma^2-\left(\gamma^2-1\right)\cos^2\theta}\; .\label{E1Leq}
\end{eqnarray}
In  Eq.~(\ref{E1Leq}) we defined:
\begin{eqnarray}
&&\rho_1=\frac{\beta}{\beta_1}\; ,\\
&&r=\sqrt{\gamma^2\left(1-\rho_1^2\right)+\cot^2\theta}\; ,
\end{eqnarray}
where $\beta_1$ is the velocity of the projectile particle in the center-of-mass (CM) frame
and $\beta$ is the velocity of this center-of-mass. Note that the kinematics 
is very different depending on the masses. If $m_1>m_2$, then both solutions in~\eq{E1Leq} are physical, 
while for $m_1<m_2$ only the ``$+$''~solution is physical, while the ``$-$''-solution is not. In our case we 
can safely assume that $m_1<m_2$. 

The reason for the difference discussed above is that when the projectile 
particle is heavier then the target one, then in the center of mass frame the projectile 
particle's velocity is smaller than the center-of-mass velocity in the laboratory frame, 
i.e $\beta_1<\beta$. 
This means that no matter in which direction the projectile particle goes in the CM 
frame, it will always move in the positive hemisphere of the laboratory frame with respect to the  
initial momentum of the projectile particle. Thus, each direction of motion of the scattered 
projectile particle in the laboratory frame corresponds to two distinct directions in the center 
of mass frame. That is the origin of two solutions from~\eq{E1Leq}. Alternatively, for $m_1<m_2$, 
we have $\beta_1>\beta$ and each direction of the projectile particle in laboratory frame comes from 
a distinct direction of this particle in the center of mass frame, leading to the corresponding 
unique energy in the laboratory frame.
 
\begin{figure}[h!]
\begin{eqnarray}
  && \begin{picture}(20,10)(80,10)
     \put(45,20){$p$}
     \mbox{\epsfxsize=3.4truecm \hbox{\epsfbox{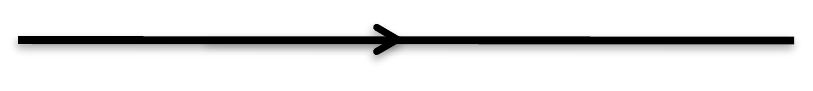}}  }
  \end{picture}\quad 
  \raise-6pt \hbox{$  = \ \mbox{\normalsize $i$}\, \frac{\nslash}{2}\:  \frac{1}{p^-+p_{\perp}^2/{p^++i\eps}}$ } 
  \nonumber\\[28pt]
   && \begin{picture}(20,10)(80,10)
     \mbox{\epsfxsize=3.4truecm \hbox{\epsfbox{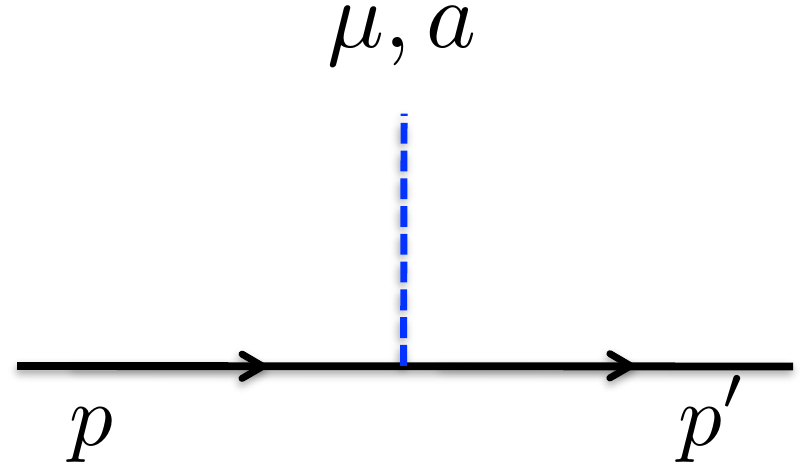}}  }
  \end{picture}\quad 
 \hbox{$  = \ \mbox{\normalsize $i g T^a$}\, \:  \left(n^{\mu}+\frac{\gamma_{\perp}^{\mu}\pslash_{\perp}}{p^+}+\frac{\pslash_{\perp}'\gamma_{\perp}^{\mu}}{p'^{+}}-\bar{n}^{\mu}\frac{\pslash_{\perp}'\pslash_{\perp}}{p'^+ p^+}\right)\frac{\bnslash}{2}$ } 
  \nonumber\\[28pt]
   && \begin{picture}(20,10)(80,10)
     \mbox{\epsfxsize=3.4truecm \hbox{\epsfbox{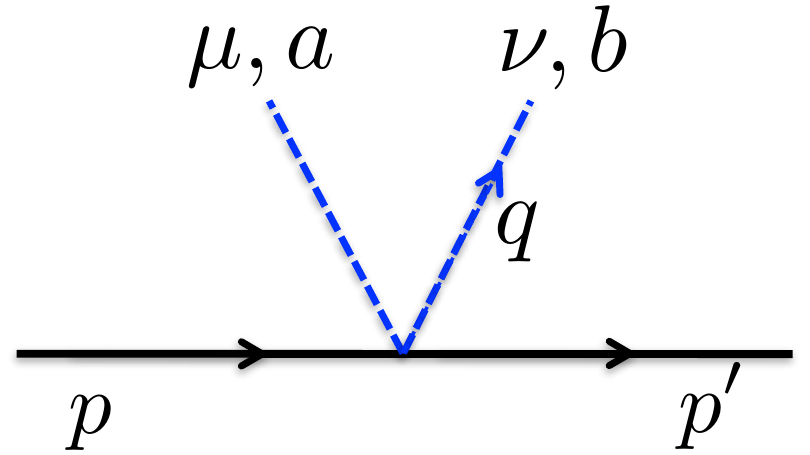}}  }
  \end{picture}\quad 
 \hbox{$  = \ \mbox{\normalsize}\, \:  \frac{i g^2\, T^a T^b}{p^+-q^+}\left[\gamma_{\perp}^{\mu}\gamma_{\perp}^{\nu}-\frac{\gamma_{\perp}^{\mu}\pslash_{\perp}}{p^+}\bar{n}^{\nu}-\frac{\pslash_{\perp}'\gamma_{\perp}^{\nu}}{p'^+}\bar{n}^{\mu}+\frac{\pslash_{\perp}'\pslash_{\perp}}{p^+p'^+}\bar{n}^{\mu}\bar{n}^{\nu}\right]\frac{\bnslash}{2}$ } \nonumber\\
    && \qquad \qquad 
 \hbox{$  \ \mbox{\normalsize{$+$}}\, \:  \frac{i g^2\, T^b T^a}{q^++p'^+}\left[\gamma_{\perp}^{\nu}\gamma_{\perp}^{\mu}-\frac{\gamma_{\perp}^{\nu}\pslash_{\perp}}{p^+}\bar{n}^{\mu}-\frac{\pslash_{\perp}'\gamma_{\perp}^{\mu}}{p'^+}\bar{n}^{\nu}+\frac{\pslash_{\perp}'\pslash_{\perp}}{p^+p'^+}\bar{n}^{\mu}\bar{n}^{\nu}\right]\frac{\bnslash}{2}$ } \nonumber
\end{eqnarray}
{\caption[1]{Feynman rules of SCET for the interactions between collinear quarks and gluons.}
\label{fig:SCETFR} }
\end{figure}

\section{Feynman rules of \text{\SCETG}}\label{Appendix:SCETG}
In this appendix we review the Feynman rules of SCET and also present all the relevant Feynman rules of 
$\text{\SCETG}$ for different gauge choices. We consider the initially static source case. 

First, we start with the SCET rules. Using the Lagrangian of SCET~\cite{Bauer:2000yr} given in~\eq{SCETL1} 
one finds the Feynman rules of interaction between the collinear quarks and gluons given in 
figure~\ref{fig:SCETFR}. These rules are given in the covariant gauge. There are additional vertices 
in this gauge when one has more than 2 collinear gluons and two collinear quarks at the same point, 
which we omit here. The same Feynman rules as in figure~\ref{fig:SCETFR} are valid in the positive 
light-cone gauge $A^+$, with the simplification that all the $\bar{n}^{\mu}, \bar{n}^{\nu}$ terms vanish. 
Also, all the mentioned omitted vertices vanish in the light-cone gauge.

\begin{figure}[h!]
\begin{eqnarray}
\qquad \qquad\qquad  && \begin{picture}(20,10)(80,10)
       \put(-19,2){\circle{22}}\put(-25,0){$R_{\xi}$}       \put(55,11){$q_1$}     \put(7,36){$p$}      \put(87,36){$p'$}  \put(40,-7){$(b_1)_{T_i}$}
     \mbox{\epsfxsize=3.4truecm \hbox{\epsfbox{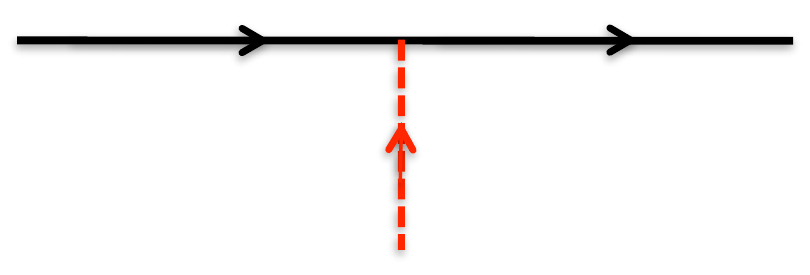}}  }
  \end{picture}\quad 
  \raise15pt \hbox{$  = \ \mbox{\normalsize $i$}\, v(q_{1\perp})\, (b_1)_R\,(b_1)_{T_i}\frac{\bnslash}{2}$ } 
  \nonumber\\[15pt]
  && \begin{picture}(20,10)(80,10)
       \put(0,15){$\mu, a$}    \put(85,15){$\nu, b$}    \put(55,11){$q_1$}     \put(7,36){$p$}      \put(87,36){$p'$}   \put(40,-7){$(c_1)_{T_i}$}
     \mbox{\epsfxsize=3.4truecm \hbox{\epsfbox{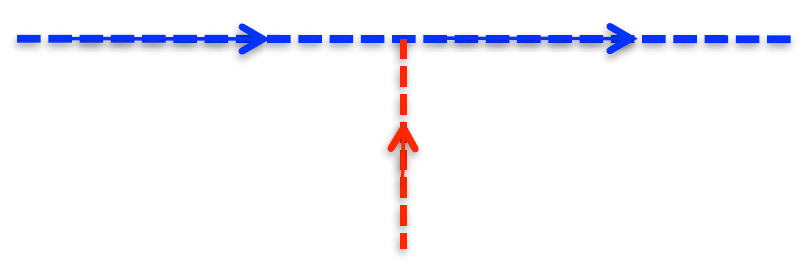}}  }
  \end{picture}\quad 
  \raise15pt \hbox{$  = \ \mbox{\normalsize }\,v(q_{1\perp})\, f^{abc_1}\,(c_1)_{T_i}\,\left[g^{\mu\nu}\,\bar{n}\mcdot p+\bar{n}^{\mu} \,q^{\nu}_{1\perp}-\bar{n}^{\nu}\,q^{\mu}_{1\perp}-\frac{1-\frac{1}{\xi}}{2}\left(\bar{n}^{\nu}p^{\mu}+\bar{n}^{\mu}p'^{\nu}\right)\right]$ } 
  \nonumber\\[20pt] 
  \hline
  \hline
  \nonumber\\[15pt]
    && \begin{picture}(20,10)(80,10)
               \put(-19,21){\circle{22}}\put(-25,19){$A^+$}     \put(55,11){$q_1$}     \put(7,36){$p$}      \put(87,36){$p'$}  \put(40,-7){$(b_1)_{T_i}$}
     \mbox{\epsfxsize=3.4truecm \hbox{\epsfbox{SCETGFR1.pdf}}  }
  \end{picture}\quad  
  \raise15pt \hbox{$  = \ \mbox{\normalsize }\,  i \,v(q_{1\perp})\, (a)_R\,(b_1)_{T_i}\left(1+\frac{p^2-p'^2}{p^+\left[q_1^+\right]}\right)\frac{\bnslash}{2}$ } 
  \nonumber\\[20pt]
  && \begin{picture}(20,10)(80,10)
   \put(0,15){$\mu, a$}    \put(85,15){$\nu, b$}   \put(55,11){$q_1$}     \put(7,36){$p$}      \put(87,36){$p'$}   \put(40,-7){$(c_1)_{T_i}$}
     \mbox{\epsfxsize=3.4truecm \hbox{\epsfbox{SCETGFR2.pdf}}  }
  \end{picture}\quad 
  \raise15pt \hbox{$  = \ \mbox{\normalsize}\, v(q_{1\perp})\,f^{abc_1}\,(c_1)_{T_i}\,\left[g^{\mu\nu}_{\perp}\,\bar{n}\mcdot p\left(1+\frac{p^2-p'^2}{p^+\left[q^+_1\right]}\right)+\frac{q^{\mu}_{1\perp}p'^{\nu}+q^{\nu}_{1\perp} p^{\mu}}{\left[q_1^+\right]}\right]$ } 
  \nonumber\\[20pt]
   && \begin{picture}(20,10)(80,10)    \put(24,11){$q_1$}   \put(62,11){$q_2$}     \put(7,40){$p$}      \put(87,40){$p'$}  \put(22,-7){$(b_1)_{T_i}$}\put(55,-7){$(b_2)_{T_j}$}
     \mbox{\epsfxsize=3.4truecm \hbox{\epsfbox{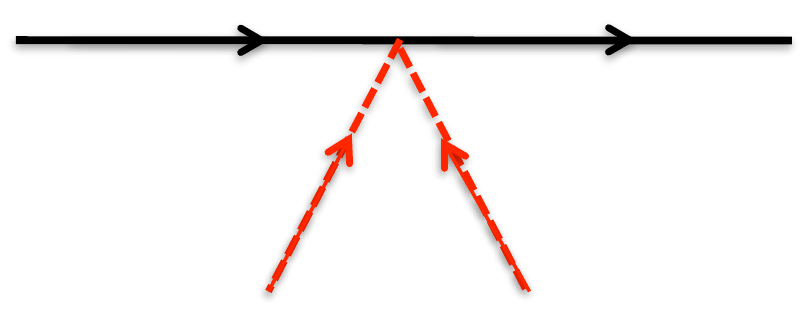}}  }
  \end{picture}\quad 
\raise20pt \hbox{$  = \ \mbox{\normalsize}\, \:  i \,v(q_{1\perp})v(q_{2\perp})\,(b_1 b_2)_R\,(b_1)_{T_i}(b_2)_{T_j}\,\frac{2\,q_{1\perp}\mcdot \,\,q_{2\perp}}{p^+\left[q_1^+\right]\left[q_2^+\right]}\frac{\bnslash}{2}\, $ } 
  \nonumber\\[15pt]
   && \begin{picture}(20,10)(80,10)   \put(24,11){$q_1$}   \put(62,11){$q_2$}     \put(7,40){$p$}      \put(87,40){$p'$}  \put(22,-7){$(c)_{T_i}$}\put(55,-7){$(d)_{T_j}$}
     \mbox{\epsfxsize=3.4truecm \hbox{\epsfbox{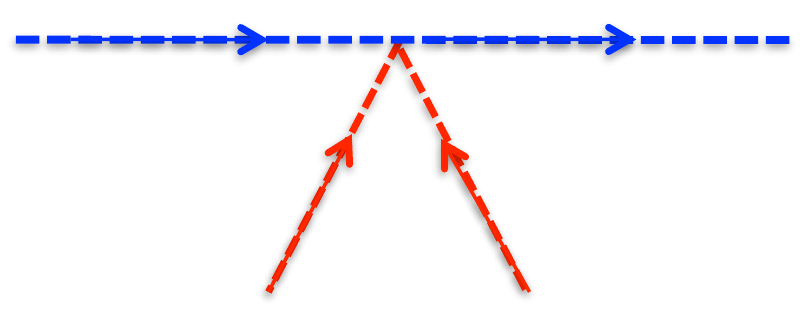}}  }
  \end{picture}\quad 
\raise20pt \hbox{$  = \ \mbox{\normalsize}\, \:  \frac{(-i)v(q_{1\perp})v(q_{2\perp})}{\left[q_1^+\right]\left[q_2^+\right]}\,(c)_{T_i}\,(d)_{T_j}\,\Bigg[f^{cax}f^{xdb}\left(q_{1\perp}\mcdot\, q_{2\perp}\, g^{\mu\nu}_{\perp}-q_{1\perp}^{\nu} q_{2\perp}^{\mu}\right)+$ } \nonumber\\
    && \qquad \qquad 
\raise20pt \hbox{$  \ \mbox{\normalsize{$+$}}\, \:  f^{cbx}f^{xad}\left(q_{1\perp}^{\mu} q_{2\perp}^{\nu}-q_{1\perp}\mcdot\, q_{2\perp}\, g^{\mu\nu}_{\perp}\right)+f^{cdx}f^{xab}\left(q_{1\perp}^{\mu} q_{2\perp}^{\nu}-q_{1\perp}^{\nu} q_{2\perp}^{\mu}\right)\Bigg]$ }\nonumber\\[10pt]
 && \begin{picture}(20,10)(80,10)     \put(55,11){$q_1$}     \put(7,36){$p$}      \put(87,36){$p'$}  \put(40,-7){$(b_1)_{T_i}$}    \put(45,52){$\mu, a$}
     \mbox{\epsfxsize=3.4truecm \hbox{\epsfbox{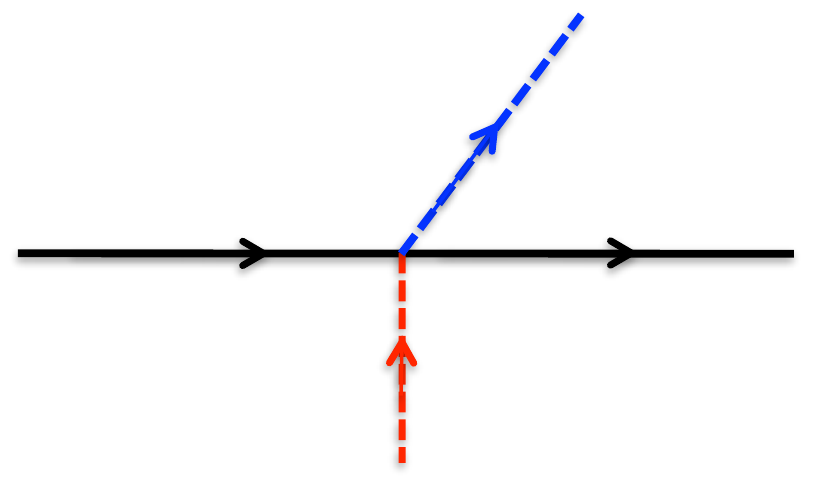}}  }
  \end{picture}\quad 
  \raise14pt \hbox{$  = \ \mbox{\normalsize}\, \:  \frac{(-i)\,v(q_{1\perp})}{\left[q^+_1\right]}\left[\frac{(ab_1)_R\, (b_1)_{T_i}}{p^+}\,\gamma_{\perp}^{\mu}\qslash_{1\perp}+\frac{(b_1)_{T_i}\,(b_1a)_R}{p'^+}\,\qslash_{1\perp}\gamma_{\perp}^{\mu}\right] \frac{\bnslash}{2}$ } 
   \nonumber\\[25pt] 
  \hline
  \hline
  \nonumber\\[20pt]
  \qquad \qquad  && \begin{picture}(20,10)(80,10)
     \put(-33,35){Hybrid}  \put(55,11){$q_1$}     \put(7,36){$p$}      \put(87,36){$p'$}  \put(40,-7){$(b_1)_{T_i}$}
     \mbox{\epsfxsize=3.4truecm \hbox{\epsfbox{SCETGFR1.pdf}}  }
  \end{picture}\quad 
  \raise15pt \hbox{$  = \ \mbox{\normalsize $i$}\, v(q_{1\perp})\, (b_1)_R\,(b_1)_{T_i}\frac{\bnslash}{2}$ } 
  \nonumber\\[20pt]
  && \begin{picture}(20,10)(80,10)
       \put(0,15){$\mu, a$}    \put(85,15){$\nu, b$}   \put(55,11){$q_1$}     \put(7,36){$p$}      \put(87,36){$p'$}   \put(40,-7){$(c_1)_{T_i}$}
     \mbox{\epsfxsize=3.4truecm \hbox{\epsfbox{SCETGFR2.pdf}}  }\nonumber
  \end{picture}\quad 
  \raise15pt \hbox{$  = \ \mbox{\normalsize }\,v(q_{1\perp})\, f^{abc_1}\,(c_1)_{T_i}\, g^{\mu\nu}_{\perp}\,\bar{n}\mcdot p$ } 
\end{eqnarray} 
{\caption[1]{Feynman rules of $\text{\SCETG}$ involving Glauber gluons. The first two vertices 
are given in the covariant gauge. Vertices from  three through seven are for the light-cone gauge  $A^+=0$. Last two vertices are for the Hybrid gauge. 
Note that the last three vertices in the light-cone gauge are power-suppressed in the covariant and Hybrid gauges.}
\label{fig:SCETGFR} }
\end{figure}

Our approach to deriving the effective Lagrangian of $\text{\SCETG}$ was to treat the effect 
from Glauber gluons on the target jet as one of an external background field, which the source 
creates. Having determined the scaling of the vector potential, we then read off the Glauber 
term from the SCET Lagrangian as it is given in~\eq{SCETL1} with a trivial addition of an 
external vector potential to the covariant derivative. We present in figure~\ref{fig:SCETGFR} the 
resulting Feynman rules. Note that we include the propagator of the Glauber gluon into the 
Feynman rule as well as the color and overall Dirac structure from the static fermionic source 
that we consider.

In the main part of the paper we consider three gauge choices.  First, we used the covariant 
gauge in the sections~\ref{collisional}, \ref{radiative}. Second, we used the light-cone gauge 
and third was the hybrid gauge in  section~\ref{invariant}. In the first case we used covariant 
gauge for both pure SCET and the Glauber interactions given in figure~\ref{fig:SCETGFR}. In the 
second case we did the same for the light-cone gauge. Finally, for the hybrid gauge, we used the 
light-cone gauge for SCET gauge fixing, while using the covariant gauge for Glauber gluons.  In all three cases we found the same results for physical quantities, 
as expected.

\section{Longitudinal integrals}\label{appendix:integrals}
In this appendix we define and calculate all longitudinal momentum integrals that are required 
for the evaluation of jet broadening and radiative energy loss. Our notation is $I^{(n)}_m$, where 
$n$ indicates number of Glauber gluon exchanges, and $m$ is the index corresponding to the 
Feynman diagram in question. We perform each integral exactly, without assuming a 
soft gluon approximation for radiative energy loss. However, from the exact results we can easily 
take the soft gluon limit.

\subsection{Jet broadening}\label{Appendix:integralsColl}
The longitudinal momentum integral that appears in the single Born diagram for quark or gluon 
jet broadening equals to the following expression:
\begin{eqnarray}
   I^{(1)}_1=\int\,\frac{d q^-_1}{2\pi}\,\e^{i q_1^-\delta z_1}\,\Delta_g(p,q_1)=
\int\,\frac{d q^-_1}{2\pi}\,\e^{i q^-_1\delta z_1}\,\frac{1}{\omega_1-q^-_1}=-i\e^{i\omega_1\delta z_1}\; ,
\label{broadI11}
\end{eqnarray}
where the inverse formation time $\omega_1$ is defined according to~\eq{GlauberProp} and~\eq{poleofGlauberProp}:
\begin{eqnarray}
\omega_1=\Omega(p,q_1)=p^--\frac{(\vc{p}_{\perp}-\vc{q}_{1\perp})^2-i\eps}{p^+} \; .
\end{eqnarray}
The double Born longitudinal integral looks like:
\begin{eqnarray}
   I^{(2)}_1&=&\int\,\frac{d q^-_1}{2\pi}\,\frac{d q^-_2}{2\pi}
\e^{i (q^-_1\delta z_1+q^-_2\delta z_2)}\,\Delta_g(p,q_2)\,\Delta_g(p,q_1+q_2)\nonumber\\
      &=&(-i)\int \frac{dq_2^{-}}{(2\pi)}\, \e^{i((\omega_{12}-q_2^-) \delta z_1+q_2^-\delta z_2)}\, 
\Delta_g(p,q_2)=-\e^{i\left(\omega_{12}\delta z_1+\omega_2(\delta z_2-\delta z_1)\right)}\; ,\label{broadI21}
\end{eqnarray}
where inverse formation time $\omega_{12}$ equals:
\begin{eqnarray}
\omega_{12}=\Omega(p,q_1+q_2)=p^--\frac{(\vc{p}_{\perp}-\vc{q}_{1\perp}-\vc{q}_{2\perp})^2-i\eps}{p^+}\; ,
\end{eqnarray}
and $\omega_2$ is identical to $\omega_1$ up to $q_1\leftrightarrow q_2$. In deriving~\eq{broadI11} we used the fact that $\omega_1$ is in the upper complex plane, which we 
should choose, since $\delta z_1>0$. Similarly in first and second step in deriving~\eq{broadI21} 
we used that $\omega_{12}$ and $\omega_{2}$ are in the upper $q^-$ complex plane. In the second step we 
also used that $\delta z_2>\delta z_1$, which is true for second order in opacity diagram (we use time ordered notation $z_0<z_1<z_2<...$). Next,
we calculate the contact limit of the integral $I_1^{(2)}$, when $\delta z_2=\delta z_1+0$:
\begin{eqnarray}
   &&I^{(2c)}_1=\int\,\frac{d q^-_1}{2\pi}\,\frac{d q^-_2}{2\pi}\e^{i (q^-_1+q^-_2)\delta z_1}\,\Delta_g(p,q_2)\,
\Delta_g(p,q_1+q_2)=(-i)\int \frac{dq_2^{-}}{2\pi}\, 
e^{i(\omega_{12}-q_2^-+q_2^-)\delta z_1}\, \Delta_g(p,q_2)\nonumber\\
   &&=-i\e^{i\omega_{12}\delta z_1}\int \frac{d q_2^-}{2\pi}\frac{1}{\omega_2-q_2^-}
=\frac{i\e^{i\omega_{12}\delta z_1}}{2\pi}\left(\ln(\infty-\omega_2)-\ln(-\infty-\omega_2)\right) 
= -\frac{1}{2}\e^{i\omega_{12}\delta z_1}\; .
\label{broadI21c}
\end{eqnarray}
As one can see, in the second step in deriving~\eq{broadI21c} we cannot use Cauchy's theorem to perform the remaining $q_2^-$ integral, since the boundary term at infinity does not vanish. Instead, we perform the integral directly. We find:
\begin{eqnarray}
   &&I^{(2c)}_1=\frac{1}{2}\,I^{(2)}_1(\delta z_2=\delta z_1)\; .\label{broadI21cfinal}
\end{eqnarray}

\subsection{Radiative energy loss}\label{appendix:integralsRad}
The longitudinal momentum integrals that appear in the first three single Born diagrams on figure~\ref{fig:BremA1} are:\footnote{The remaining two vanish when one takes the physical emitted gluon polarization vector.}
\begin{eqnarray}
    &&I_1^{(1)}=\int \frac{d q^-_1}{2\pi}\, \e^{i q^-_1\delta z_1}  \, \Delta_g(p+k, q_1)\; ,\label{I1formula}\\
      &&I_{2}^{(1)}=\int \frac{d q^-_1}{2\pi}\, \e^{i q^-_1\delta z_1}\, \Delta_g(p, q_1)\Delta_g(p+k, q_1)\; ,\\
      &&I_{3}^{(1)}=\int \frac{d q^-_1}{2\pi}\, \e^{i q^-_1\delta z_1}\, \Delta_g(k, q_1)\Delta_g(p+k, q_1)\; .\label{I13Formulas}
\end{eqnarray}
These three integrals are functions of three inverse formation times, all given by the poles 
in~\eq{poleofGlauberProp} of propagator in~\eq{GlauberProp}:
\begin{eqnarray}
&&\Omega_1=\Omega(p+k,q_1)=p^-+k^--\frac{(\vc{p}_{\perp}+\vc{k}_{\perp}-\vc{q}_{1\perp})^2-i\eps}{p^++k^+}\; ,\\
&&\Omega_2=\Omega(p,q_1)=p^--\frac{(\vc{p}_{\perp}-\vc{q}_{1\perp})^2-i\eps}{p^+}\; ,\\
&&\Omega_3=\Omega(k,q_1)=k^--\frac{(\vc{k}_{\perp}-\vc{q}_{1\perp})^2-i\eps}{k^+}\; .
\end{eqnarray}
The first integral in~\eq{I1formula} we already derived in~\eq{broadI11}, where we need to substitute $\omega_1\rightarrow \Omega_1$. The integrals $I_{2}^{(1)}$ and $I_{3}^{(1)}$ can be reduced to the same integral in~\eq{broadI11} by using the following trivial identity, which follows directly from definitions in~\eq{GlauberProp} and~\eq{poleofGlauberProp}:
\begin{eqnarray}
    \Delta_g(p_1, q_1)\,\Delta_g(p_2, q_1)=-\frac{\Delta_g(p_1, q_1)-\Delta_g(p_2, q_1)}
{\Omega(p_1, q_1)-\Omega(p_2, q_1)}\; .\label{DeltaGIdentity}
\end{eqnarray}
Note that in~\eq{DeltaGIdentity} the numerator depends on $q^-_1$, while the denominator does not. Thus,  
integrals $I_{2}^{(1)}$ and $I_{3}^{(1)}$ become a combination of two integrals in~\eq{broadI11}. 
As a result, we get the following exact expressions:
\begin{eqnarray}
      &&I_1^{(1)}=-i\,\e^{i\Omega_1\delta z_1}\; , \\
      &&I_2^{(1)}=\frac{i}{\Omega_1-\Omega_2}\left(\e^{i\Omega_1\delta z}-\e^{i\Omega_2\delta z_1}\right)\; , \\
      &&I_3^{(1)}\equiv \frac{i}{\Omega_1-\Omega_3}\left(\e^{i\Omega_1\delta z}-\e^{i\Omega_3\delta z_1}\right) \; .
\end{eqnarray}
The soft gluon emission approximation corresponds to $x \ll 1$ and is of interest in sections~\ref{collisional} and \ref{radiative}. In this limit the inverse formation times $\Omega_{1,2,3}$ reduce to only two 
non-trivial functions $\omega_0,\omega_1$. The corresponding approximate results read:
\begin{eqnarray}
&&\Omega_1\approx \frac{\vc{k}_{\perp}^2}{xp_0^+}\equiv \omega_0\; ,
\qquad\qquad \Omega_2\approx 0\; , \qquad\qquad 
\Omega_3=\frac{\vc{k}_{\perp}^2-(\vc{k}_{\perp}-\vc{q}_{1\perp})^2}{x p_0^+}\equiv \omega_0-\omega_1\; ,\\
&&I_1^{(1)}\approx -i\e^{i\omega_0\delta z_1}\; ,\qquad I_{2}^{(1)}\approx \frac{i}{\omega_0}\left[\e^{i\omega_0\delta z_1}-1\right]\; , 
\qquad I_{3}^{(1)}\approx i\frac{\e^{i\omega_0\delta z_1}}{ \omega_1}\left[1-\e^{-i\omega_1\delta z_1}\right] .
\end{eqnarray}

Next we move to the integrals in two Glauber exchange diagrams and their contact limits. The longitudinal momentum integrals that appear in the first six diagrams in figure~\ref{fig:BremA2} equal:\footnote{The remaining three amplitudes in the figure~\ref{fig:BremA2} vanish.}
\begin{eqnarray}
    &&I_1^{(2)}=\int \frac{d q_1^-}{2\pi}\,\frac{d q_2^-}{2\pi}\, \e^{i \left(q_1^-\delta z_1+q_2^-\delta z_2\right)}  \, \Delta_g(p+k, q_2)\Delta_g(p+k, q_1+q_2)\; ,\label{I21formula}\\
    &&I_2^{(2)}=\int \frac{d q_1^-}{2\pi}\,\frac{d q_2^-}{2\pi}\, \e^{i \left(q_1^-\delta z_1+q_2^-\delta z_2\right)}  \, \Delta_g(p, q_2) \Delta_g(p, q_1+q_2)\Delta_g(p+k, q_1+q_2)\; ,\label{I22formula}\\
    &&I_3^{(2)}=\int \frac{d q_1^-}{2\pi}\,\frac{d q_2^-}{2\pi}\, \e^{i \left(q_1^-\delta z_1+q_2^-\delta z_2\right)}  \, \Delta_g(k, q_2)\Delta_g(k, q_1+q_2)\Delta_g(p+k, q_1+q_2)\; ,\label{I23formula}\\
    &&I_4^{(2)}=\int \frac{d q_1^-}{2\pi}\,\frac{d q_2^-}{2\pi}\, \e^{i \left(q_1^-\delta z_1+q_2^-\delta z_2\right)}  \, \Delta_g(p, q_2)\Delta_g(k, q_1)\Delta_g(p+k, q_1+q_2)\; ,\label{I24formula}\\
    &&I_5^{(2)}=\int \frac{d q_1^-}{2\pi}\,\frac{d q_2^-}{2\pi}\, \e^{i \left(q_1^-\delta z_1+q_2^-\delta z_2\right)}  \, \Delta_g(p, q_2)\Delta_g(p+k, q_2)\Delta_g(p+k, q_1+q_2)\; ,\label{I25formula}\\
    &&I_6^{(2)}=\int \frac{d q_1^-}{2\pi}\,\frac{d q_2^-}{2\pi}\, \e^{i \left(q_1^-\delta z_1+q_2^-\delta z_2\right)}  \, \Delta_g(k, q_2)\Delta_g(p+k, q_2)\Delta_g(p+k, q_1+q_2)\; .\label{I26formula}
\end{eqnarray}
The corresponding integrals contain seven inverse formation times, which later after taking 
the contact limit and averaging over the medium states reduce to only two inverse formation times. 
We define the following seven poles $\alpha_i$: 
\begin{eqnarray}
&&\alpha_1=\Omega(p+k, q_2)\; ,\qquad\qquad \alpha_2=\Omega(p+k, q_1+q_2)\; ,
\qquad\qquad \alpha_3=\Omega(p, q_2)\; , \qquad\qquad \alpha_4=\Omega(p, q_1+q_2)\; ,\nonumber\\
&&\alpha_5=\Omega(k, q_2)\; ,\qquad\qquad \alpha_6=\Omega(k, q_1+q_2)\; , \qquad\qquad 
\alpha_7=\Omega(k, q_1)\; .
\end{eqnarray}
Next we note that all double Born integrals except for $I_4^{(2)}$ can be found using combination 
of already calculated integral in~\eq{broadI21} and the identity in~\eq{DeltaGIdentity}. 
The results are following:
\begin{eqnarray}
    &&I_1^{(2)}=-\e^{i\left(\alpha_2\delta z_1+\alpha_1(\delta z_2-\delta z_1)\right)}\; ,\qquad I_1^{(2c)}=\frac{1}{2}\,I^{(2)}_1(\delta z_2=\delta z_1) \;, \label{RI21formula}\\
    &&I_2^{(2)}=\frac{\e^{i\alpha_3\left(\delta z_2-\delta z_1\right)}\left(\e^{i\alpha_4\delta z_1}-\e^{i\alpha_2\delta z_1}\right)}{\alpha_4-\alpha_2}\; , \qquad I_2^{(2c)}=\frac{1}{2}\,I^{(2)}_2(\delta z_2=\delta z_1)\;, \label{RI22formula}\\
      &&I_3^{(2)}=\frac{\e^{i\alpha_5\left(\delta z_2-\delta z_1\right)}\left(\e^{i\alpha_6\delta z_1}-\e^{i\alpha_2\delta z_1}\right)}{\alpha_6-\alpha_2}\; , \qquad I_3^{(2c)}=\frac{1}{2}\,I^{(2)}_3(\delta z_2=\delta z_1)\;, \label{RI32formula}\\
    &&I_5^{(2)}=\frac{\e^{i\alpha_2\delta z_1}\left(\e^{i\alpha_3\left(\delta z_2-\delta z_1\right)}-\e^{i\alpha_1\left(\delta z_2-\delta z_1\right)}\right)}{\alpha_3-\alpha_1}\; , \qquad I_5^{(2c)}=0\; ,\label{RI25formula}\\
    &&I_6^{(2)}=\frac{\e^{i\alpha_2\delta z_1}\left(\e^{i\alpha_5\left(\delta z_2-\delta z_1\right)}-\e^{i\alpha_1\left(\delta z_2-\delta z_1\right)}\right)}{\alpha_5-\alpha_1}\; , \qquad I_6^{(2c)}=0\; .\label{RI26formula}
\end{eqnarray}
The final integral $I_4^{(2)}$ we just work out:
\begin{eqnarray}
    &&I_4^{(2)}=\int \frac{d q_1^-}{2\pi}\,\frac{d q_2^-}{2\pi} \,\frac{\e^{i \left(q_1^-\delta z_1+q_2^-\delta z_2\right)} }{(\alpha_3-q_2^-)(\alpha_7-q_1^-)(\alpha_2-q_1^--q_2^-)}=\int \frac{d q_2^-}{2\pi}\,\frac{\e^{i q_2^-\delta z_2}}{(\alpha_3-q_2^-)}\,\frac{i\left(\e^{i\alpha_7\delta z_1}-\e^{i(\alpha_2-q_2^-)\delta z_1}\right)}{\alpha_7-\alpha_2+q_2^-}\; ,\nonumber\\\label{RI24formulaIntermediate}
\end{eqnarray}
where in the first term of the remaining $q_2^-$ integral we have to close the contour in the 
upper complex plane, while in the second one we close above for $\delta z_2>\delta z_1$ and below 
in the opposite case. Also, we know that $\alpha_3$ is in the positive complex plane, but we have to 
figure~out the sign of imaginary part of $\left(\alpha_7-\alpha_2\right)$:
\begin{eqnarray}
    \text{Im}\left(\alpha_7-\alpha_2\right)=\frac{\eps}{k^+}-\frac{\eps}{p^++k^+}=\frac{\eps(1-x)}{p_0^+}>0 \;.
\end{eqnarray}
Thus, for physical momenta $p, k$, the second denominator in the last integral in~\eq{RI24formulaIntermediate} has a pole in the negative complex plane. With this in mind, we find the $I_4^{(2)}$ for two cases of interest:
\begin{eqnarray}
    &&I_4^{(2)}(\delta z_2>\delta z_1)=\frac{\e^{i(\alpha_3\delta z_2+\alpha_7\delta z_1)}-\e^{i\left(\alpha_2\delta z_1+\alpha_3(\delta z_2-\delta z_1)\right)}}{\alpha_7-\alpha_2+\alpha_3}\; ,\\
      &&I_4^{(2)}(\delta z_2<\delta z_1)=\frac{\e^{i(\alpha_3\delta z_2+\alpha_7\delta z_1)}-\e^{i\left(\alpha_2\delta z_1+(\alpha_2-\alpha_7)(\delta z_2-\delta z_1)\right)}}{\alpha_7-\alpha_2+\alpha_3}\; .
\end{eqnarray}
Finally, taking the contact limit of the last integration in~\eq{RI24formulaIntermediate} we get:
\begin{eqnarray}
    &&I_4^{(2c)}=I_4^{(2)}(\delta z_2=\delta z_1+0)=I_4^{(2)}(\delta z_2=\delta z_1-0)\; .
\end{eqnarray}
Next, in the contact limit and taking the average over medium, we have $\vc{q}_{1\perp}+\vc{q}_{2\perp}=\vc{0}$. This makes some of the $\alpha_i$ trivial. Directly from their definition we get:
\begin{eqnarray}
  \alpha_2=\frac{(p+k)^2}{p_0^+}\; ,\qquad\qquad   \alpha_4=\frac{p^2}{p^+}=0\; ,\qquad\qquad   \alpha_6=\frac{k^2}{k^+}=0\; .
\end{eqnarray}
Taking these equations into account we list all the contact limit integrals:
\begin{eqnarray}
    &&I_1^{(2c)}=-\frac{1}{2}\e^{i\alpha_2\delta z_1}\; ,\label{I21cformula}\\
    &&I_2^{(2c)}=\frac{1}{2}\frac{\e^{i\alpha_2\delta z_1}-1}{\alpha_2}\; ,\label{I22cformula}\\
    &&I_3^{(2c)}=\frac{1}{2}\frac{\e^{i\alpha_2\delta z_1}-1}{\alpha_2}\; ,\label{I23cformula}\\
    &&I_4^{(2c)}=\frac{\e^{i\alpha_2\delta z_1}}{\alpha_7-\alpha_2+\alpha_3}\,\left(\e^{i\left(\alpha_7-\alpha_2+\alpha_3\right)\delta z_1}-1\right)\; ,\label{I24cformula}\\
    &&I_5^{(2c)}=0\; ,\label{I25cformula}\\
    &&I_6^{(2c)}=0\; .\label{I26cformula}
\end{eqnarray}
As discussed above, the seven inverse formation times reduced to only two combinations after 
averaging over the medium. We define these two inverse formation times in the following way:
\begin{eqnarray}
  &&\Omega_4=\alpha_2\; , \qquad \qquad  \Omega_5=\alpha_7+\alpha_3\; .
\end{eqnarray}
Note that up until now we never used the soft gluon approximation, and all integrals and contact limits were exact. Finally, taking the soft gluon approximation gives:
\begin{eqnarray}
  &&\Omega_4\approx \omega_0\; ,\qquad\qquad\Omega_5\approx \omega_0-\omega_1\; .
\end{eqnarray}

\section{$T$ Wilson line}\label{appendix:TLine}

\begin{figure}[h!]
\begin{center}
\epsfig{file=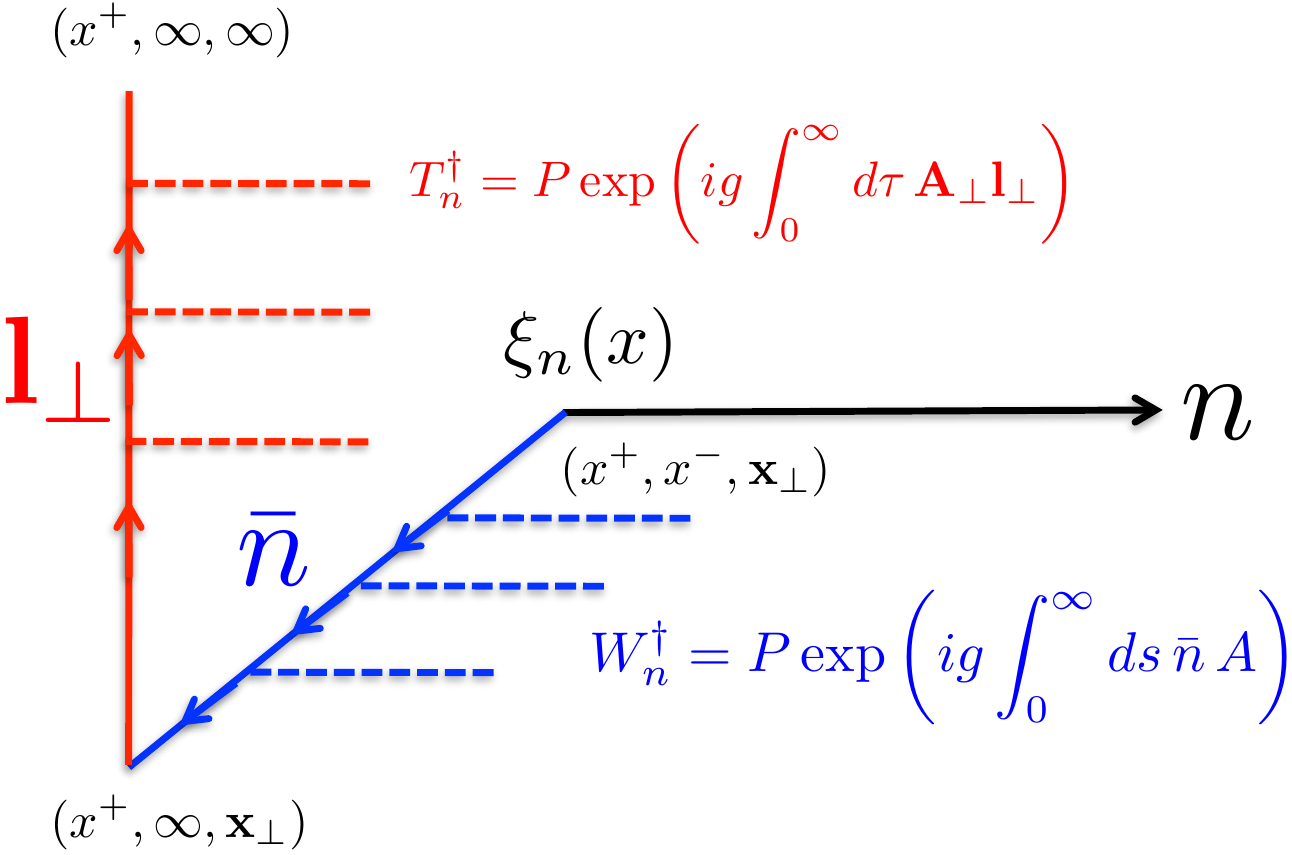, width=5cm}\qquad\epsfig{file=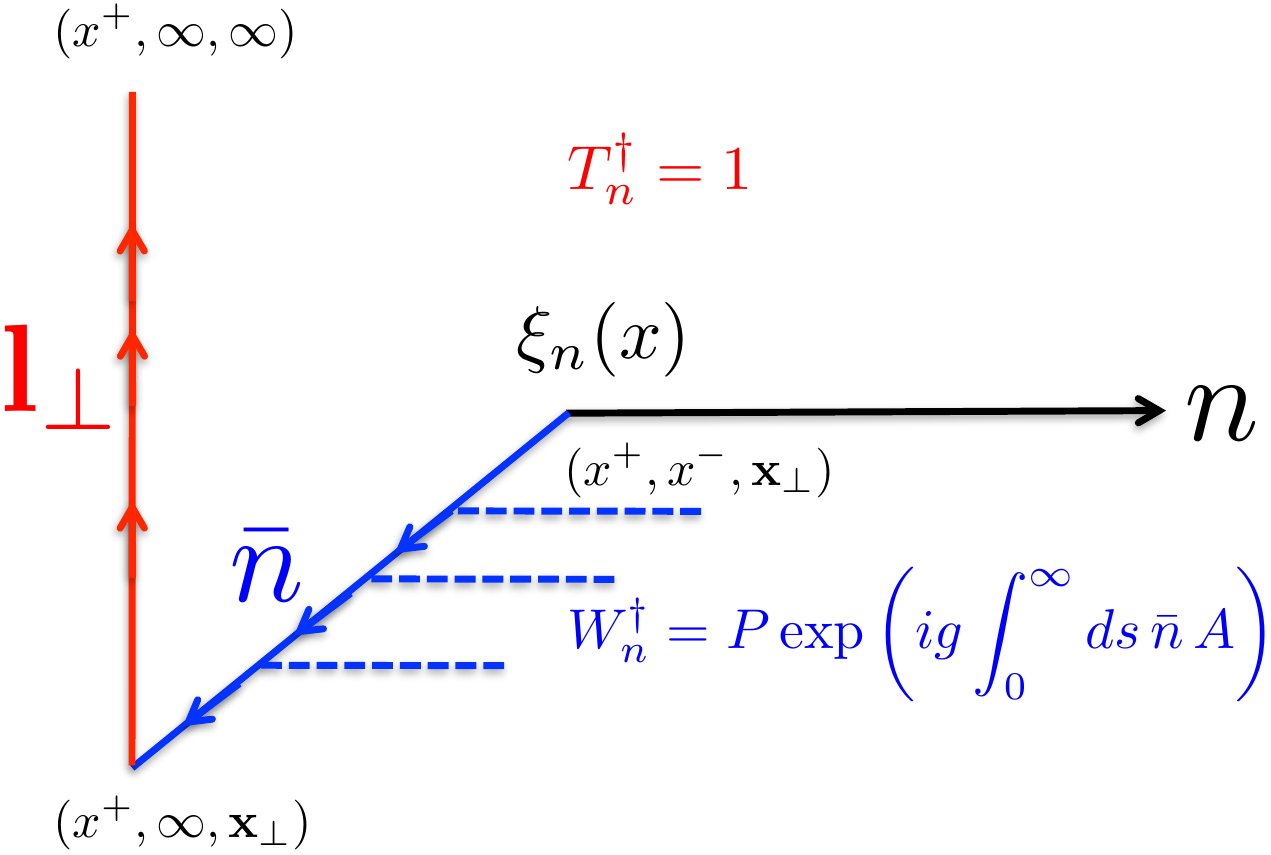, width=5cm}
\qquad\epsfig{file=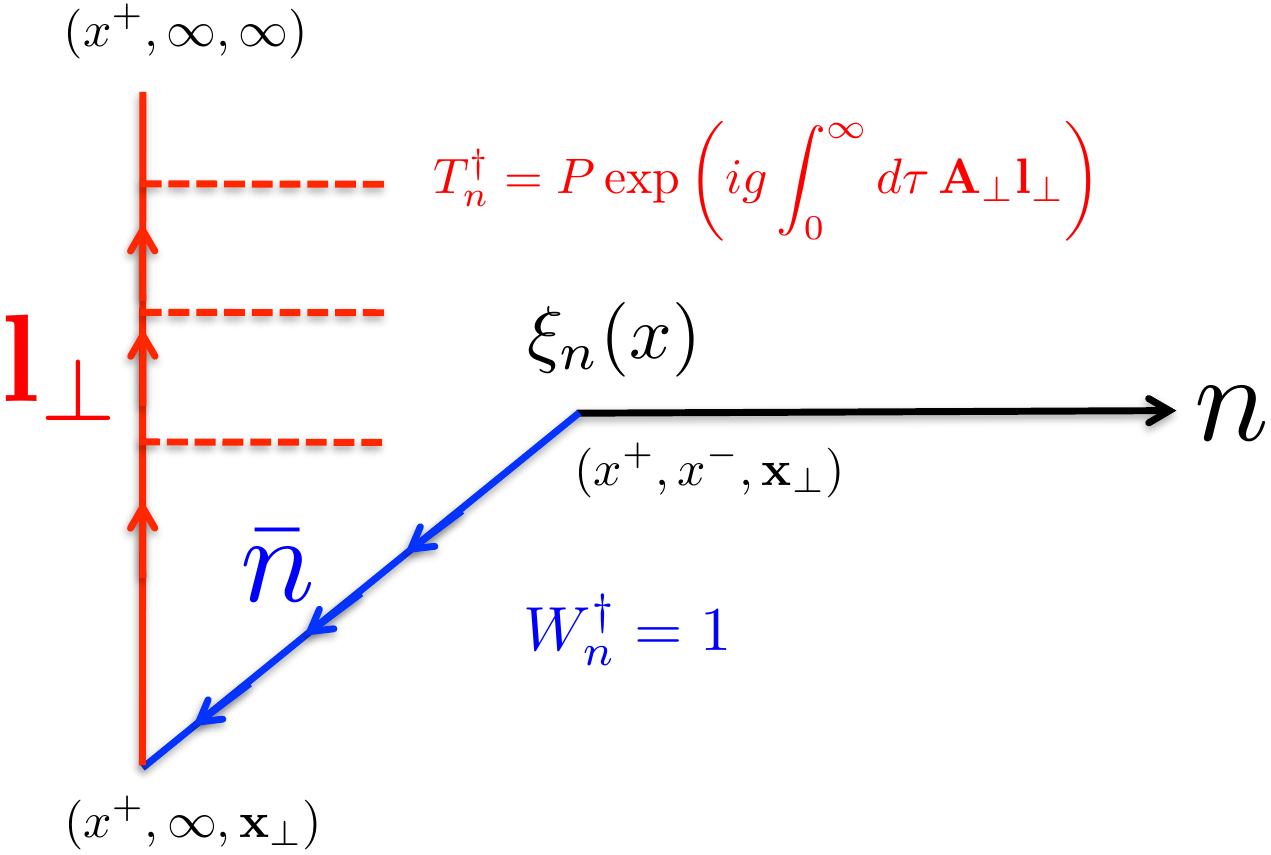, width=5cm}
\caption{\label{fig:WilsonFig} The collinear and transverse gauge links. Left: the generic gauge, 
middle: regular covariant gauge and right: singular light-cone gauge.} 
 \end{center}
\label{splits}
\end{figure}

Wilson lines are a common ingredient for building gauge-invariant objects in field theory. For example, 
the collinear Wilson line $W_n$ is used in SCET to dress the collinear quark field $\xi_n$ to form a 
gauge-invariant jet field $\chi_n= W_n^{\dagger}\,\xi_n$. However, in the light-cone gauge the collinear 
Wilson line disappears $W_n^{\dagger}=1$. In the context of SDIS it has been realized ~\cite{Ji:2002aa}, 
\cite{Belitsky:2002sm} that additional transverse  gauge link  is required in this gauge in order to 
describe the final state interactions in a gauge invariant way.

More recently, it has been proposed that SCET has to be expanded~\cite{Idilbi:2010im} by including a 
transverse gauge link  for gauge invariance of certain non-perturbative matrix elements, like the 
transverse momentum-dependent parton distribution functions. Also, it has been shown in  
Ref.~\cite{Idilbi:2010im} that the SCET jet function calculated in the light-cone gauge in the presence 
of the transverse gauge link, which the authors denote as the transverse Wilson line $T_n^{\dagger}$, is 
independent of the light-cone prescription, which unambiguously cancels between the gluon 
propagator and the gauge link, and leads to a result which is the same as in the covariant gauge.
\begin{eqnarray}
  && \chi_{n,p}= T_n^{\dagger}W_n^{\dagger}\,\xi_{n,p}\; ,\label{GIQdef}\\
   &&T_n^{\dagger}=\mathcal{P}\exp{\left(ig\int_{0}^{\infty}
\,d\tau\,\vc{A}_{\perp}\mcdot\vc{l}_{\perp}\right)}\; ,\label{Tndef}
\end{eqnarray}
where $\mathcal{P}$ denotes the path ordering.
As one can see from the figure~\ref{fig:WilsonFig} the effect of the transverse gauge link vanishes 
in the covariant gauge, since the gauge field is zero at infinity, while in the singular light-cone 
gauge this gauge link is non-zero depending on the boundary conditions, similar to the gluon propagator  
prescription in the light-cone gauge. 

Our calculation of the single and double Born amplitudes in the light-cone gauge in $\text{\SCETG}$ 
in the absence of the $T_n$ Wilson line leads to the same result as in the covariant gauge, but only 
in the $-i\eps$ prescription. In all other prescriptions we get a different result, as one can see 
from~\eq{LCGEq1234} and table~\ref{tb2}. In this appendix we show that once the $T_n$ Wilson line 
is introduced into the calculation, the prescription dependence cancels and the gauge invariance is 
restored similarly to the result in \cite{Idilbi:2010im} for the SCET jet function calculated 
in the light-cone gauge. In the remainder of this section we assume the light-cone gauge, so 
$W_n=1$ and we derive the Feynman rules for $T_n$ Wilson line emission.

\begin{figure}[h!]
\begin{eqnarray}
  && \begin{picture}(10,0)(10,0)
\mbox{\epsfxsize=5truecm \hbox{\epsfbox{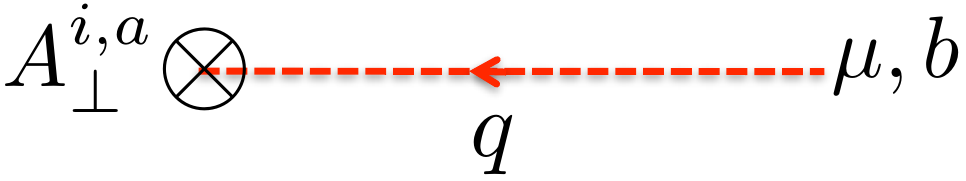}}  }
  \end{picture}\quad \Large 
  \raise15pt \hbox{$  \qquad \qquad \qquad \qquad \qquad= \qquad i\,\delta^{ab}\,
\frac{\bar{n}^{\mu}q^i}{q^2+i\eps}C_{\infty}^{\text{(Pres)}}
\left(\frac{1}{q^++i\eps}-\frac{1}{q^+-i\eps}\right)$\; , } 
\nonumber
\end{eqnarray}
{\caption[1]{The propagator of the $T$ Wilson line emitted gluon.}
\label{fig:TWLineProp} }
\end{figure}

A useful expression for the propagator of the gluon emitted from the transverse gauge 
link was derived in ~\cite{Idilbi:2010im} and we show it in Figure \ref{fig:TWLineProp}, where the prescription dependence is encoded into the number $C_{\infty}^{\text{(Pres)}}$.

\noindent  Its dependence on the prescription is given in the table~\ref{tb3}. Note that the difference of this table and 
the propagator in figure~\ref{fig:TWLineProp} from \cite{Idilbi:2010im} is due to the different notation of 
ingoing vs outgoing momentum flow into the vertex.

\begin{table}[h!]
\begin{center}
\begin{tabular}{||c||c||c|c|}
	\hline
Prescription& $\frac{1}{[q^+]}$&  $C_{\infty}^{\text{(Pres)}}$  \\
	\hline \hline
$+i\eps$& $\frac{1}{q^++i\eps}$&  $1$    \\
$-i\eps$& $\frac{1}{q^+-i\eps}$&  $0$    \\
PV& $\frac{1}{2}\left(\frac{1}{q^++i\eps}+\frac{1}{q^+-i\eps}\right)$&  $\frac{1}{2}$       \\
ML& $\frac{1}{q^++i\eps \text{sign}(q^-)}$&  $\theta(q^-)$      \\	
	\hline
\end{tabular}
\caption{Dependence of  $C_{\infty}^{\text{(Pres)}}$ on the light-cone prescription.}
\end{center}
\label{tb3}
\end{table}

First, from the form of the propagator in figure~\ref{fig:TWLineProp} one can see that the $T_n$ Wilson 
line cannot produce physical gluons in the final state, since $\bar{n}\mcdot \eps=0$ in the 
light-cone gauge. However, this propagator contracted with the static source term $v_{\mu}$ doesn't 
vanish and is leading order in the effective theory power counting. In order to derive the Feynman 
rules of $T_n$ emission from the quark line we use the definition of the gauge invariant quark field 
\eq{GIQdef} and the explicit expression for the transverse gauge link in ~\eq{Tndef}. 
Finally we include the propagator in figure~\ref{fig:TWLineProp} to obtain first two Feynman rules 
in the figure~\ref{fig:TWLine} below.

\begin{figure}[t!]
\begin{eqnarray}
 \qquad \qquad\qquad&& \begin{picture}(20,10)(80,10)
     \mbox{\epsfxsize=3.4truecm \hbox{\epsfbox{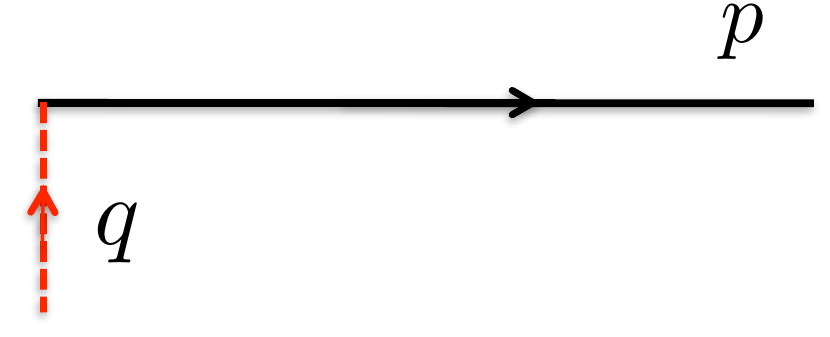}}  }
  \end{picture}\quad 
  \raise15pt \hbox{$  = (b)_r(b)_{T_i}\,v(q_{\perp}) C_{\infty}^{\text{(Pres)}}\left(\frac{1}{q^++i\eps}-\frac{1}{q^+-i\eps}\right)$ } 
  \nonumber\\[15pt]
  && \begin{picture}(20,10)(80,10)
     \mbox{\epsfxsize=3.4truecm \hbox{\epsfbox{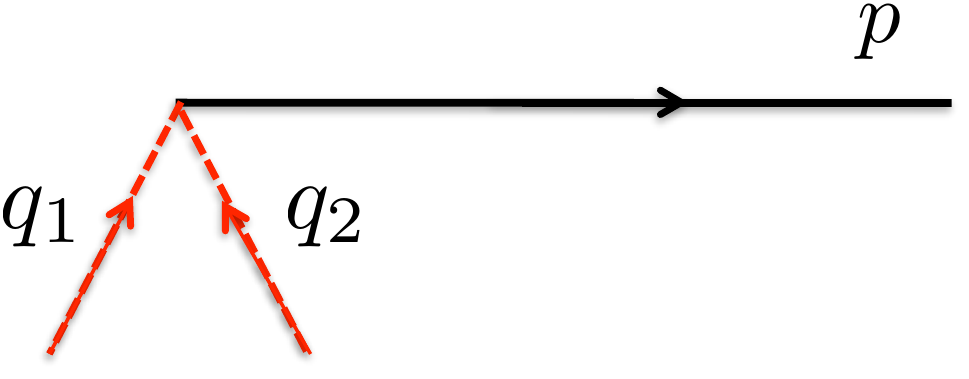}}  }
  \end{picture}\quad 
  \raise15pt \hbox{$  =-\frac{1}{2} \frac{l_{\perp}q_{1\perp}(b_1 b_2)_r+l_{\perp}q_{2\perp}(b_2 b_1)_r}{l_{\perp}q_{1\perp}+l_{\perp}q_{2\perp}}\,(b_1)_{T_i}(b_2)_{T_j}\,v(q_{1\perp})v(q_{2\perp}) $ } \nonumber\\
    &&   \raise15pt \hbox{$ \qquad\qquad\qquad \times\, \left[C_{\infty}^{\text{(Pres)}}\right]^2\left(\frac{1}{q_1^++i\eps}-\frac{1}{q_1^+-i\eps}\right)\left(\frac{1}{q_2^++i\eps}-\frac{1}{q_2^+-i\eps}\right)$}
  \nonumber\\[15pt]
    && \begin{picture}(20,10)(80,10)
            \put(85,15){$\mu, a$}
     \mbox{\epsfxsize=3.4truecm \hbox{\epsfbox{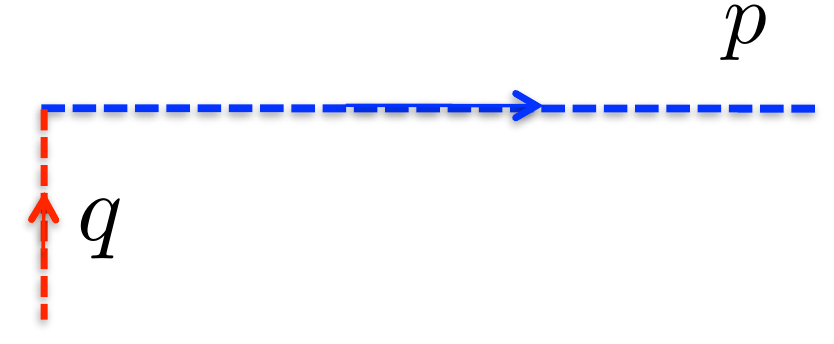}}  }
  \end{picture}\quad  
  \raise15pt \hbox{$  =-if^{abc}\,(-g_{\perp}^{\mu\nu}) \,(c)_r(b)_{T_i}\,v(q_{\perp}) C_{\infty}^{\text{(Pres)}}\left(\frac{1}{q^++i\eps}-\frac{1}{q^+-i\eps}\right)$ } 
  \nonumber\\[15pt]
  && \begin{picture}(20,10)(80,10)
    \put(85,15){$\mu, a$}
     \mbox{\epsfxsize=3.4truecm \hbox{\epsfbox{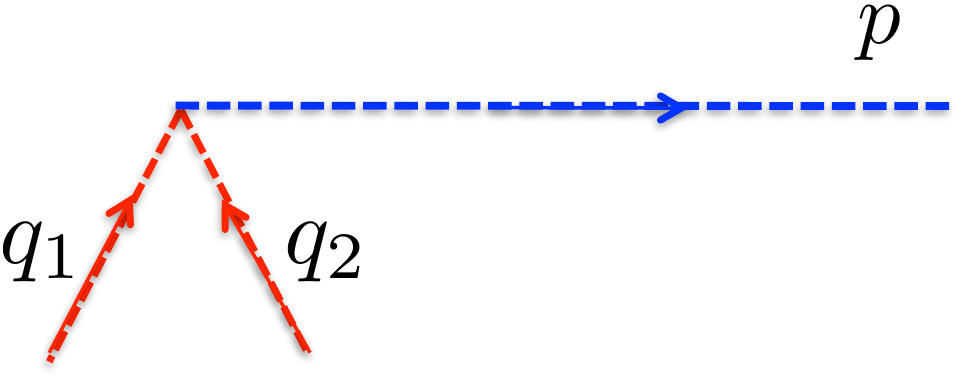}}  }
  \end{picture}\quad 
  \raise15pt \hbox{$  =-\frac{1}{2} \,(-g_{\perp}^{\mu\nu}) \,\frac{l_{\perp}q_{1\perp}\,f^{ab_1c_1}f^{c_1 b_2 e}+l_{\perp}q_{2\perp}\,f^{ab_2c_1}f^{c_1 b_1 e}}{l_{\perp}q_{1\perp}+l_{\perp}q_{2\perp}}\,(e)_r\,(b_1)_{T_i}(b_2)_{T_j}\,v(q_{1\perp})v(q_{2\perp})$ } \nonumber\\ [5pt]
  &&   \raise15pt \hbox{$ \qquad\qquad\qquad \times\,\left[C_{\infty}^{\text{(Pres)}}\right]^2\left(\frac{1}{q_1^++i\eps}-\frac{1}{q_1^+-i\eps}\right)\left(\frac{1}{q_2^++i\eps}-\frac{1}{q_2^+-i\eps}\right)$}
  \nonumber
\end{eqnarray}
{\caption[1]{Feynman rules for the gluon emission from the transverse Wilson line $T_n$ for single and double gluon emission from a quark and a gluon line.}
\label{fig:TWLine} }
\end{figure}

In order to derive similar Feynman rules with the gluon line we need the expression of the gauge 
invariant gluon field which is a straightforward generalization of the previous definition 
\eq{GIBDef}, now including the $T$ Wilson line:
\begin{eqnarray}
\mathcal{B}^{\mu}_n=\frac{1}{g}\left[T_n^{\dagger}W_n^{\dagger}\,\left(i \partial_n^{\mu}+gA_n^{\mu}\right) \,W_n T_n\right], \label{GIBnewdef}
\end{eqnarray}
where the square brackets indicate that the derivative acts only within the brackets. From the expression~\eq{GIBnewdef} and again using the definition~\eq{Tndef} and the propagator from figure~\ref{fig:TWLineProp} 
we obtain the last two Feynman rules in figure~\ref{fig:TWLine}.\newpage
\bibliography{bibliography}

\providecommand{\href}[2]{#2}\begingroup\raggedright\begin{thebibliography}{10}

\bibitem{Sterman:1977wj}
G.~Sterman and S.~Weinberg, {\it {Jets from Quantum Chromodynamics}},  {\em
  Phys. Rev. Lett.} {\bf 39} (1977) 1436.

\bibitem{Feynman:1978dt}
R.~P. Feynman, R.~D. Field, and G.~C. Fox, {\it {Quantum-chromodynamic approach
  for the large-transverse- momentum production of particles and jets}},  {\em
  Phys. Rev.} {\bf D18} (1978) 3320.

\bibitem{Ellis:2007zzc}
J.~R. Ellis, {\it {Beyond the standard model with the LHC}},  {\em Nature} {\bf
  448} (2007) 297--301.

\bibitem{Aad:2009wy}
{\bf The ATLAS} Collaboration, G.~Aad {\em et~al.}, {\it {Expected Performance
  of the ATLAS Experiment - Detector, Trigger and Physics}},
  \href{http://arxiv.org/abs/0901.0512}{{\tt arXiv:0901.0512}}.

\bibitem{Ball:2007zza}
{\bf CMS} Collaboration, G.~L. Bayatian {\em et~al.}, {\it {CMS technical
  design report, volume II: Physics performance}},  {\em J. Phys.} {\bf G34}
  (2007) 995--1579.

\bibitem{Collins:1981ta}
J.~C. Collins and G.~F. Sterman, {\it {SOFT PARTONS IN QCD}},  {\em Nucl.
  Phys.} {\bf B185} (1981) 172.

\bibitem{Collins:1989gx}
J.~C. Collins, D.~E. Soper, and G.~F. Sterman, {\it {Factorization of Hard
  Processes in QCD}},  {\em Adv. Ser. Direct. High Energy Phys.} {\bf 5} (1988)
  1--91, [\href{http://arxiv.org/abs/hep-ph/0409313}{{\tt hep-ph/0409313}}].

\bibitem{Bauer:2000ew}
C.~W. Bauer, S.~Fleming, and M.~E. Luke, {\it {Summing Sudakov logarithms in B
  --> X/s gamma in effective field theory}},  {\em Phys. Rev.} {\bf D63} (2000)
  014006, [\href{http://arxiv.org/abs/hep-ph/0005275}{{\tt hep-ph/0005275}}].

\bibitem{Bauer:2000yr}
C.~W. Bauer, S.~Fleming, D.~Pirjol, and I.~W. Stewart, {\it {An effective field
  theory for collinear and soft gluons: Heavy to light decays}},  {\em Phys.
  Rev.} {\bf D63} (2001) 114020,
  [\href{http://arxiv.org/abs/hep-ph/0011336}{{\tt hep-ph/0011336}}].

\bibitem{Bauer:2001ct}
C.~W. Bauer and I.~W. Stewart, {\it {Invariant operators in collinear effective
  theory}},  {\em Phys. Lett.} {\bf B516} (2001) 134--142,
  [\href{http://arxiv.org/abs/hep-ph/0107001}{{\tt hep-ph/0107001}}].

\bibitem{Bauer:2001yt}
C.~W. Bauer, D.~Pirjol, and I.~W. Stewart, {\it {Soft-Collinear Factorization
  in Effective Field Theory}},  {\em Phys. Rev.} {\bf D65} (2002) 054022,
  [\href{http://arxiv.org/abs/hep-ph/0109045}{{\tt hep-ph/0109045}}].

\bibitem{Bauer:2002nz}
C.~W. Bauer, S.~Fleming, D.~Pirjol, I.~Z. Rothstein, and I.~W. Stewart, {\it
  {Hard scattering factorization from effective field theory}},  {\em Phys.
  Rev.} {\bf D66} (2002) 014017,
  [\href{http://arxiv.org/abs/hep-ph/0202088}{{\tt hep-ph/0202088}}].

\bibitem{Manohar:2003vb}
A.~V. Manohar, {\it {Deep inelastic scattering as x --> 1 using soft-collinear
  effective theory}},  {\em Phys. Rev.} {\bf D68} (2003) 114019,
  [\href{http://arxiv.org/abs/hep-ph/0309176}{{\tt hep-ph/0309176}}].

\bibitem{Becher:2006mr}
T.~Becher, M.~Neubert, and B.~D. Pecjak, {\it {Factorization and momentum-space
  resummation in deep- inelastic scattering}},  {\em JHEP} {\bf 01} (2007) 076,
  [\href{http://arxiv.org/abs/hep-ph/0607228}{{\tt hep-ph/0607228}}].

\bibitem{Fleming:2007xt}
S.~Fleming, A.~H. Hoang, S.~Mantry, and I.~W. Stewart, {\it {Top Jets in the
  Peak Region: Factorization Analysis with NLL Resummation}},  {\em Phys. Rev.}
  {\bf D77} (2008) 114003, [\href{http://arxiv.org/abs/0711.2079}{{\tt
  arXiv:0711.2079}}].

\bibitem{Bauer:2008dt}
C.~W. Bauer, S.~P. Fleming, C.~Lee, and G.~F. Sterman, {\it {Factorization of
  e+e- Event Shape Distributions with Hadronic Final States in Soft Collinear
  Effective Theory}},  {\em Phys. Rev.} {\bf D78} (2008) 034027,
  [\href{http://arxiv.org/abs/0801.4569}{{\tt arXiv:0801.4569}}].

\bibitem{Becher:2009th}
T.~Becher and M.~D. Schwartz, {\it {Direct photon production with effective
  field theory}},  {\em JHEP} {\bf 02} (2010) 040,
  [\href{http://arxiv.org/abs/0911.0681}{{\tt arXiv:0911.0681}}].

\bibitem{Mantry:2009qz}
S.~Mantry and F.~Petriello, {\it {Factorization and Resummation of Higgs Boson
  Differential Distributions in Soft-Collinear Effective Theory}},  {\em Phys.
  Rev.} {\bf D81} (2010) 093007, [\href{http://arxiv.org/abs/0911.4135}{{\tt
  arXiv:0911.4135}}].

\bibitem{Stewart:2009yx}
I.~W. Stewart, F.~J. Tackmann, and W.~J. Waalewijn, {\it {Factorization at the
  LHC: From PDFs to Initial State Jets}},  {\em Phys. Rev.} {\bf D81} (2010)
  094035, [\href{http://arxiv.org/abs/0910.0467}{{\tt arXiv:0910.0467}}].

\bibitem{Fleming:2007qr}
S.~Fleming, A.~H. Hoang, S.~Mantry, and I.~W. Stewart, {\it {Jets from massive
  unstable particles: Top-mass determination}},  {\em Phys. Rev.} {\bf D77}
  (2008) 074010, [\href{http://arxiv.org/abs/hep-ph/0703207}{{\tt
  hep-ph/0703207}}].

\bibitem{Becher:2007ty}
T.~Becher, M.~Neubert, and G.~Xu, {\it {Dynamical Threshold Enhancement and
  Resummation in Drell- Yan Production}},  {\em JHEP} {\bf 07} (2008) 030,
  [\href{http://arxiv.org/abs/0710.0680}{{\tt arXiv:0710.0680}}].

\bibitem{Becher:2008cf}
T.~Becher and M.~D. Schwartz, {\it {A Precise determination of $\alpha_s$ from
  LEP thrust data using effective field theory}},  {\em JHEP} {\bf 07} (2008)
  034, [\href{http://arxiv.org/abs/0803.0342}{{\tt arXiv:0803.0342}}].

\bibitem{Ahrens:2008nc}
V.~Ahrens, T.~Becher, M.~Neubert, and L.~L. Yang, {\it {Renormalization-Group
  Improved Prediction for Higgs Production at Hadron Colliders}},  {\em Eur.
  Phys. J.} {\bf C62} (2009) 333--353,
  [\href{http://arxiv.org/abs/0809.4283}{{\tt arXiv:0809.4283}}].

\bibitem{Hornig:2009vb}
A.~Hornig, C.~Lee, and G.~Ovanesyan, {\it {Effective Predictions of Event
  Shapes: Factorized, Resummed, and Gapped Angularity Distributions}},  {\em
  JHEP} {\bf 05} (2009) 122, [\href{http://arxiv.org/abs/0901.3780}{{\tt
  arXiv:0901.3780}}].

\bibitem{Vitev:2008rz}
I.~Vitev, S.~Wicks, and B.-W. Zhang, {\it {A theory of jet shapes and cross
  sections: from hadrons to nuclei}},  {\em JHEP} {\bf 11} (2008) 093,
  [\href{http://arxiv.org/abs/0810.2807}{{\tt arXiv:0810.2807}}].

\bibitem{Vitev:2009rd}
I.~Vitev and B.-W. Zhang, {\it {Jet tomography of high-energy nucleus-nucleus
  collisions at next-to-leading order}},  {\em Phys. Rev. Lett.} {\bf 104}
  (2010) 132001, [\href{http://arxiv.org/abs/0910.1090}{{\tt
  arXiv:0910.1090}}].

\bibitem{Renk:2009hv}
T.~Renk, {\it {Medium-modified Jet Shapes and other Jet Observables from
  in-medium Parton Shower Evolution}},  {\em Phys. Rev.} {\bf C80} (2009)
  044904, [\href{http://arxiv.org/abs/0906.3397}{{\tt arXiv:0906.3397}}].

\bibitem{Neufeld:2010fj}
R.~B. Neufeld, I.~Vitev, and B.~W. Zhang, {\it {The physics of
  $Z^0/\gamma^*$-tagged jets at the LHC}},
  \href{http://arxiv.org/abs/1006.2389}{{\tt arXiv:1006.2389}}.

\bibitem{Salur:2009vz}
S.~Salur, {\it {Full Jet Reconstruction in Heavy Ion Collisions}},  {\em Nucl.
  Phys.} {\bf A830} (2009) 139c--146c,
  [\href{http://arxiv.org/abs/0907.4536}{{\tt arXiv:0907.4536}}].

\bibitem{Lai:2009zq}
{\bf PHENIX} Collaboration, Y.-S. Lai, {\it {Probing medium-induced energy loss
  with direct jet reconstruction in p+p and Cu+Cu collisions at PHENIX}},  {\em
  Nucl. Phys.} {\bf A830} (2009) 251c--254c,
  [\href{http://arxiv.org/abs/0907.4725}{{\tt arXiv:0907.4725}}].

\bibitem{Ploskon:2009zd}
{\bf STAR} Collaboration, M.~Ploskon, {\it {Inclusive cross section and
  correlations of fully reconstructed jets in 200 GEV Au+Au and p+p
  collisions}},  {\em Nucl. Phys.} {\bf A830} (2009) 255c--258c,
  [\href{http://arxiv.org/abs/0908.1799}{{\tt arXiv:0908.1799}}].

\bibitem{Aamodt:2010jd}
{\bf ALICE Collaboration} Collaboration, K.~Aamodt {\em et~al.}, {\it
  {Suppression of Charged Particle Production at Large Transverse Momentum in
  Central Pb--Pb Collisions at $\sqrt{s_{_{NN}}} = 2.76$ TeV}},  {\em
  Phys.Lett.} {\bf B696} (2011) 30--39,
  [\href{http://arxiv.org/abs/1012.1004}{{\tt arXiv:1012.1004}}]. * Temporary
  entry *.

\bibitem{Chatrchyan:2011sx}
{\bf CMS Collaboration} Collaboration, S.~Chatrchyan {\em et~al.}, {\it
  {Observation and studies of jet quenching in PbPb collisions at
  nucleon-nucleon center-of-mass energy = 2.76 TeV}},
  \href{http://arxiv.org/abs/1102.1957}{{\tt arXiv:1102.1957}}. * Temporary
  entry *.

\bibitem{Aad:2010bu}
{\bf Atlas Collaboration} Collaboration, G.~Aad {\em et~al.}, {\it {Observation
  of a Centrality-Dependent Dijet Asymmetry in Lead-Lead Collisions at
  sqrt(S(NN))= 2.76 TeV with the ATLAS Detector at the LHC}},
  \href{http://arxiv.org/abs/1011.6182}{{\tt arXiv:1011.6182}}. * Temporary
  entry *.

\bibitem{Idilbi:2008vm}
A.~Idilbi and A.~Majumder, {\it {Extending Soft-Collinear-Effective-Theory to
  describe hard jets in dense QCD media}},  {\em Phys. Rev.} {\bf D80} (2009)
  054022, [\href{http://arxiv.org/abs/0808.1087}{{\tt arXiv:0808.1087}}].

\bibitem{DEramo:2010ak}
F.~D'Eramo, H.~Liu, and K.~Rajagopal, {\it {Transverse Momentum Broadening and
  the Jet Quenching Parameter, Redux}},
  \href{http://arxiv.org/abs/1006.1367}{{\tt arXiv:1006.1367}}.

\bibitem{Gyulassy:2002yv}
M.~Gyulassy, P.~Levai, and I.~Vitev, {\it {Reaction operator approach to
  multiple elastic scatterings}},  {\em Phys. Rev.} {\bf D66} (2002) 014005,
  [\href{http://arxiv.org/abs/nucl-th/0201078}{{\tt nucl-th/0201078}}].

\bibitem{Qiu:2003pm}
J.-w. Qiu and I.~Vitev, {\it {Transverse momentum diffusion and broadening of
  the back- to-back di-hadron correlation function}},  {\em Phys. Lett.} {\bf
  B570} (2003) 161--170, [\href{http://arxiv.org/abs/nucl-th/0306039}{{\tt
  nucl-th/0306039}}].

\bibitem{Baier:2000mf}
R.~Baier, D.~Schiff, and B.~G. Zakharov, {\it {Energy loss in perturbative
  QCD}},  {\em Ann. Rev. Nucl. Part. Sci.} {\bf 50} (2000) 37--69,
  [\href{http://arxiv.org/abs/hep-ph/0002198}{{\tt hep-ph/0002198}}].

\bibitem{Gyulassy:2000er}
M.~Gyulassy, P.~Levai, and I.~Vitev, {\it {Reaction operator approach to
  non-Abelian energy loss}},  {\em Nucl. Phys.} {\bf B594} (2001) 371--419,
  [\href{http://arxiv.org/abs/nucl-th/0006010}{{\tt nucl-th/0006010}}].

\bibitem{Wang:2001ifa}
X.-N. Wang and X.-f. Guo, {\it {Multiple parton scattering in nuclei: Parton
  energy loss}},  {\em Nucl. Phys.} {\bf A696} (2001) 788--832,
  [\href{http://arxiv.org/abs/hep-ph/0102230}{{\tt hep-ph/0102230}}].

\bibitem{Arnold:2002ja}
P.~B. Arnold, G.~D. Moore, and L.~G. Yaffe, {\it {Photon and Gluon Emission in
  Relativistic Plasmas}},  {\em JHEP} {\bf 06} (2002) 030,
  [\href{http://arxiv.org/abs/hep-ph/0204343}{{\tt hep-ph/0204343}}].

\bibitem{Djordjevic:2009cr}
M.~Djordjevic, {\it {Theoretical formalism of radiative jet energy loss in a
  finite size dynamical QCD medium}},  {\em Phys.Rev.} {\bf C80} (2009) 064909,
  [\href{http://arxiv.org/abs/0903.4591}{{\tt arXiv:0903.4591}}].

\bibitem{StewartTalk}
I.~Stewart and I.~Rothstein, {\it Talk presented at scet2010 in ringberg,
  germany}, .

\bibitem{Luke:1999kz}
M.~E. Luke, A.~V. Manohar, and I.~Z. Rothstein, {\it {Renormalization group
  scaling in nonrelativistic QCD}},  {\em Phys. Rev.} {\bf D61} (2000) 074025,
  [\href{http://arxiv.org/abs/hep-ph/9910209}{{\tt hep-ph/9910209}}].

\bibitem{Bauer:2010cc}
C.~W. Bauer, B.~O. Lange, and G.~Ovanesyan, {\it {On Glauber modes in
  Soft-Collinear Effective Theory}},
  \href{http://arxiv.org/abs/1010.1027}{{\tt arXiv:1010.1027}}.

\bibitem{Collins:1982wa}
J.~C. Collins, D.~E. Soper, and G.~Sterman, {\it {FACTORIZATION FOR ONE LOOP
  CORRECTIONS IN THE DRELL-YAN PROCESS}},  {\em Nucl. Phys.} {\bf B223} (1983)
  381.

\bibitem{Bodwin:1984hc}
G.~T. Bodwin, {\it {Factorization of the Drell-Yan Cross-Section in
  Perturbation Theory}},  {\em Phys. Rev.} {\bf D31} (1985) 2616 [Erratum:
  Phys. Rev. D 34 (Dec, 1986), 3932].

\bibitem{Collins:1985ue}
J.~C. Collins, D.~E. Soper, and G.~F. Sterman, {\it {Factorization for Short
  Distance Hadron - Hadron Scattering}},  {\em Nucl. Phys.} {\bf B261} (1985)
  104.

\bibitem{Abbott:1981ke}
L.~F. Abbott, {\it {Introduction to the Background Field Method}},  {\em Acta
  Phys. Polon.} {\bf B13} (1982) 33.

\bibitem{Vitev:2007ve}
I.~Vitev, {\it {Non-Abelian energy loss in cold nuclear matter}},  {\em
  Phys.Rev.} {\bf C75} (2007) 064906,
  [\href{http://arxiv.org/abs/hep-ph/0703002}{{\tt hep-ph/0703002}}].

\bibitem{Baumgart:2010qf}
M.~Baumgart, C.~Marcantonini, and I.~W. Stewart, {\it {Parton Shower with NLO
  Kinematic Power Corrections}},  \href{http://arxiv.org/abs/1007.0758}{{\tt
  arXiv:1007.0758}}.

\bibitem{Bauer:2008qu}
C.~W. Bauer, O.~Cata, and G.~Ovanesyan, {\it {On different ways to quantize
  Soft-Collinear Effective Theory}},
  \href{http://arxiv.org/abs/0809.1099}{{\tt arXiv:0809.1099}}.

\bibitem{Idilbi:2010im}
A.~Idilbi and I.~Scimemi, {\it {Singular and Regular Gauges in Soft Collinear
  Effective Theory: The Introduction of the New Wilson Line T}},
  \href{http://arxiv.org/abs/1009.2776}{{\tt arXiv:1009.2776}}. * Temporary
  entry *.

\bibitem{Vitev:2008vk}
I.~Vitev and B.-W. Zhang, {\it {A Systematic study of direct photon production
  in heavy ion collisions}},  {\em Phys.Lett.} {\bf B669} (2008) 337--344,
  [\href{http://arxiv.org/abs/0804.3805}{{\tt arXiv:0804.3805}}].

\bibitem{Sharma:2009hn}
R.~Sharma, I.~Vitev, and B.-W. Zhang, {\it {Light-cone wave function approach
  to open heavy flavor dynamics in QCD matter}},  {\em Phys.Rev.} {\bf C80}
  (2009) 054902, [\href{http://arxiv.org/abs/0904.0032}{{\tt
  arXiv:0904.0032}}].

\bibitem{Zhang:2003yn}
B.-W. Zhang and X.-N. Wang, {\it {Multiple parton scattering in nuclei: Beyond
  helicity amplitude approximation}},  {\em Nucl.Phys.} {\bf A720} (2003)
  429--451, [\href{http://arxiv.org/abs/hep-ph/0301195}{{\tt hep-ph/0301195}}].

\bibitem{Neufeld:2010dz}
R.~Neufeld, I.~Vitev, and B.-W. Zhang, {\it {Toward a determination of the
  shortest radiation length in nature}},
  \href{http://arxiv.org/abs/1010.3708}{{\tt arXiv:1010.3708}}.

\bibitem{Vitev:2010ci}
I.~Vitev, {\it {A brief overview of fixed-order perturbative QCD calculations
  of jet production in heavy-ion collisions}},
  \href{http://arxiv.org/abs/1010.5803}{{\tt arXiv:1010.5803}}. * Temporary
  entry *.

\bibitem{Ji:2002aa}
X.-d. Ji and F.~Yuan, {\it {Parton distributions in light cone gauge: Where are
  the final state interactions?}},  {\em Phys.Lett.} {\bf B543} (2002) 66--72,
  [\href{http://arxiv.org/abs/hep-ph/0206057}{{\tt hep-ph/0206057}}].

\bibitem{Belitsky:2002sm}
A.~V. Belitsky, X.~Ji, and F.~Yuan, {\it {Final state interactions and gauge
  invariant parton distributions}},  {\em Nucl.Phys.} {\bf B656} (2003)
  165--198, [\href{http://arxiv.org/abs/hep-ph/0208038}{{\tt hep-ph/0208038}}].

\end{thebibliography}\endgroup

\end{document}